\documentclass[]{aa}
\usepackage[varg]{txfonts}
\usepackage{hyperref}
\usepackage{amsmath}
\usepackage{amssymb}
\usepackage{graphicx}
\usepackage{bm}
\usepackage[fleqn]{cases}
\usepackage{scalerel}
\bibpunct{(}{)}{;}{a}{}{,}

\renewcommand*\d{\mathop{}\!\mathrm{d}}
\newcommand{\dint}{\displaystyle\int}
\newcommand{\rhom}{\overline{\rho}}
\newcommand{\pgm}{\overline{p}}
\newcommand{\um}{\widetilde{u}}
\newcommand{\om}{\widetilde{\omega}}
\newcommand{\enm}{\widetilde{e}}
\newcommand{\fe}{F^\mathrm{(e)}}
\newcommand{\xpart}{\mathbf{x}^{\ast (k)}}
\newcommand{\vpart}{\mathcal{V}^{\ast (k)}}
\newcommand{\opart}{\omega^{\ast (k)}}
\newcommand{\ebot}{e_\mathrm{bot}}
\newcommand{\cmat}{\left( \mathcal{C}^{-1} \right)}

\begin{document}

\title{A new particle-based code for Lagrangian stochastic models applied to stellar turbulent convection}
\titlerunning{Lagrangian stochastic models applied to stellar turbulent convection}
\author{J. Philidet\inst{\ref{inst1}} \and K. Belkacem\inst{\ref{inst1}}}
\institute{LIRA, Observatoire de Paris, Université PSL, Sorbonne Université, Université Paris Cité, CY Cergy Paris Université, CNRS, 92190 Meudon, France \label{inst1}}
\date{Received 29 July 2025 / Accepted 26 October 2025}

\abstract
{The inclusion of convection in stellar evolution models, mostly based on Mixing Length Theories, lacks realism, especially near convective-radiative interfaces. Furthermore, the interaction of convection with oscillations is poorly understood, giving rise to surface effects that currently prevent us from accurately predicting seismic frequencies, and therefore from fully exploiting the asteroseismic data of low-mass stars.}
{We aim to develop a new formalism to model the one-point statistics of stellar convection, to implement it in a new numerical code, and to validate this implementation against benchmark cases.}
{This new formalism is based on Lagrangian Probability Density Function (PDF) methods, where a Fokker-Planck equation for the PDF of particle-based turbulent properties is integrated in time. The PDF equation is established so that the underlying transport equations for all first- and second-order moments of the turbulent flow are identical to the exact ones stemming from first principles. We then develop a Monte-Carlo implementation of this method, where the flow is represented by a large number of notional particles acting as realisations of the PDF. Notional particles interact with each other through the time- and space-dependent mean flow, which is estimated from the particle realisations through a scheme similar to Smoothed Particle Hydrodynamics.}
{We establish a model for the evolution of turbulent properties along Lagrangian trajectories applicable to stellar turbulent convection, with only a minimal number of physical assumptions necessary to close the system. In particular, no closure is needed for the non-linear advection terms, which are included exactly through the Lagrangian nature of formalism. The numerical implementation of this new formalism allows us to extract time-dependent maps of the statistical properties of turbulent convection in a way which is not possible in grid-based large-eddy simulations, in particular the turbulent pressure, Reynolds stress tensor, internal energy variance and convective flux.}
{}


\maketitle

\nolinenumbers

\section{Introduction}

    Convection is ubiquitous in stellar interiors. It causes large-scale motions that efficiently mix energy, angular momentum and chemical elements within stars. As such, it plays a key role in shaping the structure and evolution of the stellar interiors \citep{BohmVitense1992}.
    
    For low-mass stars exhibiting stochastically excited oscillations, the modelling of the surface is particularly crucial for inferring stellar properties through asteroseismology. Indeed, the characterisation of asteroseismic targets is heavily reliant on stellar evolution models, whose shortcomings lead to biases in predicted seismic frequencies, and therefore biases in radius, mass and age determination \citep[e.g.][]{Lebreton2014a,Lebreton2014b}. Currently, most of the limitations come from our inability to correctly model the very turbulent superficial layers of low-mass stars harbouring a convective envelope. These surface effects \citep[e.g.][]{JCD1997,Rosenthal1999} force us to only consider surface-independent seismic indicators, such as frequency separation ratios \citep{Roxburgh2003,OtiFloranes2005}, and therefore prevents us from exploiting all the information contained in the observed frequencies. This is particularly limiting in a time when asteroseismic observations, thanks to missions like CoRoT \citep{Baglin2006a,Baglin2006b}, \textit{Kepler} \citep{Borucki2010}, or the future mission PLATO \citep{Rauer2025}, have reached a level of precision that clearly highlights the limitations of theory.
    
    The inclusion of convective transport in stellar models invariably relies on Mixing Length Theories (MLT), the basics of which have been devised by \citet{BohmVitense1958} \citep[see also][]{Kippenhahn1967}. This description is founded on an overly simplified picture of convection, where convective elements are displaced upwards over a unique travel distance before depositing all their entropy in the surrounding medium. By contrast, 3D hydrodynamic simulations show the non-linear, multi-scale nature of stellar convection, not captured by Mixing-Length Theories \citep[see][for a comprehensive review]{Kupka2017}. The diffusive hypothesis behind MLT is at odds with the advective and non-local nature of actual convection. Standard MLT, in its current implementation in stellar evolution codes, also has the greatest difficulties accounting for radiative/convective interfaces: local models of convection cannot describe the penetration of convective plumes into neighbouring stably stratified radiative regions \citep{Zahn1991}. This has important implications: for example, the uncertainty on the age determination of stars with convective cores is dominated by our inability to quantify the increase in effective core size brought about by convective penetration. Prescriptions for angular momentum transport across interface layers are also challenging to derive, and the inadequacy of those implemented in stellar evolution codes is partly responsible for the overestimation of the spin-up incurred by the core of evolved stars as they ascend the Red Giant Branch \citep{Mosser2012a}.
    
    Standard MLT is also ill-suited to describe surface convection in stars harbouring convective envelopes. Because the convective, thermal and dynamical timescales all coincide in the uppermost layers of these stars, the description of convection must account for the time-dependence of the properties of the medium from which the convective elements arise. Many attempts to incorporate a time dependence in MLT have been made \citep[e.g.][]{Unno1967,Gough1977a,Balmforth1992a,Grigahcene2005}, and subsequently exploited to try to model asteroseismic surface effects \citep{Balmforth1992b,Grigahcene2012,Houdek2017,Sonoi2017}. So far, however, the success of this approach has been limited. Several mechanisms are at play in surface effects, including for example the mechanical work exerted by turbulent pressure on the oscillations, or the wave-induced modulation of the heat exchange stemming from turbulent dissipation. Different time-dependent mixing-length formalisms have provided conflicting results regarding the nature of the dominant mechanism, even in the case of the Sun \citep[see][for reviews]{Belkacem2013,Houdek2015,Samadi2015}. Furthermore, the free parameters of time-dependent MLT are hard to relate to the physics of the stellar interior, because of the phenomenological nature of the formalism. As a result, agreement between the modelled and observed surface effect in the Sun can only be obtained at the price of tuning these parameters \citep[e.g.][]{Belkacem2012,Houdek2017,Houdek2019}. Extensions of MLT to include higher-order moments of the flow variables have been developed \citep{Canuto1997,Xiong1997}, and subsequently applied to the stellar case to investigate the effect of convection on pulsation stability \citep{Xiong2000}, but this approach suffers essentially from the same limitations, and did not allow to solve the surface effect problem. We note however the recent promising results obtained by \citep{Kupka2022,Ahlborn2022} for treating interfaces in the limit of efficient convection.
    
    On the other hand, 3D Large-Eddy Simulations (LES) have been exploited with success to refine the theoretical description of stellar convection \citep[see][]{Kupka2017}. In particular, many works have been devoted to exploiting 3D LES to investigate how convection impacts oscillations. \citet{Belkacem2019} have studied the damping of acoustic modes by convection, and shown that 3D LES could help disentangle the importance of the various mechanism by which the damping can occur. \citet{Samadi2001a,Samadi2001b} have developed a formalism for the excitation of solar-like oscillations by surface convection, and used hydrodynamical simulations to show that the formalism explains the observed mode amplitudes. Conversely, \citet{Zhou2019} have used hydrodynamical simulations to assess the efficiency of acoustic mode excitation by convection throughout the Hertzsprung-Russell (HR) diagram. Also, the matter of surface effects on the frequencies have been investigated with such simulations in the case of the Sun by \citet{Schou2020}. While their approach proves very promising, it suffers from limitations as regards its use to model surface effects for a large variety of stars. For instance, the limited spatial extent of these simulations means that only a few modes are present, which complicates the study of how the different contributions to surface effects depend on frequency. The prohibitive computational cost of these simulations also prevents the creation of extensive grids of models, spanning a wide range of masses or effective temperatures. A more fundamental problem, however, is that 3D LES cannot correctly include turbulent dissipation of energy into heat \citep[e.g.][]{Kupka2017}. Indeed, the very high Reynolds numbers characterising stellar turbulent convection means that all relevant scales cannot be resolved in 3D hydrodynamical simulations. As a result, the effect of the small sub-grid scales on the large-scale motions is included in a drastically simplified manner, either by letting the numerical viscosity of the spatial discretisation scheme dominate, or by artificially enhancing dissipative effects, for example with hyperviscosity \citep{Borue1995}. Either way, 3D LES fail to correctly include turbulent dissipation. This is particularly problematic for surface effects, for which it is expected to play a major role \citep[e.g.][]{Grigahcene2005,Belkacem2019,Belkacem2021}.

    In a context where hydrodynamical LES show unavoidable limitations in their ability to deal with turbulent closure, Probability Density Function (PDF) methods have emerged in the past decades as an alternative approach \citep[e.g.][]{Pope2000}. It consists in solving a transport equation for the one-point, one-time Eulerian joint PDF of all quantities describing the dynamical, thermodynamic, and possibly chemical state of a flow. Since their first developments \citep{Dopazo1975,Janicka1979,Pope1981}, PDF methods have found a considerable echo in the engineering fluid dynamics community, especially to model turbulent combustion \citep[see][for a thorough review]{Haworth2010}, but also in atmospheric science \citep{Rodean1996}. Since the seminal work of \citet{Pope1979}, the largely predominant approach in solving PDF transport equations has consisted in implementing particle methods, where the flow is viewed in a purely Lagrangian point of view, and discretised into fluid parcels. This has led to many developments to Lagrangian stochastic models, both for incompressible turbulence \citep{Haworth1987,Pope1990,Pope1994b,Dreeben1998,VanSlooten1998,Bakosi2008} and compressible turbulence \citep{Delarue1997,Delarue1998,Das2005,Almeida2021}. These PDF methods offer strikingly compelling advantages for modelling stellar turbulent convection. In particular, the adopted Lagrangian point of view provides a natural way to include the advective transport of momentum and energy, exactly and without having to model it. This is a key advantage of Lagrangian PDF methods over traditional LES, because most of the difficulty in modelling stellar convection stems precisely from non-linear advection. While this does not entirely eliminate the need for turbulent closure, it nevertheless makes the closing procedure much easier. It also allows to include turbulent dissipation in a physical and controlled manner, by solving for the joint velocity-energy-dissipation PDF, something that LES do not allow to do.
    
    Numerical solutions designed to implement PDF methods are predominantly based on Lagrangian-particle/Eulerian-mesh methods, where solving the particle equations requires knowledge of the means of the flow (density, velocity, etc.), and a Eulerian, mesh-based, finite-volume scheme is still needed to solve the (linear) transport equations for those means \citep{Haworth2010}. However, such methods would not do in our case, because it would require modelling separately the large-scale mean flow and the small-scale turbulent fluctuations, and would therefore defeat one of the primary purposes of this work, namely studying the active impact of convection on waves. In this work, we thus make use of stand-alone particle methods instead, where the means of the flow are derived from the particle properties through filtered ensemble averages. Such an approach, based on kernel-estimation techniques, was developed by \citet{Welton1997,Welton1998}, and presents the advantage of simultaneously and consistently modelling both the mean flow quantities and their turbulent fluctuations. We had previously shown \citep{Philidet2021,Philidet2022} how this approach can be exploited to establish a physical, semi-analytical model for the impact of convection on the main properties of the global acoustic modes of a star (its frequency, i.e. the surface effects, but also its damping rate and its excitation rate). However, this was done with an oversimplified model for the PDF transport equation.
    
    The primary contributions of this work are two-fold. First, we establish a Lagrangian particle description adapted to stellar turbulent convection, in a way which exempts us from having to close the non-linear advection terms, from which stems most of the difficulty in devising MLT formulations. The analytical developments leading to this model are presented in Sect.~\ref{sec:model}. Second, we present a numerical implementation of this method based on a Monte-Carlo stand-alone particle algorithm, where a kernel estimation scheme is used to extract the mean flow quantities from the particles. This implementation is presented in Sect.~\ref{sec:numerical}, and validated against benchmark cases in Sect.~\ref{sec:benchmark}. Finally, we showcase results pertaining to the statistics of stellar convection in Sect.~\ref{sec:results}, and we offer conclusions and perspectives in Sect.~\ref{sec:conclusion}.

\section{The Lagrangian stochastic model \label{sec:model}}

The Monte-Carlo Lagrangian-particle method that we exploit to model stellar turbulent convection relies on the concept of ``notional'' particles, a large number of which can be used to describe realisations of the flow, and thus allowing to reconstruct the one-point, one-time joint PDF of all the flow variables. These notional particles evolve in time according to a set of stochastic differential equations (SDE), leading to Lagrangian stochastic models for turbulence \citep[see][for reviews and books on the subject]{Pope1994a,Pope2000,Haworth2010}. In this section, we establish the SDE describing turbulent convection in the stellar context, and outline the main physical assumptions involved in the derivation. Details of the derivation are given in App.~\ref{app:lagrangian_model}.

\subsection{General framework}

    Since we wish to model the joint velocity-energy-dissipation PDF of the flow, each notional particle must be described by their position $\mathbf{x}^\ast$, velocity $\mathbf{u}^\ast$, internal energy $e^\ast$ and turbulent dissipation rate $\epsilon^\ast$ (the latter being replaced by the turbulent frequency $\omega^\ast \equiv \epsilon^\ast / k$, where $k$ is the turbulent kinetic energy). Each of these properties evolve in time according to SDEs that, in their most general form, can be expressed as \citep[e.g.][]{Pope2000,Heinz2003}
    \begin{align}
        & \d \mathbf{x}^\ast = \mathbf{u}^\ast ~ \d t ~, \label{eq:SDE_position} \\
        & \d \mathbf{u}^\ast = \mathbf{a_u}\left( \mathbf{x}^\ast, t; \mathbf{u}^\ast, e^\ast, \omega^\ast \right) ~ \d t + b_u\left( \mathbf{x}^\ast, t; \mathbf{u}^\ast, e^\ast, \omega^\ast \right) \d \mathbf{W_u} ~, \label{eq:SDE_velocity} \\
        & \d e^\ast = a_e\left( \mathbf{x}^\ast, t; \mathbf{u}^\ast, e^\ast, \omega^\ast \right) ~ \d t + b_e\left( \mathbf{x}^\ast, t; \mathbf{u}^\ast, e^\ast, \omega^\ast \right) \d W_e ~, \label{eq:SDE_energy} \\
        & \d \omega^\ast = a_\omega \left( \mathbf{x}^\ast, t; \mathbf{u}^\ast, e^\ast, \omega^\ast \right) ~ \d t + b_\omega \left( \mathbf{x}^\ast, t; \mathbf{u}^\ast, e^\ast, \omega^\ast \right) \d W_\omega \label{eq:SDE_frequency} ~,
    \end{align}
    each of which has a deterministic part (the first term on the right-hand side, $a_X$ with $X=\{u,e,\omega\}$), called the drift term, and a stochastic part (the second term on the right-hand side, $b_X$), called the diffusion term. The quantities $\d W_X$ are increments over the time step $\d t$ of mutually independent Wiener processes, meaning that
    \begin{align*}
        & \overline{W_i(t)} = 0~, \\
        & \overline{W_i(t) W_j(t+\d t)} = \delta(\d t) ~ \delta_{ij}~,
    \end{align*}
    where $\delta$ is the Dirac distribution, $\delta_{ij}$ the Kronecker symbol, and the notation $\overline{X}$ refers to the ensemble average of $X$.

    \subsection{Velocity and energy equations}
        
        The derivation of the drift and diffusion coefficients is detailed in App.~\ref{app:lagrangian_model}. The main idea is ensure that the exact Fokker-Planck equation derived from the equations of hydrodynamics is equivalent to the Fokker-Planck equation derived from Eqs.~\ref{eq:SDE_position}, \ref{eq:SDE_velocity}, \ref{eq:SDE_energy} and \ref{eq:SDE_frequency}. This procedure is relatively standard for obtaining Lagrangian stochastic models, and has been the subject of considerable efforts by the fluid dynamics community \citep{Haworth2010}. In this work, we adapt this method to stellar convection, where two physical ingredients are particularly important, namely (i) the compressibility of the flow, and (ii) the convective flux which plays a central role in buoyancy-driven convection. The non-linear advection terms in the momentum and energy equations are modelled exactly, without having to be included explicitly by hand. This is one of the key advantages of this class of models over MLT. Other non-linear turbulent terms, however, do require to be closed. As described in Sect.~\ref{app:lagrangian_model_velocity_energy}, a number of assumption are required to obtained a tractable model. The main hypotheses are as follows: \textbf{(1)} that the fluctuations of density $\rho'$, gas pressure $p'$ and specific internal energy $e''$ are related to each other through a polytropic relation given by Eq.~\ref{eq:polytropic_relation}, characterised by a polytropic index $n$, \textbf{(2)} that the compressibility of the turbulent velocity is uncorrelated with other turbulent fluctuations, so that $\overline{p' \partial u_i'' / \partial x_i} = \overline{e'' \partial u_i'' / \partial x_i} = \overline{p' e'' \partial u_i'' / \partial x_i} = 0$, and \textbf{(3)} that the effect of the viscous stress tensor $\sigma_{ij}$ can be neglected, except in the term $\overline{\sigma_{ik}' \partial u_j'' / \partial x_k} = \rhom \epsilon / 3 ~ \delta_{ij}$. Under these assumptions, the drift and diffusion coefficients in the velocity and energy equations read (see Sect.~\ref{app:lagrangian_model_velocity_energy})
        \begin{align}
            & a_{u,i} = G_{0i} + G_{ij} \left( u_j^\ast - \widetilde{u_j} \right) + G_{ei} \left( e^\ast - \enm \right) ~, \label{eq:drift_velocity} \\
            & a_e = K_0 + K_j \left( u_j^\ast - \widetilde{u_j} \right) + K_e \left( e^\ast - \enm \right) ~, \label{eq:drift_energy} \\
            & b_u = \left( C_0 \omega^\ast k \right)^{1/2} ~, \label{eq:diffusion_velocity} \\
            & b_e = \left( C_1 \omega^\ast k_e \right)^{1/2} ~, \label{eq:diffusion_energy}
        \end{align}
        where
        \begin{equation*}
            G_{0i} = -\dfrac{1}{\rhom} \dfrac{\partial \pgm}{\partial x_i} + g_i ~,
        \end{equation*}
        \begin{equation*}
            G_{ij} = -\left( \dfrac{1}{2} + \dfrac{3}{4} C_0 \right) \om \delta_{ij} - \dfrac{1}{2} C_0 \left( \omega^\ast - \om \right) k \cmat_{ij} ~,
        \end{equation*}
        \begin{equation*}
            G_{ei} = \dfrac{\Gamma_1 - 1}{n - 1} \dfrac{1}{\pgm} \dfrac{\partial \pgm}{\partial x_i} - \dfrac{1}{2} C_0 \left( \omega^\ast - \om \right) k \cmat_{i0} ~,
        \end{equation*}
        \begin{multline}
            K_0 = -\dfrac{\pgm}{\rhom} \dfrac{\partial \um_i}{\partial x_i} - \dfrac{\Gamma_1 - 1}{n - 1} \left( \dfrac{1}{\rhom} \dfrac{\partial \rhom \fe_i}{\partial x_i} - \dfrac{\fe_i}{\pgm} \dfrac{\partial \pgm}{\partial x_i} \right) \\
            + \kappa \dfrac{\partial^2 \enm}{\partial x_i \partial x_i} + \om k ~,
            \label{eq:K0}
        \end{multline}
        \begin{equation*}
            K_i = -\dfrac{1}{2} C_1 \left( \omega^\ast - \om \right) k_e \cmat_{0i} ~,
        \end{equation*}
        \begin{multline*}
            K_e = -\left( \dfrac{1}{2} C_1 \om + (\Gamma_1 - 1) \dfrac{\partial \um_i}{\partial x_i} + \dfrac{\om k}{\enm (n-1)} \right) \\
            - \dfrac{1}{2} C_1 \left( \omega^\ast - \om \right) k_e \cmat_{00}~,
        \end{multline*}
        the notation $\widetilde{X} \equiv \overline{\rho X} / \rhom$ refers to the density-weighted (or Favre) ensemble average of $X$, $\rhom$ and $\pgm$ are the Reynolds average of the density and gas pressure, $\mathbf{g}$ is the gravitational acceleration (considered constant and uniform), $\Gamma_1$ is the first adiabatic exponent, $n$ is the polytropic index, $\kappa$ is the radiative diffusivity, and the coefficients of the $4 \times 4$ covariance matrix $\mathcal{C}$ are defined by Eq.~\ref{eq:covariance_matrix} (the index $0$ refers to internal energy, and each of the indices $i = 1,2,3$ to the three components of the velocity). The mean pressure is given by the ensemble average of the ideal gas law
        \begin{equation*}
            \pgm = (\Gamma_1 - 1) \rhom \, \enm ~.
        \end{equation*}
        We emphasise that all averages depend on time, as well as the position of the particle. We also introduced the second order moments characterising the velocity-energy turbulent fluctuations
        \begin{align*}
            & R_{ij} \equiv \widetilde{u_i'' u_j''} ~, \qquad \mathrm{where} ~ \mathbf{u}'' \equiv \mathbf{u} - \mathbf{\um} ~, \\
            & k \equiv R_{ii} / 2 ~, \\
            & k_e \equiv \widetilde{e''^2} ~, \qquad \mathrm{where} ~ e'' \equiv e - \enm ~, \\
            & \mathbf{\fe} \equiv \widetilde{e'' \mathbf{u}''} ~,
        \end{align*}
        which are respectively the Reynolds stress tensor, the turbulent kinetic energy, the internal energy variance, and the internal energy flux. Repeated indices correspond to Einstein summation.

        The constant $C_0$ appearing in Eq.~\ref{eq:diffusion_velocity} is the Kolmogorov constant, the universality of which has been widely discussed and investigated \citep[see][for a discussion]{Heinz2003}. It is now understood that the value that should be adopted for $C_0$ is model-dependent, and may only be estimated by comparing model results with DNS or experimental data obtained in similar conditions. For the model considered here, the value is $C_0 = 3.5$ \citep{Pope1991}, and is found to be rather independent of the Reynolds number. Since the constant $C_1$ plays the same role as $C_0$, but for the specific internal energy, we adopted the same value $C_1 = 3.5$.

    \subsection{Turbulent frequency equation}

        A stochastic model is added for the turbulent frequency of the notional particles, which allows to include the effect of internal intermittency on the turbulent time scale of velocity and energy fluctuations. While early models were constructed which led to the experimentally observed log-normal distribution of turbulent dissipation \citep[e.g.][]{Pope1990,Pope1991}, these are computationally expensive because of the long tails of the distribution. An alternative has been developed by \citet{VanSlooten1998}, where the coefficients in Eq.~\ref{eq:SDE_frequency} are given by (see App.~\ref{app:lagrangian_model_omega})
        \begin{align}
            & a_\omega = -\Omega \left( \omega^\ast - \om \right) - \om \omega^\ast S_\omega ~, \label{eq:drift_frequency} \\
            & b_\omega = \left( 2 \sigma^2 \omega^\ast \om \Omega \right)^{1/2} \label{eq:diffusion_frequency} ~,
        \end{align}
        where $\Omega$ is the conditional mean turbulent frequency \citep{VanSlooten1998}
        \begin{equation}
            \Omega \equiv C_\Omega \dfrac{\overline{\rho\omega | \omega > \om }}{\overline{\rho \left| \omega > \om \right.}} ~,
            \label{eq:conditional_turbulent_frequency}
        \end{equation}
        the fixed parameter $\sigma^2$ controls the variance of the $\omega^\ast$ distribution and $S_\omega$ is a source term for turbulent frequency.
        
        The fundamental idea behind the introduction of this conditional frequency is to include external intermittency: if the fluid only has a certain probability of being turbulent at a given position in the flow, then the appropriate inverse mean turbulent time scale is not given by $\om$ but by $\Omega$. The constant $C_\Omega$ is chosen so that $\Omega = \om$ in a fully turbulent region. For homogeneous turbulence, without the source term, the stochastic model leads to a stationary distribution in the form of a $\Gamma$ function, with mean $\om$ and variance $\sigma^2 \om^2$. In order to be consistent with DNS data for moderate Reynolds number \citep{Yeung1989}, \citet{VanSlooten1998} indicated that the adopted value for the turbulent frequency variance should be $\sigma^2 = 1/4$, in which case $C_\Omega = 0.69$. We adopt the same values in this paper.

        In their approach, the inverse turbulent time scale in the velocity equation is given by $\Omega$, which is a mean quantity, instead of $\omega^\ast$, which is a stochastic quantity. This solves the issue about long distribution tails increasing the computational cost. Here, however, we only introduce $\Omega$ as a turbulent time scale in Eq.~\ref{eq:SDE_frequency}, and we keep $\omega^\ast$ in the velocity and energy equations (see Eqs.~\ref{eq:diffusion_velocity} and \ref{eq:diffusion_energy}). This allows us to still include the effect of internal intermittency. By contrast with the experimentally verified log-normal distribution for $\omega$, the $\Gamma$ distribution has a much shorter tail, which alleviates at least part of the computational cost issue outlined above.
        
        The quantity $S_\omega$ represent a source term for the turbulent frequency. In the $k-\epsilon$ standard model of turbulence, it is given by \citep{Launder1974a}
        \begin{equation}
            S_\omega = (C_{\epsilon 2} - 1) - (C_{\epsilon 1} - 1) \dfrac{\mathcal{P}}{\om k} ~,
            \label{eq:turbulent_frequency_source}
        \end{equation}
        where $\mathcal{P}$ is the rate of production of turbulent kinetic energy. The first constant term represents the decay of turbulent energy in the absence of a source, and the second term gathers all the sources of turbulent energy. The model constants are given by $C_{\epsilon 1} = 1.45$ and $C_{\epsilon 2} = 1.92$ \citep{Launder1974b}. In incompressible turbulence, the kinetic energy production rate is given by $\mathcal{P} = 2 C_\mu k \mathcal{S} / \omega$, where $\mathcal{S}$ is the contracted shear tensor
        \begin{equation*}
            \mathcal{S} \equiv \dfrac{1}{4}\left( \dfrac{\partial \um_i}{\partial x_j} +\dfrac{\partial \widetilde{u_j}}{\partial x_i} \right) \left( \dfrac{\partial \um_i}{\partial x_j} + \dfrac{\partial \widetilde{u_j}}{\partial x_i} \right) ~,
        \end{equation*}
        and $C_\mu = 0.09$ \citep{Launder1974b}. In compressible turbulence, an extra source term due to buoyancy must be added, in the form of $\mathcal{P}_B = -\beta \widetilde{u_i'' T''} g_i$ \citep{Rodi1980}, where $\beta = -(1/\rho) (\partial \rho / \partial T)_P$ is the coefficient of thermal expansion. With our notations, the total turbulent energy rate of production reads
        \begin{equation}
            \mathcal{P} = C_\mu \dfrac{2 k}{\om} \mathcal{S} + \dfrac{\Gamma_1 - 1}{n - 1} \dfrac{1}{\pgm} \dfrac{\partial \pgm}{\partial x_i} \fe_i ~.
            \label{eq:turbulent_production}
        \end{equation}

\section{Numerical setup \label{sec:numerical}}

    In this section, we present a two-dimensional implementation of the Lagrangian stochastic model, introduced in Sect.~\ref{sec:model}, with the objective of simulating a near-surface convective region.

    \subsection{Setup, boundary conditions, and initial conditions}
    
        We only follow two components of the fluid particles position -- the component $x$ corresponding to the vertical coordinate, in the opposite direction of gravity, and the component $y$ representing one of the horizontal directions. However, all three components of the velocity are modelled, with the assumption of homogeneous turbulence in the $z$ direction. Our model is therefore 2.5-dimensional, thus increasing the realism of the simulation while keeping the computational cost manageable. The modelled variables are $x^\ast(t)$, $y^\ast(t)$, $u_x^\ast(t)$, $u_y^\ast(t)$, $u_z^\ast(t)$, $e^\ast(t)$ and $\omega^\ast(t)$ for a set of $N$ fluid parcels, for a total of $7N$ variables which depend on time $t$ only. The uniform gravitational acceleration $\bm{g}$ and radiative diffusivity $\kappa$ are fixed parameters.
        
        In this Lagrangian stochastic framework, boundary conditions are enforced at particle level, every time a fluid particle exits the domain at the end of a time step. In order to conserve the total mass, we keep the number of particles constant, which means we reintroduce as many fluid particles as the ones exiting the domain. If a particle exits the domain through the top or bottom boundary, it is reintroduced with a reversed vertical velocity $u_x^\ast$, and its vertical position $x^\ast$ is mirrored with respect to the boundary (similar to a wall boundary condition). The turbulent frequency $\omega^\ast$, horizontal velocity $u_y^\ast$ and horizontal position $y^\ast$ of the particle are not modified. The internal energy is treated differently: in order to impose a given energy flux at the bottom and top edges of the domain, a linear mean internal energy profile $\enm(x)$ is imposed in two narrow bands at the top and bottom of the domain, spanning only a few percents of the total volume of the domain. In practice, at the end of each time step, each particle located in either of those bands is assigned an internal energy $e^\ast = e^\mathrm{(bottom)} + \nabla_e x^\ast$, where $e^\mathrm{(bottom)}$ is the imposed internal energy at the bottom of the simulation, and $\nabla_e = -F^{(bottom)} / \kappa$, where $F^{(bottom)}$ is the imposed energy flux. In these two narrow bands, the internal energy equation is not solved explicitly.
        
        Periodic boundary conditions are enforced in the horizontal direction: a particle exiting through the left side of the box is reintroduced on the right side, and vice versa. The same periodic conditions are used in the averaging procedure (see Sect.~\ref{sec:SPH}): particles on the right side of the box contribute to the weighted average Eq.~\ref{eq:SPH_Reynolds_average} computed on the left side, and vice-versa, in such a way that the domain folds back on itself horizontally.
        
        For the initial state, we impose hydrostatic and radiative equilibrium, characterised by the following mean density and internal energy profiles
        \begin{align}
            & \enm(x) = e^\mathrm{(bottom)} + \nabla_e x ~, \label{eq:initial_energy} \\
            & \rhom(x) = \rho^\mathrm{(bottom)} \left(\dfrac{\enm(x)}{\ebot}\right)^{-\left(1 + \dfrac{g}{\nabla_e (\Gamma_1-1)}\right)} ~,
            \label{eq:initial_density}
        \end{align}
        where $\rho^\mathrm{(bottom)}$ is the initial mean density imposed at the bottom of the simulation. This amounts to a polytrop with index
        \begin{equation}
            n_\mathrm{init} = \left( 1 + \dfrac{\nabla_e (\Gamma_1 - 1)}{g} \right)^{-1} ~.
            \label{eq:initial_polytrop}
        \end{equation}
        We impose that $n_\mathrm{init}$ is identical to the polytropic index $n$ introduced in the Lagrangian stochastic model, which constrains the value of $\nabla_e$.
        
        The vertical position of each particle is initialised randomly according to the PDF $\rhom(x)$ given by Eq.~\ref{eq:initial_density}. The particles are also assigned a random horizontal position chosen uniformly between the two horizontal boundaries of the domain. Once the position of each particle is chosen, they are assigned a random velocity according to a normal distribution with zero mean and a variance chosen arbitrarily (this arbitrary variance defines the initial Reynolds stress tensor); a random specific energy according to a normal distribution with a mean given by Eq.~\ref{eq:initial_energy}, and an arbitrary variance (which defines the initial value of $k_e$); and a random turbulent frequency according to a $\Gamma$ distribution with shape parameter $k = 4$ and a scale parameter $\theta = \omega_\mathrm{init} / 4$, where $\omega_\mathrm{init}$ is the initial mean turbulent frequency, chosen arbitrarily. This initial turbulent frequency distribution corresponds to the solution of Eq.~\ref{eq:SDE_frequency} in the stationary, homogeneous case.
        
    \subsection{Time integration scheme}

        The $7N$ variables of the system evolve according to Eqs.~\ref{eq:SDE_position}, \ref{eq:SDE_velocity}, \ref{eq:SDE_energy} and \ref{eq:SDE_frequency}, which are stochastic differential equations of time only: one strength of this formalism is that there is no need to discretise the coordinate axes. We integrate the system forward in time using a stochastic predictor/corrector integration scheme, which is an algorithm of (weak) order $1.5$ \citep{Platen1995}. The scheme works in two steps. If we formally write the system of SDE in multivariate form $\d \mathbf{X} = \mathbf{a}(\mathbf{X}, t) \d t + \mathcal{B}(\mathbf{X}, t) \cdot \d \mathbf{W}$, we first predict the state of the system after one time step $\Delta t$
        \begin{equation*}
            \widehat{\mathbf{X}}_{k+1} = \mathbf{X}_k + \mathbf{a}(\mathbf{X}_k, t_k) \Delta t + \mathcal{B}(\mathbf{X}_k, t_k) \cdot \Delta \mathbf{W}_k~,
        \end{equation*}
        where $\Delta \mathbf{W}_k \equiv \mathbf{W}(t_{k+1}) - \mathbf{W}(t_k)$ is a set of independent random variables of mean $0$ and variance $\Delta t$. Then this prediction is corrected according to \citep{Platen1995} 
        \begin{multline*}
            \mathbf{X}_{k+1} = \mathbf{X}_k + \left( \alpha \mathbf{a}_\beta(\mathbf{X}_k, t_k) + (1 - \alpha) \mathbf{a}_\beta \left(\widehat{\mathbf{X}}_{k+1}, t_{k+1}\right) \right) \Delta t \\
            + \left( \vphantom{a_\beta} \beta \mathcal{B}(\mathbf{X}_k, t_k) + (1 - \beta) \mathcal{B}(\widehat{\mathbf{X}}_{k+1}, t_{k+1}) \right) \Delta W_k~.
        \end{multline*}
         The coefficients $\alpha$ and $\beta$ control the level of implicitness of the time integration scheme. They must be chosen between $0$ and $1$, with a value of $1$ corresponding to a fully explicit scheme. In this study, we consistently chose $\alpha = \beta = 1/2$. The modified drift coefficient is defined by $\mathbf{a}_\beta \equiv \mathbf{a} - \beta \mathcal{B} \cdot \left( \bm{\nabla}_\mathbf{X} : \mathcal{B} \right)$ and only differs from the actual drift coefficient if the diffusion matrix of the process explicitly depends on the stochastic process itself. In our case, this will only affect the equation on $\omega^\ast$.

         The usual Courant–Friedrichs–Lewy (CFL) stability condition is not relevant in the present scheme, since the coordinate axes are not discretised. Here, the time step is chosen based on accuracy rather than stability considerations, and we impose that the particle Courant number of each notional particle must not exceed unity. This can be written \citep*[see Eq. (106) of][]{Haworth2010}
         \begin{equation*}
             \Delta t < \mathrm{min}_{1 \leqslant k \leqslant N} \left( \dfrac{1}{\opart} ~,~ \dfrac{h_x}{\left| u_x^\ast \right|} ~,~ \dfrac{h_y}{\left| u_y^\ast \right|} ~,~ \dfrac{h_x^2 + h_y^2}{\kappa} \right) ~,
         \end{equation*}
         where $h_x$ and $h_y$ are the vertical and horizontal widths of the kernel function which is used to compute the local mean flow quantities from the set of notional particles (see Sect.~\ref{sec:SPH}).

    \subsection{Ensemble average estimation using kernels \label{sec:SPH}}

        \subsubsection{Standard kernel estimates \label{sec:standard_SPH}}

            The Lagrangian stochastic model  described by Eqs.~\ref{eq:SDE_position}, \ref{eq:SDE_velocity}, \ref{eq:SDE_energy} and \ref{eq:SDE_frequency} requires knowledge of the means (density $\rhom$, gas pressure $\pgm$, velocity $\mathbf{\um}$, internal energy $\enm$, turbulent frequency $\om$) and covariance (Reynolds stress tensor $R_{ij}$, energy variance $k_e$, energy flux $\mathbf{\fe}$) of the turbulent properties of the flow, at any given point of the domain. This information is extracted by averaging the appropriate particle-based stochastic variables over all neighbouring notional particles. To that end, as proposed by \citet{Welton1997}, we couple the Lagrangian stochastic model with a kernel estimation of the means \citep[in a fashion similar to Smoothed Particle Hydrodynamics, see][for details]{Liu2004}. 
            
            To filter local notional particles prior to averaging, we use a kernel function $K(\mathbf{r})$, where $\mathbf{r}$ is the position vector, which is centred on the location where the mean flow properties are evaluated. This function must satisfy some elementary conditions \citep[e.g.][]{Liu2004}: it must peak at $\mathbf{r} = \mathbf{0}$, decrease outwards, have a compact support, and be normalized so that its integral equals unity. We also choose -- although it is not mandatory -- a function which is symmetric, piecewise-quartic, possesses continuous first and second derivatives. Finally, with the objective of using an efficient algorithm (see below), we make the function separable so that
            \begin{equation*}
                K(\mathbf{r}) = K(r_x, h_x) ~ K(r_y, h_y) ~,
            \end{equation*}
            where $r_x$ and $r_y$ are the vertical and horizontal component of the 2D position vector $\mathbf{r}$, and $h_x$ and $h_y$ are the vertical and horizontal extents of the kernel function. The following function \citep*[also adopted by][]{Welton1998} fulfil all requirements
            \begin{subnumcases}{\label{eq:kernel_function} K(r, h) =}
                \dfrac{5}{4h} \left(1 + 3\dfrac{|r|}{h} \right) \left(1 - \dfrac{|r|}{h} \right)^3 & ~, if $r < h$ ~, \\
                0 & ~, if $r > h$ ~.
            \end{subnumcases}
    
            The ensemble average of any particle-based quantity $Q^\ast$ at any position $\mathbf{x}$ and time $t$ can then be estimated via
            \begin{equation}
                \overline{Q}(\mathbf{x}, t) = \sum_{k = 1}^N \vpart(t) ~ Q^{\ast (k)}(t) ~ K\left(\xpart(t) - \mathbf{x} ~,~ h \right)~,
                \label{eq:SPH_Reynolds_average}
            \end{equation}
            where $\vpart \equiv \Delta m / \rho^{\ast (k)}$ is the lumped volume of fluid that particle $k$ is meant to represent. The difficulty, then, is that $\rho^{\ast (k)}$ is not a modelled quantity. For that reason, the kernel estimation is much more suited to yield density-averaged means
            \begin{equation}
                \widetilde{Q}(\mathbf{x}, t) \equiv \dfrac{\overline{\rho Q}}{\overline{\rho}} = \dfrac{1}{\overline{\rho}} \sum_{k = 1}^N \Delta m ~ Q^{\ast (k)}(t) ~ K\left(\xpart(t) - \mathbf{x}, h \right)~,
                \label{eq:SPH_Favre_average}
            \end{equation}
            where the mean density is given by Eq.~\ref{eq:SPH_Reynolds_average} with $Q = \rho$
            \begin{equation*}
                \overline{\rho}(\mathbf{x}, t) = \sum_{k = 1}^N \Delta m ~ K\left(\xpart(t) - \mathbf{x}, h \right)~.
            \end{equation*}
            The gradient of $\overline{Q}$ can also be estimated using this kernel method: because the position variable $\mathbf{x}$ only appears in the kernel function $K$, we have
            \begin{equation*}
                \bm{\nabla} \overline{Q} = -\sum_{k = 1}^N \vpart(t) ~ Q^{\ast (k)}(t) ~ \bm{\nabla} K\left(\xpart(t) - \mathbf{x}, h \right)~, 
            \end{equation*}
            and higher-order spatial derivatives may be computed in the same way.
                    
            From a practical point of view, directly computing the means and their derivatives would be very time-consuming as it would be a $O(N^2)$ algorithm. To overcome this issue, we adopt the algorithm developed by \cite{Welton1998}, which mainly consists in projecting the kernel function in Fourier space. This leads to a $O(M^2 N)$ algorithm, where $N$ is the number of notional particles, and $M$ the number of Fourier modes on which the kernel is projected. In practice, we adopt a value $M = 10$, which is enough to accurately represent the kernel given by Eq.~\ref{eq:kernel_function} up to its second derivative, while leading to a drastic decrease in computation time compared to the brute-force $O(N^2)$.
        
        \subsubsection{Correcting for kernel estimation inconsistency}
        
            Any estimator for a mean field $\overline{Q}(\mathbf{x},t)$ is said to have $k$-th order consistency if it can exactly reproduce a polynomial function of the coordinates up to degree $k$. The standard formalism presented in Sect.~\ref{sec:standard_SPH} fails this condition. There are two reasons. First, the averaging kernel function $K$ is truncated by the edges of the domain: when trying to apply Eq.~\ref{eq:SPH_Reynolds_average} to a position $\mathbf{x}$ close to the boundary of the domain, the lack of particles on a certain portion of the kernel support will artificially decrease $\overline{Q}$. The second reason is that an irregular particle distribution can introduce a bias not only in $\overline{Q}$, but especially in its gradient. There are several ways of correcting the standard scheme presented in Sect.~\ref{sec:standard_SPH} so as to restore its consistency. Here, we follow \citet{Korzilius2016}, who provide a scheme with $1^\mathrm{st}$ order consistency for the means $\overline{Q}$, their gradient and their $2^\mathrm{nd}$ order derivatives. We describe this Corrective Smoothed Particle Method (CSPM) in App.~\ref{app:corrected_SPH}.

\section{Benchmarking and validation\label{sec:benchmark}}

    To validate the numerical implementation described in Sect.~\ref{sec:numerical}, we benchmarked the code against standard test cases used for particle-based hydrodynamics codes \citep{Liu2004}. These benchmark cases involve incompressible flows; in this code, the incompressible condition is enforced by setting the lumped volume $\vpart$ of each particle to the total volume of the domain divided by the total number of particles, at each time step, regardless of particle motion.

    \subsection{Deterministic cases: plane Poiseuille and Couette flows}
    
        These benchmark cases allow us to test specifically the accuracy of the kernel estimation scheme presented in Sect.~\ref{sec:SPH}.
    
        The plane Poiseuille flow is obtained by treating the top and bottom boundaries as two parallel plates with no-slip boundary conditions, and by exerting a constant and uniform body-force on the flow, parallel to the plates. We initially placed a square of $150 \times 150$ particles in the domain, regularly spaced. The particles are evolved according to the following set of deterministic differential equations
        \begin{align}
            & \d x = u_x ~ \d t ~, \label{eq:poiseuille_eq_1} \\
            & \d y = u_y ~ \d t ~, \\
            & \d u_x = \left( -\dfrac{1}{\rhom} \dfrac{\partial \pgm}{\partial x} + \nu \Delta \overline{u_x} \right) \d t ~, \\
            & \d u_y = \left( -\dfrac{1}{\rhom} \dfrac{\partial \pgm}{\partial y} + \nu \Delta \overline{u_y} + F \right) \d t ~,
            \label{eq:poiseuille_eq_2}
        \end{align}
        and the gas pressure is related to the density through $\pgm = \rhom c_s^2$. The exact solution can be expressed as an infinite series
        \begin{multline}
            u_y(x, t) = \dfrac{F x (x - L_x)}{2\nu} + \sum_{n=0}^{+\infty} \dfrac{4FL_x^2}{\nu \pi^3 (2n+1)^3} \sin\left( \dfrac{\pi x (2n+1)}{L_x} \right) \\
            \times \exp\left( -\dfrac{(2n+1)^2 \pi^2 \nu}{L_x^2} t \right) ~.
            \label{eq:exact_poiseuille}
        \end{multline}
    
\begin{figure}
            \centering
            \includegraphics[width=\linewidth]{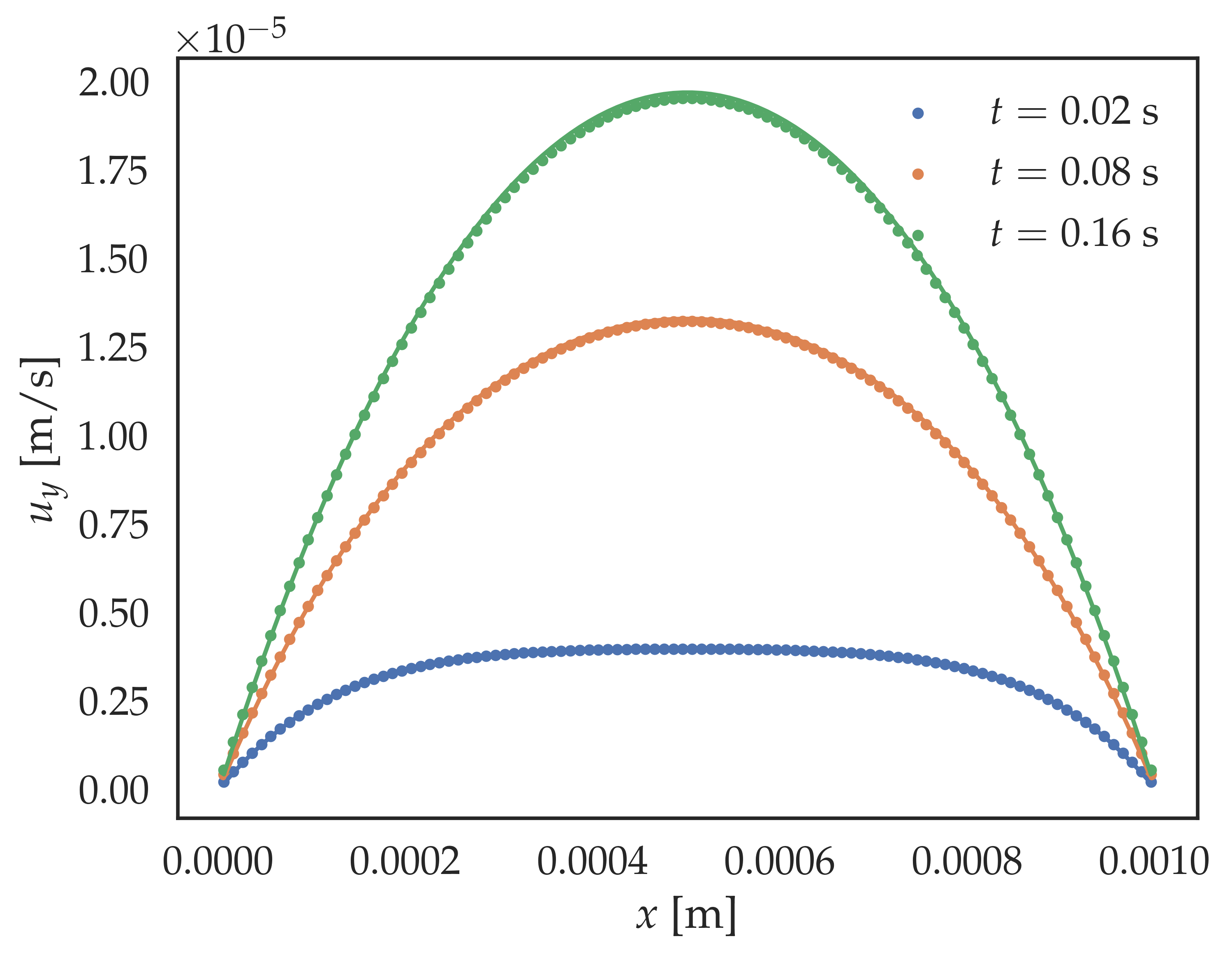}
            \caption{Incompressible Poiseuille flow benchmark case. The coloured dots show snapshots of the Poiseuille velocity profile at different times. The solid lines show the exact analytical solution given by Eq.~\ref{eq:exact_poiseuille}. Relative errors remain under $1 \%$ throughout the simulation.}
            \label{fig:poiseuille}
        \end{figure}

        Here, the code described in Sect.~\ref{sec:numerical} is implemented without the stochastic terms, thus only keeping the deterministic part of the equations in the form of Eqs.~\ref{eq:poiseuille_eq_1} to \ref{eq:poiseuille_eq_2}. The domain is chosen to be $10^{-3}$ m in height and in width. The mass of each particle is chosen so that the flow has a uniform density $\rho = 10^3$ kg/m$^2$. The kinematic viscosity is set to $\nu = 10^{-6}$ m$^2$/s, and the driving body acceleration to $F = 2 \times 10^{-4}$ m/s$^2$. We assumed a time step $\Delta t = 10^{-4}$ s, and we run the simulation for $5000$ time steps. The size of the kernel function is set to $3 \%$ of the domain size, both horizontally and vertically. Figure~\ref{fig:poiseuille} shows the horizontal velocity profile, as a function of vertical coordinate, at different times in the simulation. The departure from the exact solution, given by Eq.~\ref{eq:exact_poiseuille}, remains smaller than $1 \%$ throughout the whole simulation.
    
        The Couette flow benchmark case is similar to the Poiseuille flow, the difference being that no body-force is exerted on the flow, and the motion is driven instead by moving one of the plates at a constant velocity $v_0$. The Couette flow simulation we run has the same setup as the Poiseuille flow, except that the no-slip condition at the top boundary is replaced by a constant horizontal velocity condition. We adopted $v_0 = 2.5 \times 10^{-5}$ m/s.
        
\begin{figure}
            \centering
            \includegraphics[width=\linewidth]{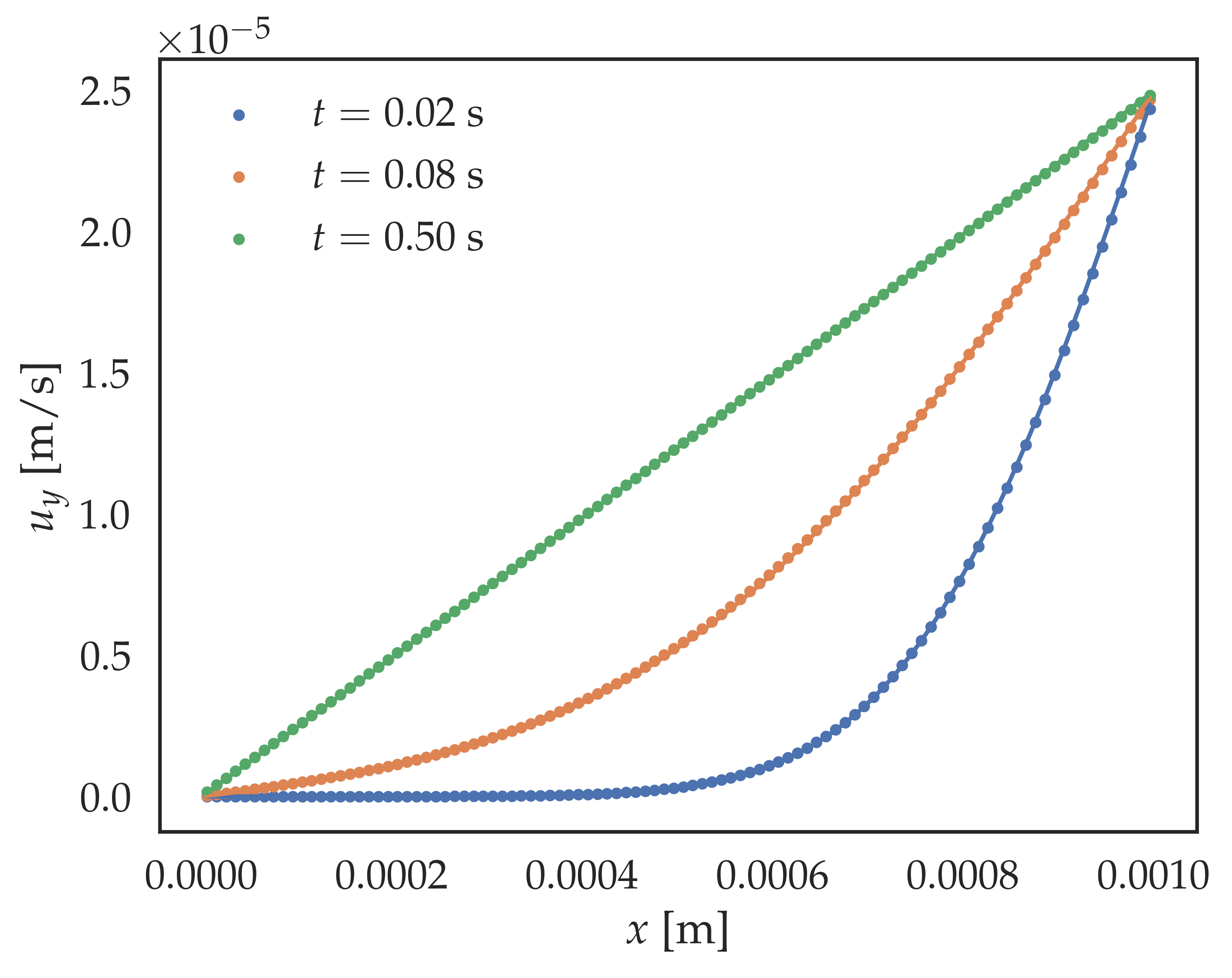}
            \caption{Incompressible Couette flow benchmark case. The symbols are the same as in Fig.~\ref{fig:poiseuille}, with the exact analytical solution being given by Eq.~\ref{eq:exact_couette}. Relative errors remain under $0.6 \%$ throughout the simulation.}
            \label{fig:couette}
        \end{figure}

        The exact solution can be expressed as an infinite series
        \begin{equation}
            u_y(x, t) = \dfrac{v_0 x}{L_x} + \sum_{n=1}^{+\infty} (-1)^n \dfrac{2 v_0}{n\pi} \sin\left( \dfrac{n \pi x}{L_x} \right) \exp\left( -\dfrac{n^2 \pi^2 \nu}{L_x^2} t \right) ~.
            \label{eq:exact_couette}
        \end{equation}
        Figure~\ref{fig:couette} shows the Couette flow horizontal velocity profile, in the same way as Fig.~\ref{fig:poiseuille} for the Poiseuille case. The departure from the exact solution, given by Eq.~\ref{eq:exact_couette}, remains smaller than $0.6 \%$ throughout the whole simulation.
        
        The topology of the plane Poiseuille and plane Couette flows is mainly controlled by the viscous force. As such, the very good agreement found in these two cases showcases not only the validity of the time integration scheme implemented in the code, but also the accuracy of the corrective kernel estimation scheme up to second spatial derivative.
    
    \subsection{Stochastic case: incompressible Simplified Langevin Model\label{sec:incomp_SLM}}
    
        In order to test the accuracy of the stochastic scheme for particle evolution along with the kernel-estimation of the means, we benchmarked the code against the incompressible Simplified Langevin Model, where the particles evolve according to the following set of stochastic differential equations
        \begin{align*}
            & \d x = u_x ~ \d t ~, \quad \d y = u_y ~ \d t ~, \\
            & \d u_\alpha = - \left( \dfrac{1}{2} + \dfrac{3}{4} C_0 \right) \omega \left( u_\alpha - \widetilde{u_\alpha} \right) \d t + \sqrt{C_0 \omega k} \d W_\alpha ~, 
        \end{align*}
        where $\alpha=\{x,y,z\}$, and $\omega$ is the fixed turbulent frequency. We recall that $\widetilde{u_\alpha}$ is the local density-averaged velocity, $k$ is the local turbulent kinetic energy, and $C_0$ is the Kolmogorov constant. This corresponds to the model presented in Sect.~\ref{sec:model}, in the limit where every particle's turbulent frequency $\omega$ and internal energy $e$ is kept constant (we set $\omega = 0.1$, and $e$ is irrelevant in this model) so that only their velocity and position is allowed to evolve. In this model, the turbulence is homogeneous, and the turbulent kinetic energy $k$ decays with time at a rate equals to $\omega$ \citep[e.g.][]{Pope2000}
        \begin{equation}
            k(t) = k_0 \exp^{-\omega t} ~,
            \label{eq:decay_law_slm}
        \end{equation}
        where $k_0$ is the initial turbulent kinetic energy at $t = 0$. We ran the simulation with $300,000$ particles, with a time step $\Delta t = 1 / (10 \omega)$, for a total duration of  $100 / \omega$. The kernel size was set to $3 \%$ of the domain size, both in the vertical and horizontal directions.
        
\begin{figure}
            \centering
            \includegraphics[width=\linewidth]{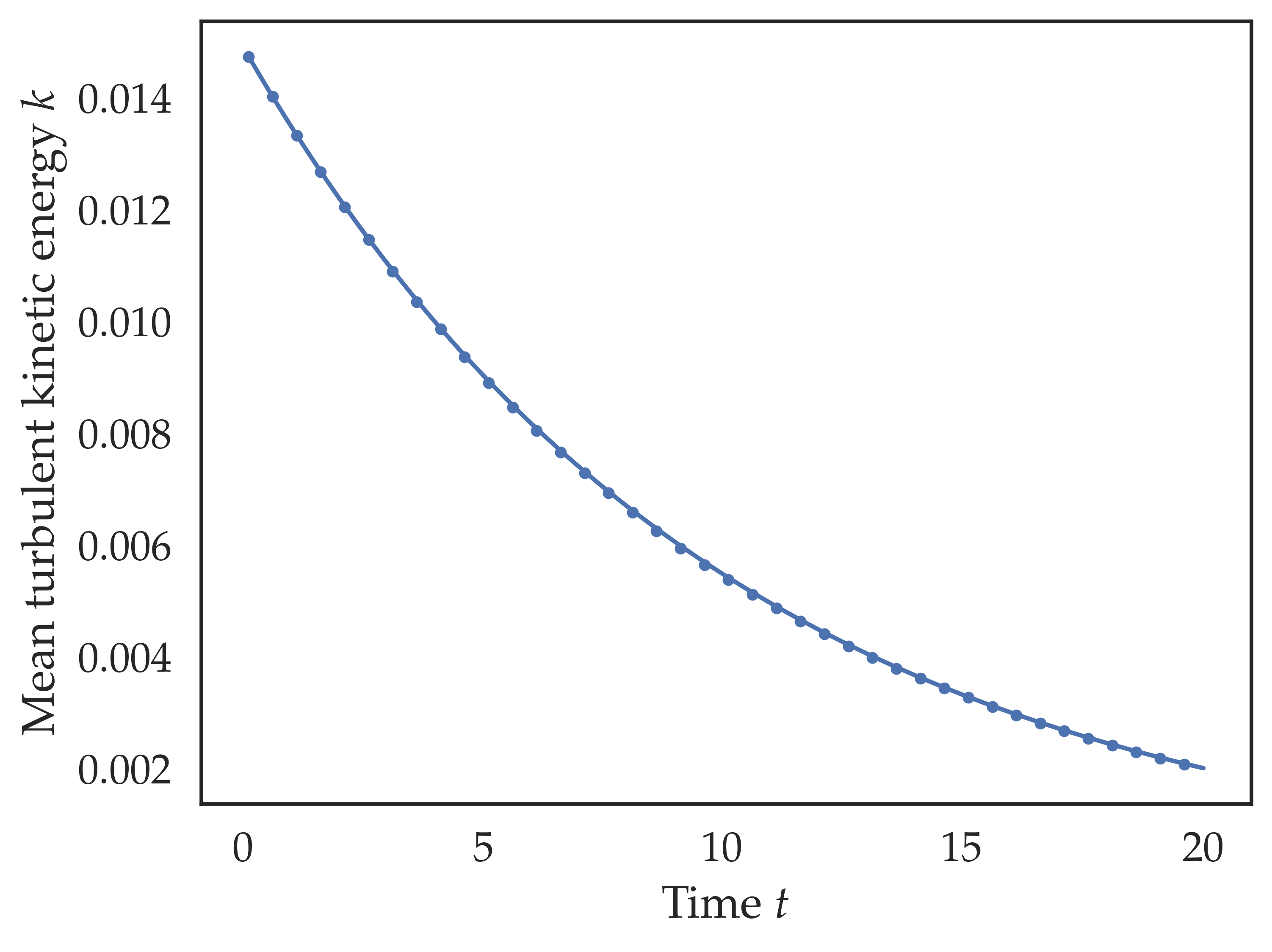}
            \caption{Evolution of the spatially averaged turbulent kinetic energy $k$ with time, for the benchmark incompressible Simplified Langevin Model case. The dots show the simulation output, while the solid line shows the analytical solution given by Eq.~\ref{eq:decay_law_slm}. Relative errors remain under $0.35 \%$ throughout the simulation.}
            \label{fig:incomp_SLM}
        \end{figure}

        We show in Fig.~\ref{fig:incomp_SLM} the evolution of the spatially averaged turbulent kinetic energy $k$ with time in the simulation output, compared to the expected result given by Eq.~\ref{eq:decay_law_slm}. The relative error never exceeds $0.35 \%$. This shows that the validity of the kernel estimation scheme extends to mean flow properties -- and especially to centered second-order moments -- in the presence of turbulence, and with the particles evolving according to stochastic equations rather than deterministic ones.
        
    \subsection{Convergence\label{sec:convergence}}

        In order to determine the appropriate number $N$ of particles needed to correctly represent the flow, as well as the appropriate kernel size $h$ and time step $\Delta t$, we ran the compressible Simplified Langevin Model case (i.e. the same model as in Sect.~\ref{sec:incomp_SLM}, but estimating mean flow density from particle positions on the fly, rather than setting a fixed value), with the same setup as in Sect.~\ref{sec:incomp_SLM}.

\begin{figure}
            \centering
            \begin{tabular}{c}
                \includegraphics[width=0.9\linewidth]{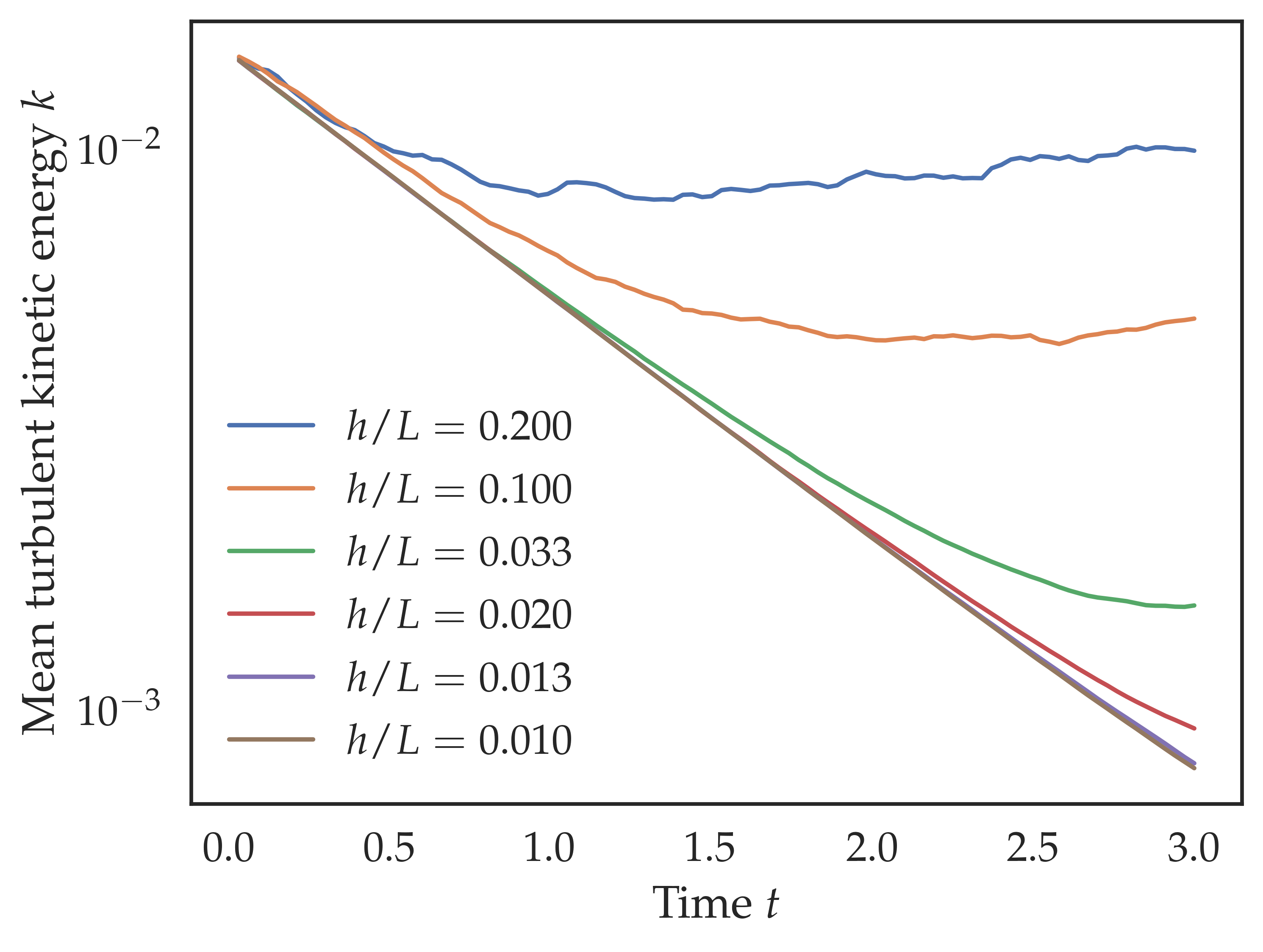} \\
                \includegraphics[width=0.9\linewidth]{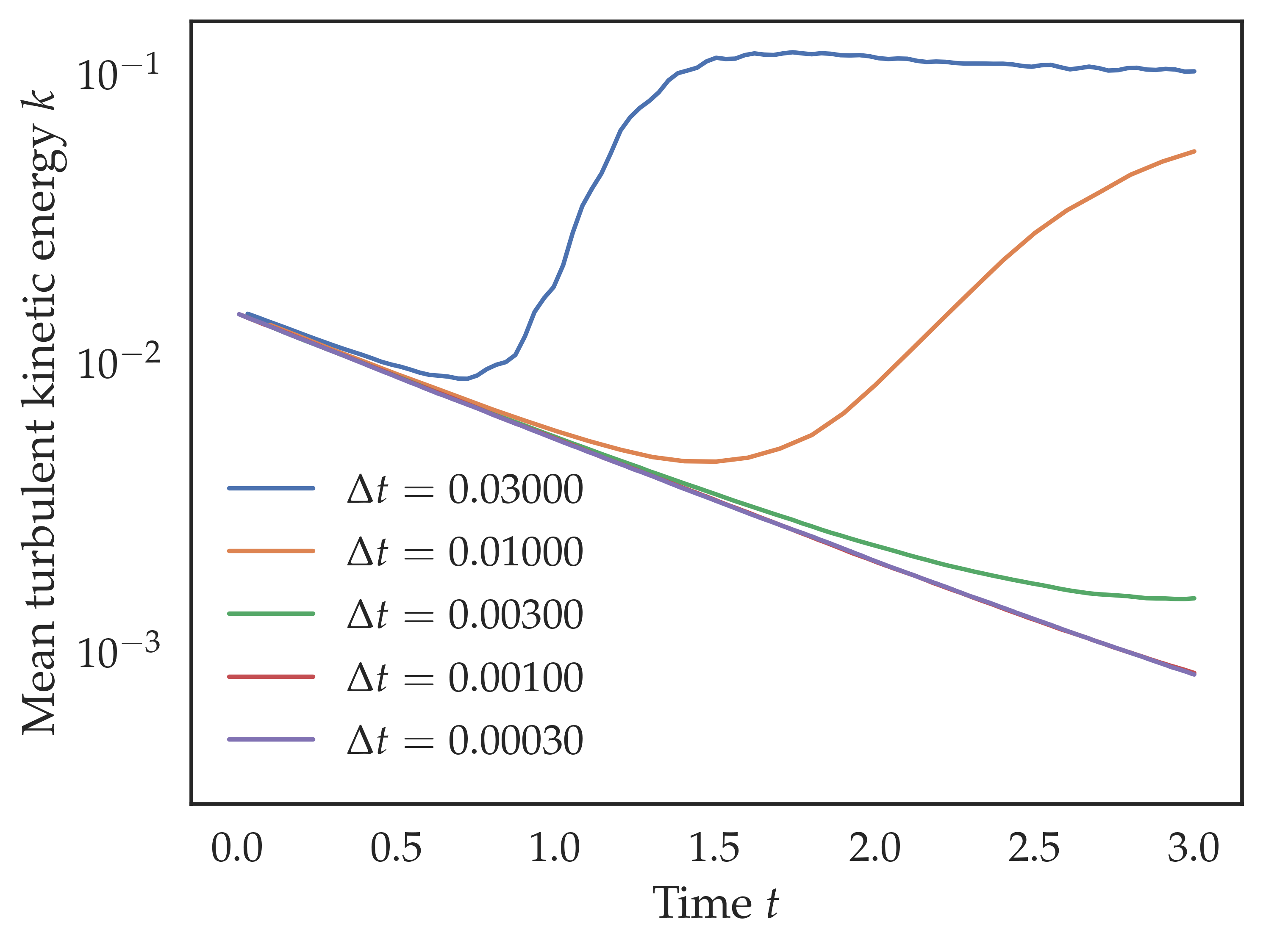} \\
                \includegraphics[width=0.9\linewidth]{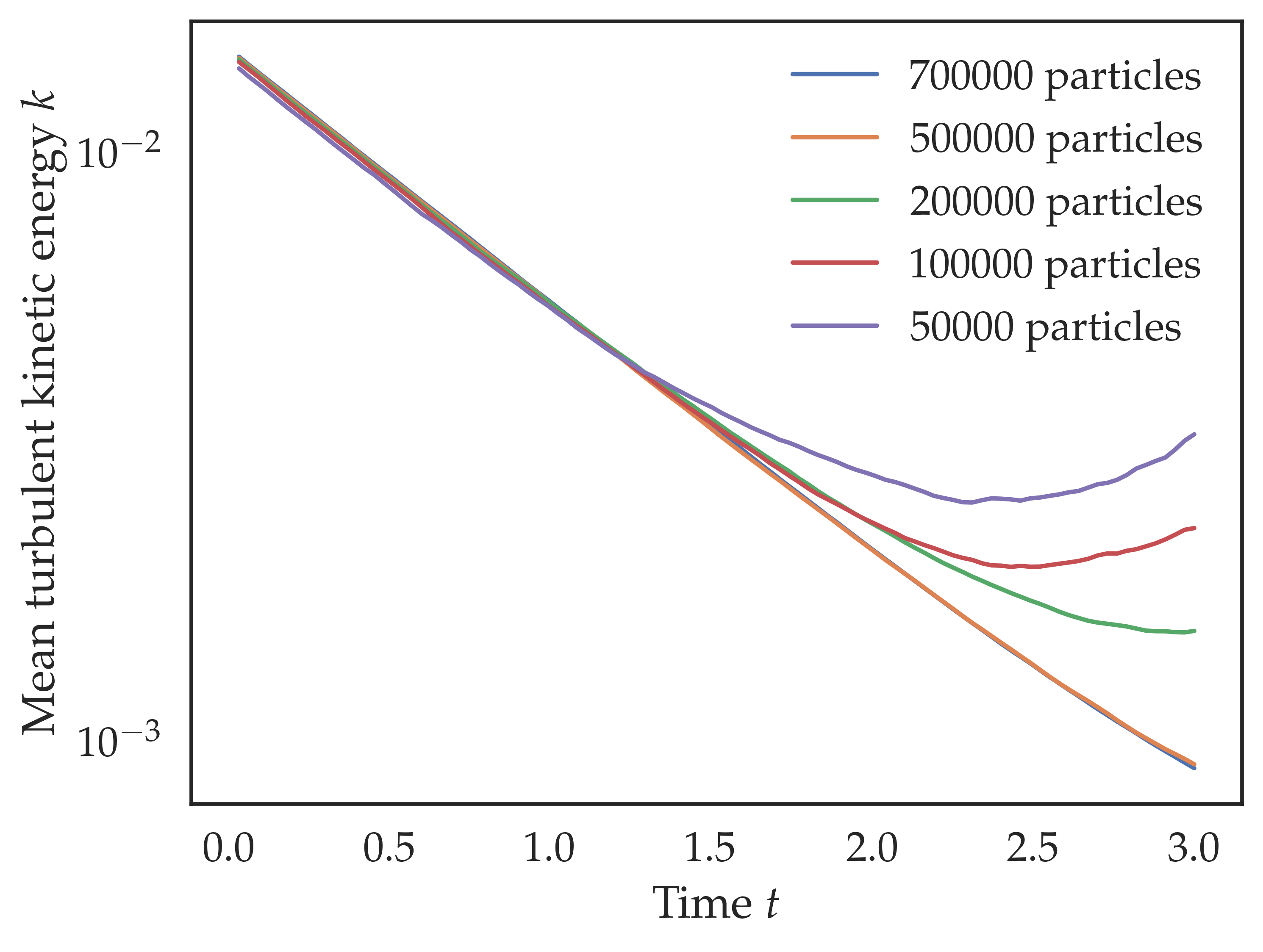}
            \end{tabular}
            \caption{Evolution of the spatially averaged turbulent kinetic energy $k$ with time, for the compressible Simplified Langevin Model case, for different run parameters. \textbf{Top:} comparison of different kernel sizes $h$. \textbf{Middle:} comparison of different time steps $\Delta t$. \textbf{Bottom:} comparison of different particle numbers $N$.}
            \label{fig:convergence}
        \end{figure}

        We first ran simulations with different kernel sizes $h$, while keeping the time step constant ($\Delta t = 0.003$), but we varied $N$ alongside $h$ in such a way that the average number of particles counting towards a kernel average is fixed. This amounts to $N \propto 1/h^2$, with a base value of $N = 200000$ for $h = L/30$, meaning that there is $\sim 220$ particles per kernel average. We then extracted the spatial average of the turbulent kinetic energy $k$, as a function of time, for each simulation. The top panel of Fig.~\ref{fig:convergence} shows that the results are independent of the kernel size if the kernel size remains smaller than about $2 \%$ of the total domain size.
        
        We did a similar comparison by varying the time step $\Delta t$, while keeping the number of particles ($N = 200,000$) and the kernel size ($h = L / 30$) fixed. The middle panel of Fig.~\ref{fig:convergence} shows that the results are independent of the time step for $\Delta t < 0.001$.
        
        Finally, ran different simulations with varying numbers of particles $N$, while keeping the time step ($\Delta t = 0.003$) and the kernel size ($h = L / 30$) fixed. The bottom panel of Fig.~\ref{fig:convergence} shows that the results are independent of the particle number for $N > 500,000$.

    Based on these results, in the rest of this study, we will adopt a kernel size of $h = L/50$, a time step of $\Delta t = 0.001$, and a particle number of $N = 500,000$.

\section{2D convectively unstable simulation\label{sec:results}}

    To showcase the capabilities of our code, we simulated a $0.45 \times 2.7$ Mm convectively unstable domain, using the Lagrangian stochastic model developed in Sect.~\ref{sec:model} and the numerical scheme described in Sect.~\ref{sec:numerical}. We considered a uniform polytropic index $n = 9$, as well as a uniform radiative diffusivity $\kappa = 23$ km$^2$/s, throughout the entire domain. The first adiabatic index is set to $\Gamma_1 = 5/3$, the gravitational acceleration to the solar surface gravity $g = 275$ m/s$^2$. Based on the results outlined in Sect.~\ref{sec:convergence}, we adopted a time step of $\Delta t = 0.001$ (in dimensionless units, corresponding to a physical time step of about $0.1$ s), the kernel size is $2 \%$ of the domain size in both directions, and we used $500,000$ particles to represent the flow.
    
    It should be noted that in mesh-free, particle methods, the choice of the kernel size $h$ is dictated not by requirements of stability (as is the grid resolution in Eulerian methods), but by requirements of accuracy and convergence (specifically weak convergence in our case). In this matter, the important parameter is the average number of particles in each kernel estimation, which is equal to $N_{pc} = N (h/L)^2$. With the values adopted above, we have $N_{pc} = 200$, which amounts to statistical errors in the estimation of the means of about $3 \%$ \citep[e.g.][]{Haworth2010}.

    \subsection{Particle properties\label{sec:particle_properties}}

\begin{figure}
            \centering
            \begin{tabular}{c}
                \includegraphics[width=\linewidth]{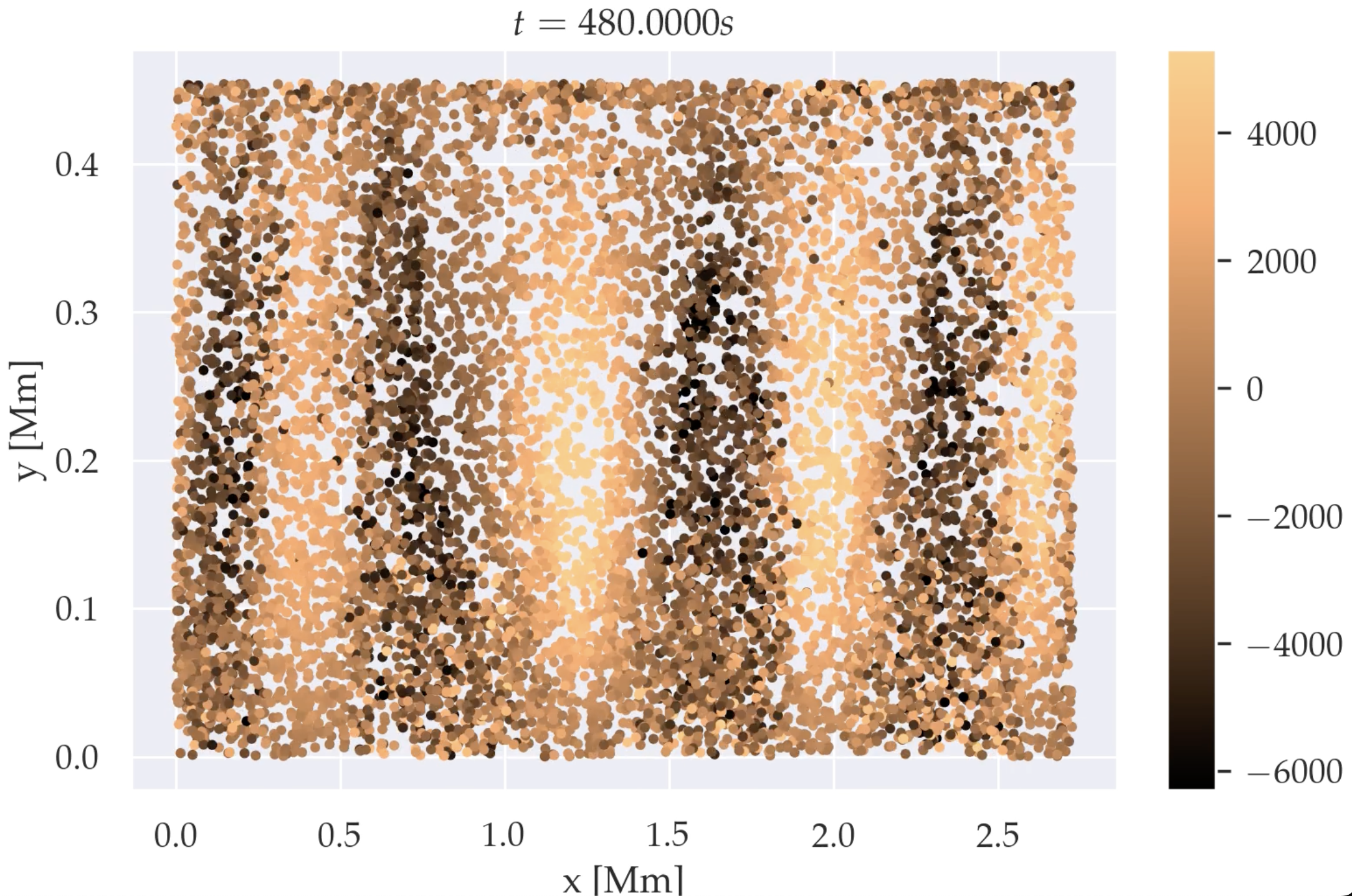} \\
                \includegraphics[width=\linewidth,trim={0 0.6cm 0 0}]{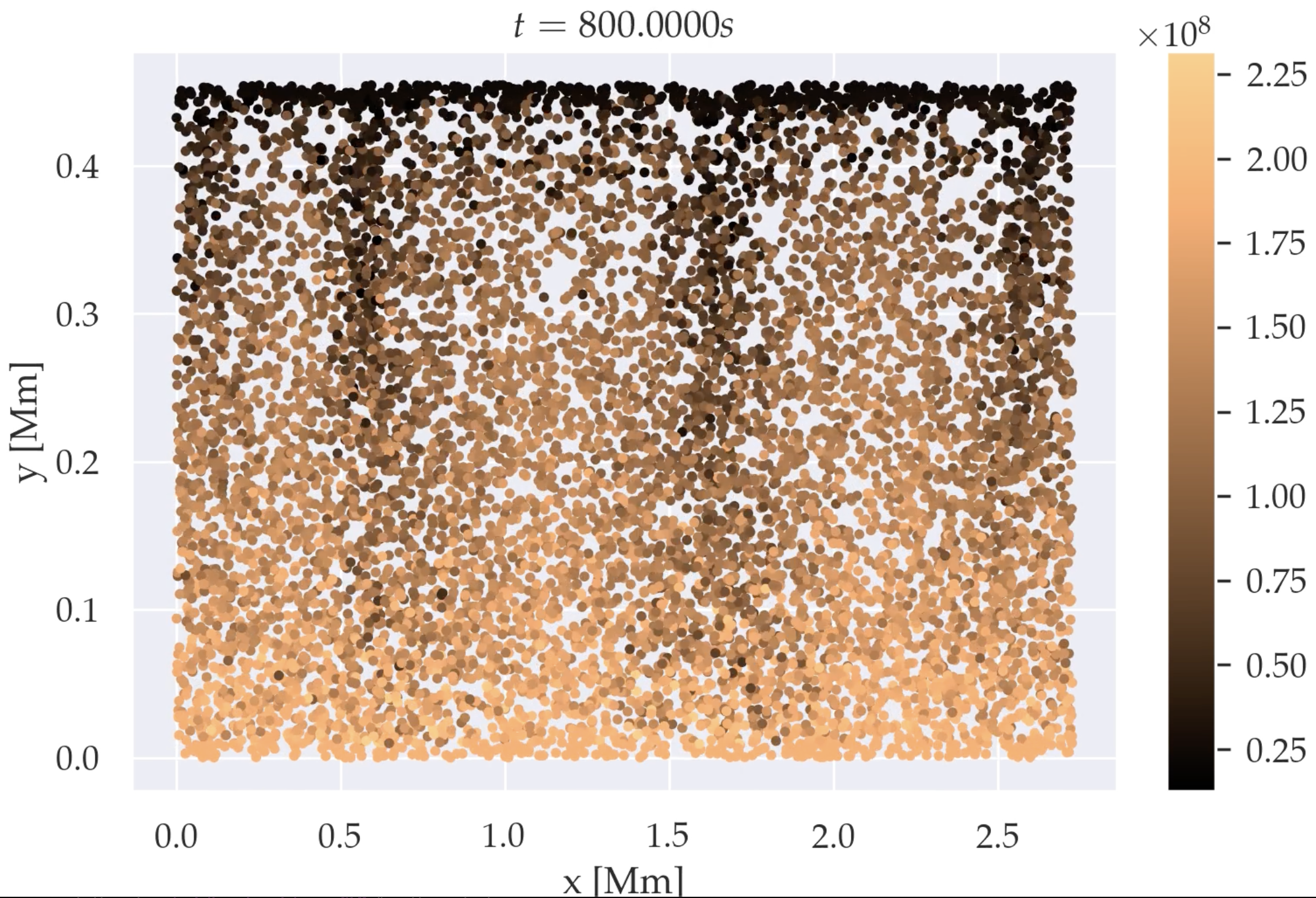} \\
                \includegraphics[width=\linewidth]{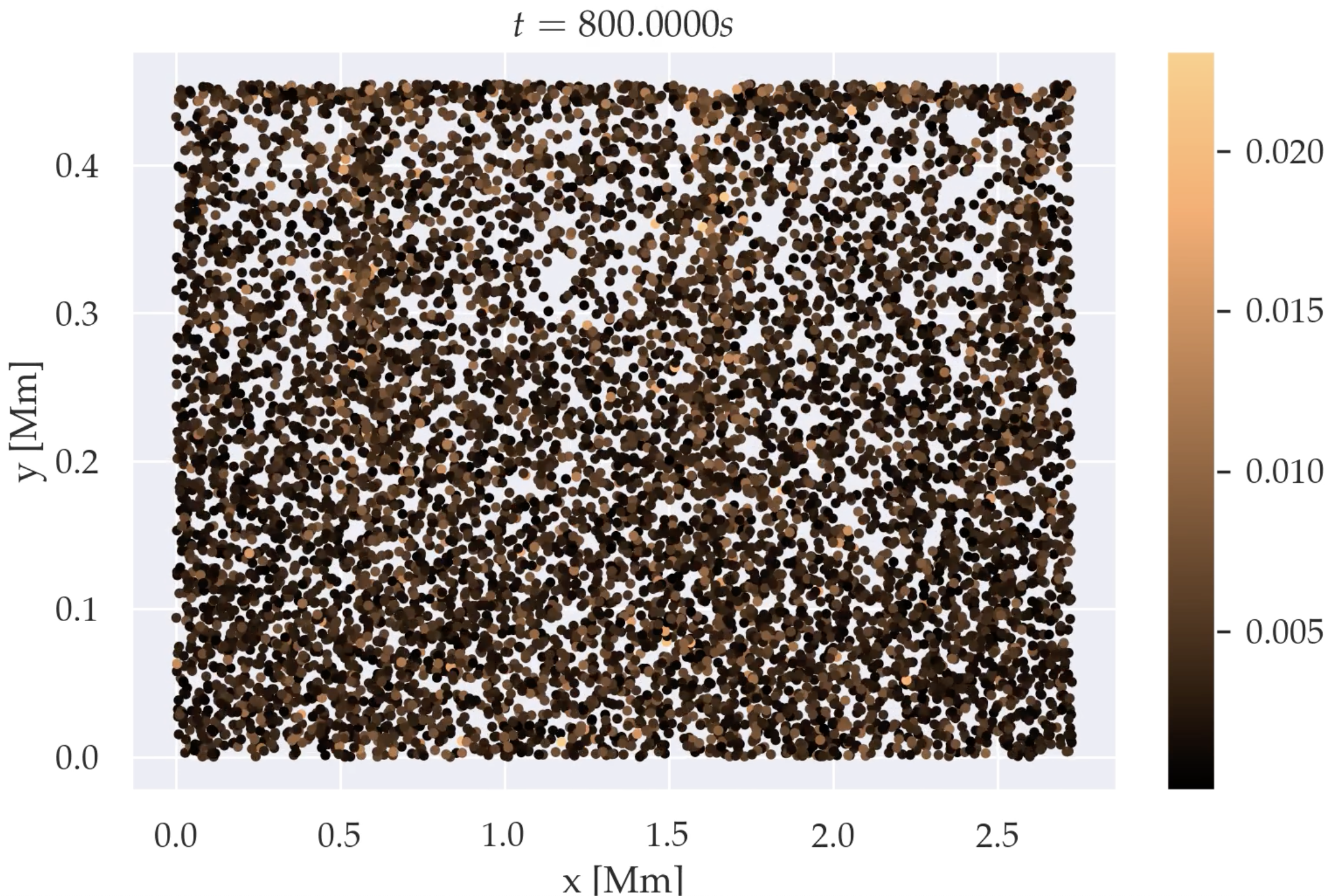}
            \end{tabular}
            \caption{Snapshot of the particle properties for the convectively unstable simulation described in Sect.~\ref{sec:results}. Each dot represents one particle, and the color code refers to the individual vertical velocity $u_x^\ast$ (\textbf{top}), specific internal energy $e^\ast$ (\textbf{middle}), and turbulent frequency $\omega^\ast$ (\textbf{bottom}) of each particle. Only $10,000$ particles are shown for readability, whereas the simulation contains $500,000$.}
            \label{fig:snapshots_particles}
        \end{figure}

        The primary output of the simulation consists of the physical properties of each particle (position, velocity, specific internal energy and turbulent frequency) as a function of time only. A snapshot of the simulation is shown in Fig.~\ref{fig:snapshots_particles}, where each dot represents a particle. We note that, for better readability, only $10,000$ notional particles ($5 \%$ of the total) are shown. The simulated flow clearly features a convectively unstable structure (especially visible in the vertical velocity and internal energy), with alternating hot up-flows and cold down-flows.

    \subsection{Mean flow properties}
    
        Using the kernel estimation scheme described in Sect.~\ref{sec:SPH}, we can extract mean flow properties from the particles. The novelty of the present particle-based approach lies in the fact that it yields proper ensemble averages, which are neither time averages nor horizontal averages. That way, we do not rely on the usual proxies for estimating ensemble averages of turbulent properties, and we can estimate mean flow properties that depend both on spatial coordinates and on time.
        
        \begin{figure}
            \centering
            \begin{tabular}{c}
                \includegraphics[width=\linewidth]{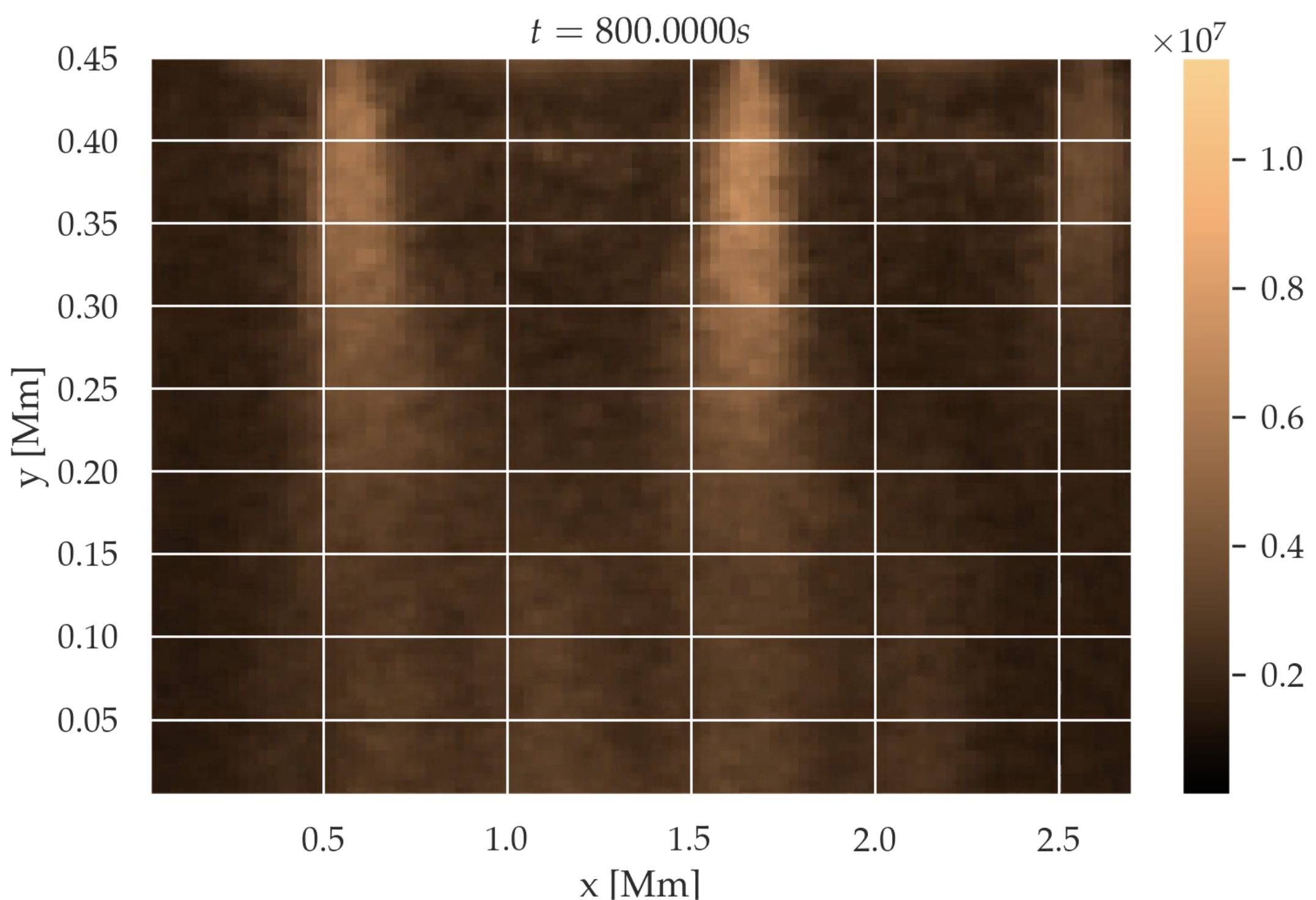} \\
                \includegraphics[width=\linewidth,trim={0 0.6cm 0 0}]{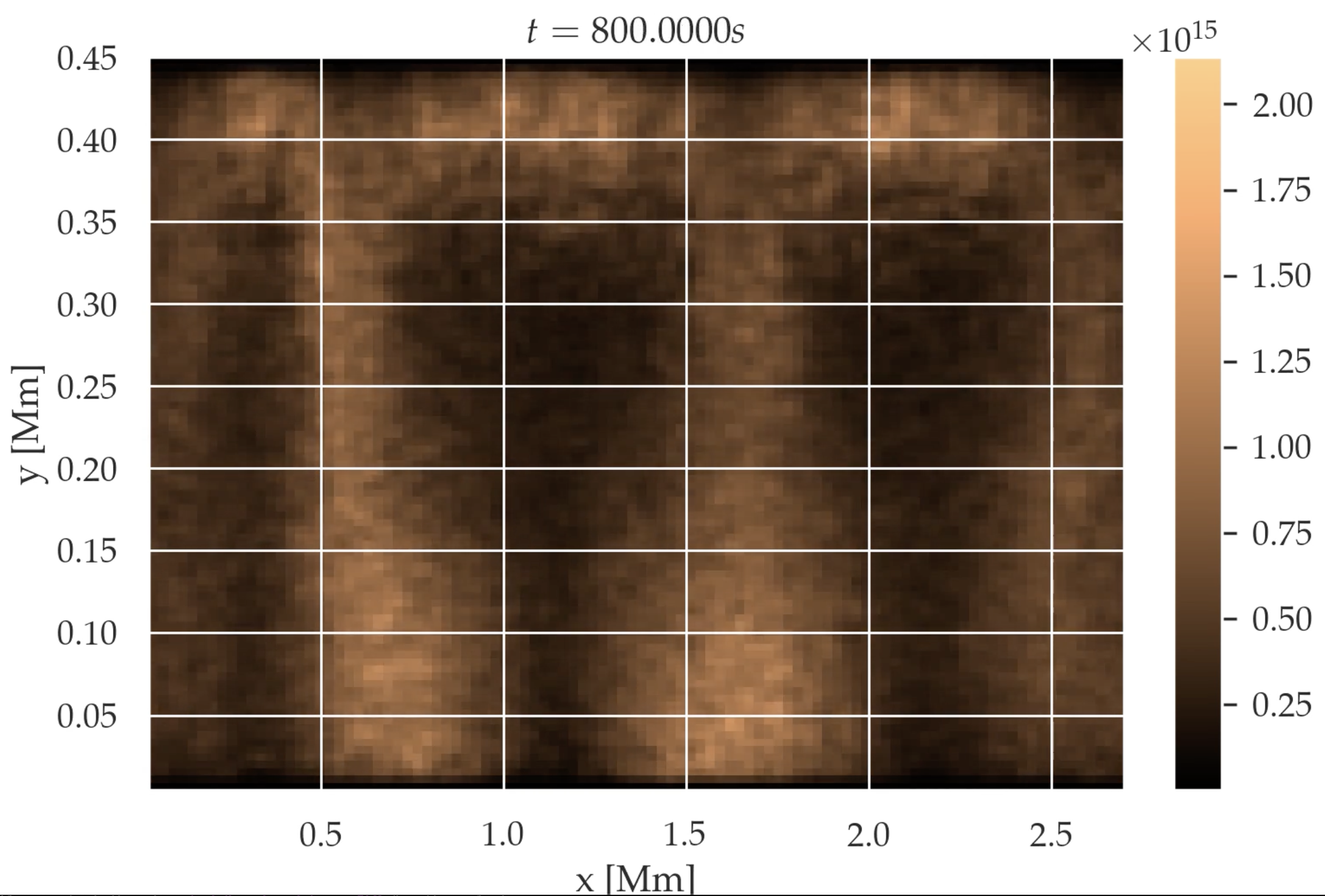} \\
                \includegraphics[width=\linewidth]{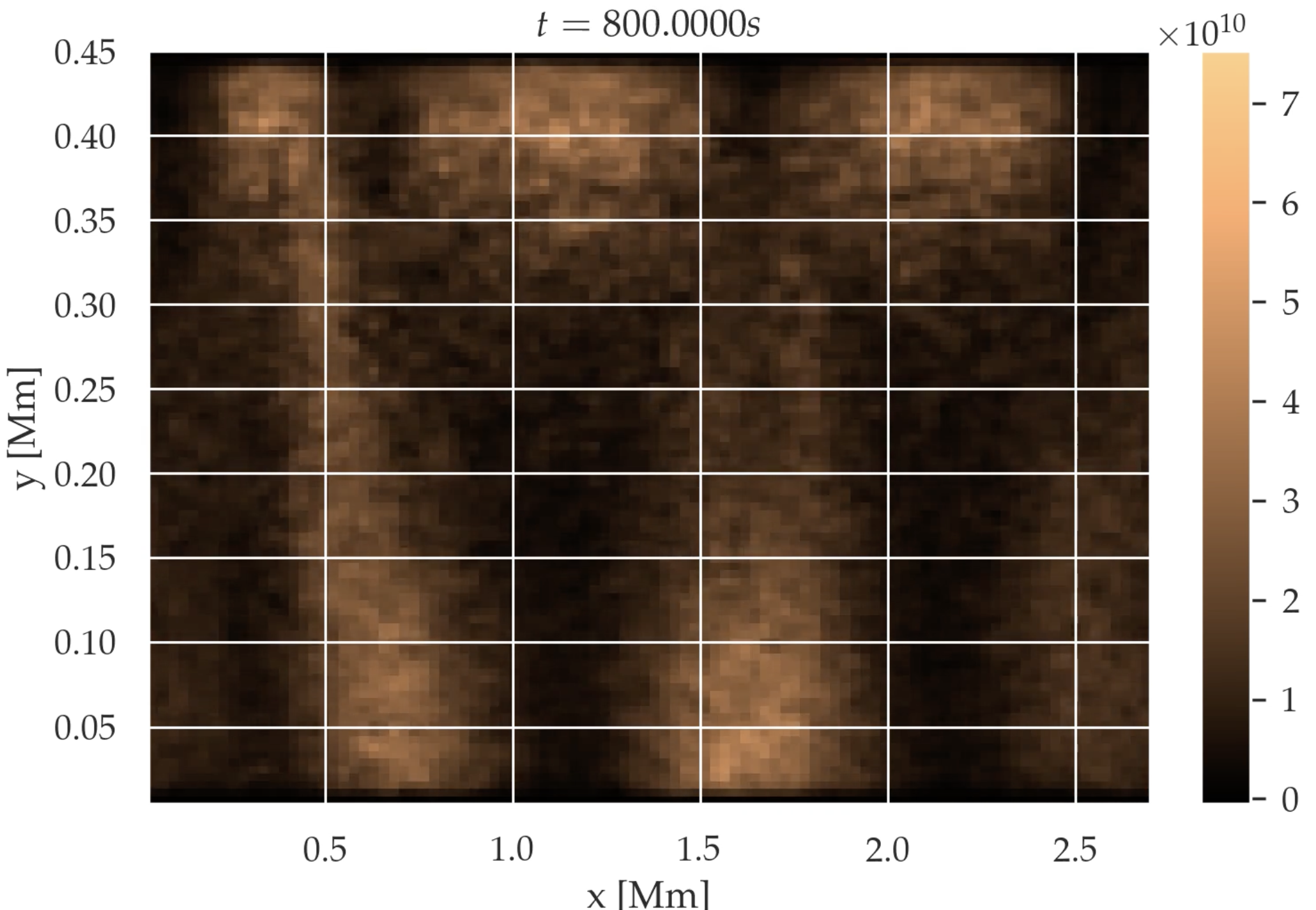}
            \end{tabular}
            \caption{Snapshot of the second-order moments maps for the convectively unstable simulation described in Sect.~\ref{sec:results}, for the same time step as Fig.~\ref{fig:snapshots_particles}. \textbf{Top:} turbulent kinetic energy $k = \widetilde{u_i'' u_i''} / 2$. \textbf{Middle:} internal energy variance $k_e = \widetilde{e''^2}$. \textbf{Bottom:} vertical internal energy flux $\fe_x = \widetilde{u_x'' e''}$.}
            \label{fig:snapshots_mean}
        \end{figure}
        
        The top, middle, and bottom panels of Fig.~\ref{fig:snapshots_mean} show maps of the turbulent kinetic energy $k \equiv \widetilde{u_i'' u_i''} / 2$, internal energy variance $k_e \equiv \widetilde{e''^2}$, and vertical internal energy flux $\fe_x \equiv \widetilde{e'' u_x''}$ respectively, for the same time step as in Fig.~\ref{fig:snapshots_particles}. The second-order statistics of the flow clearly have a different depth dependence in the upflows and downflows. In particular, the turbulent kinetic energy is higher in the downflows than in the upflows, which leads to a negative overall kinetic energy flux throughout the convectively-unstable region. This is in agreement with the known behaviour of stellar surface convection, from 3D hydrodynamic simulations \citep{Kupka2017}.
        
    \subsection{Energy flux budget\label{sec:energy_flux_budget}}
        
\begin{figure}[!ht]
            \centering
            \begin{tabular}{c}
                \includegraphics[width=\linewidth]{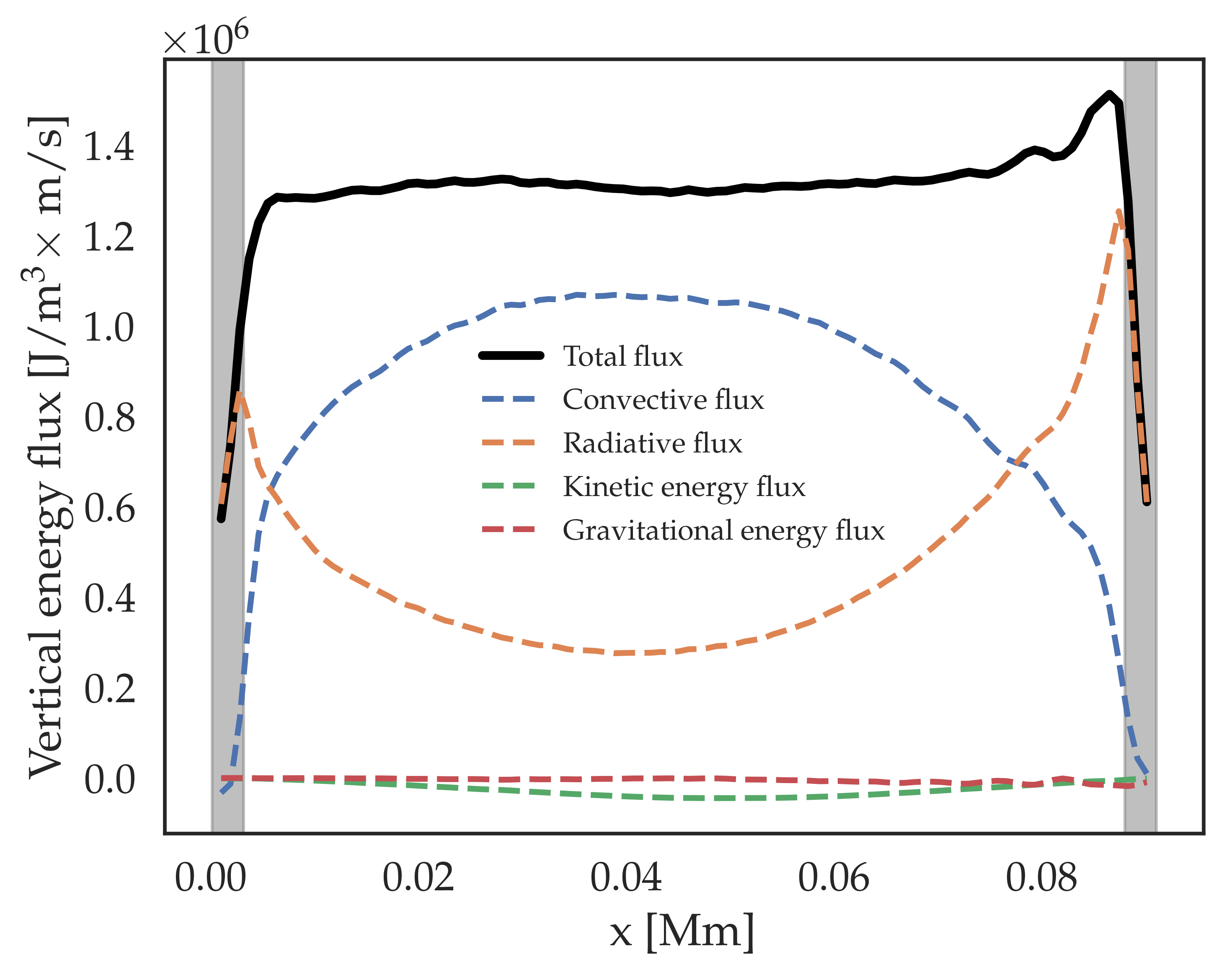} \\
                \includegraphics[width=\linewidth]{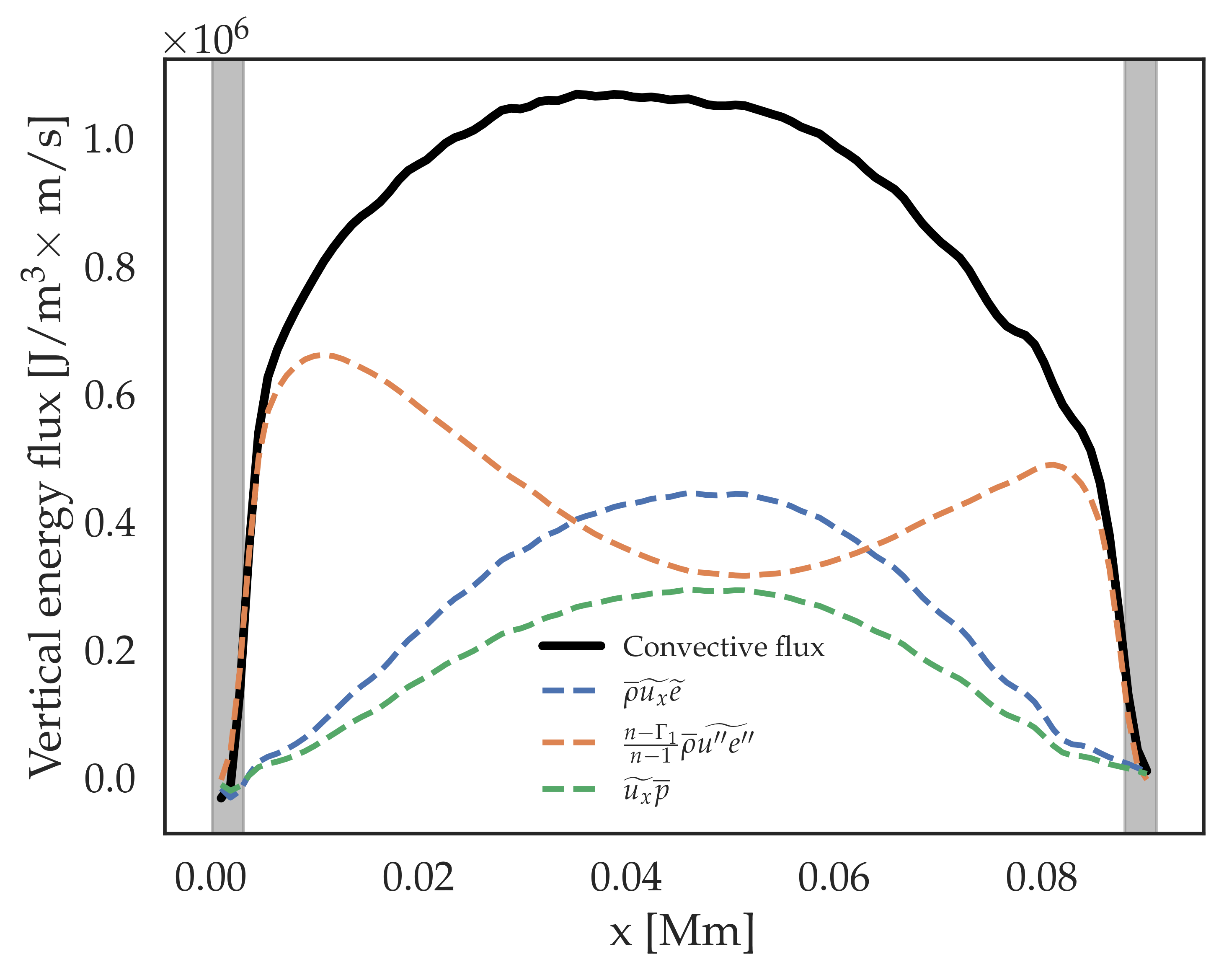} \\
                \includegraphics[width=\linewidth]{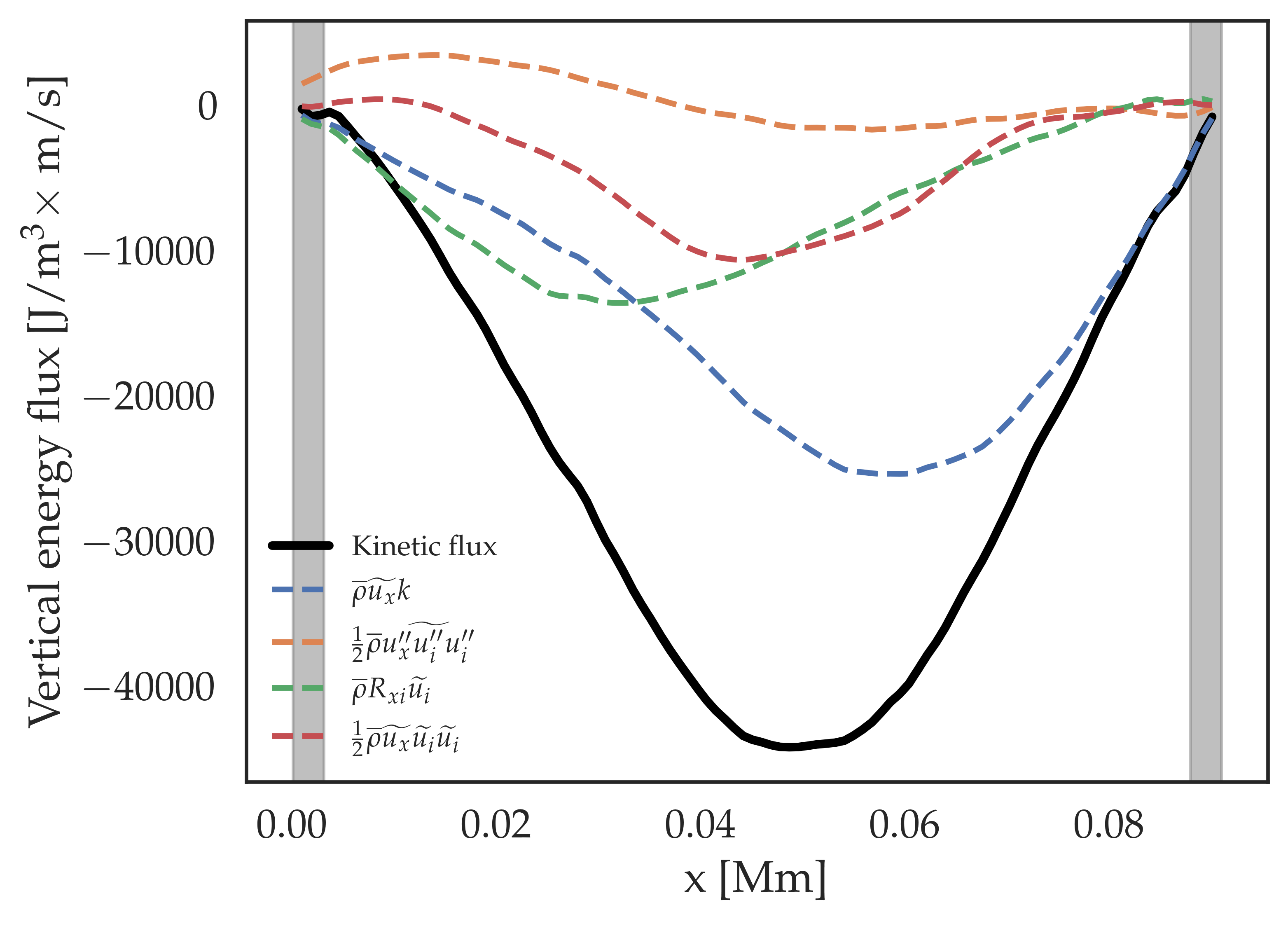}
            \end{tabular}
            \caption{Horizontal and time average of the energy flux budget. The grey areas mark the two narrow regions at the top and bottom of the domain where the internal energy profile is imposed, and the particle-based internal energy equation is not solved. \textbf{Top:} Total energy flux budget, defined in Eq.~\ref{eq:budget_total}. \textbf{Middle:} Convective energy flux budget, defined in Eq.~\ref{eq:budget_convective}. \textbf{Bottom:} Kinetic energy flux budget, defined in Eq.~\ref{eq:budget_kinetic}.}
            \label{fig:energy_flux_budget}
        \end{figure}

        Adding together the equation on large-scale kinetic energy -- stemming from Eq.~\ref{eq:mean_velocity} --, the equation on small-scale kinetic energy -- i.e. the trace of Eq.~\ref{eq:reynolds_stress} -- and the equation on mean internal energy Eq.~\ref{eq:mean_energy}, we obtain the following total energy equation
        \begin{equation}
            \dfrac{\partial}{\partial t} \left( \rhom \enm + \rhom k + \dfrac{1}{2} \rhom \widetilde{u_i} \widetilde{u_i} \right) + \dfrac{\partial}{\partial x_i} \left( \fe_i + F_i^{\mathrm{(kin)}} + F_i^{\mathrm{(rad)}} + F_i^{\mathrm{(grav)}} \right) = 0 ~,
            \label{eq:budget_total}
        \end{equation}
        where
        \begin{align}
            & \fe_i = \rhom \enm \widetilde{u_i} + \dfrac{n - \Gamma_1}{n - 1} \rhom \widetilde{e'' u_i''} + \pgm \widetilde{u_i} ~, \label{eq:budget_convective} \\
            & F_i^{\mathrm{(kin)}} = \rhom k \widetilde{u_i} + \dfrac{1}{2} \rhom \overset{\hstretch{5.0}{\sim}}{u_i'' u_j'' u_j''} + \rhom R_{ij} \widetilde{u_j} + \dfrac{1}{2} \rhom \widetilde{u_i} \widetilde{u_j} \widetilde{u_j} ~, \label{eq:budget_kinetic} \\
            & F_i^{\mathrm{(rad)}} = -\rhom \kappa \dfrac{\partial \enm}{\partial x_i} ~, \\
            & F_i^{\mathrm{(grav)}} = \rhom \widetilde{u_i} \Psi ~.
        \end{align}
        In the stationary state, the sum of all energy fluxes should be conserved. This is illustrated in Fig.~\ref{fig:energy_flux_budget}, which show the vertical profiles (i.e. the time and horizontal average) of the different contributions to the energy flux budget equation Eq.~\ref{eq:budget_total}, excluding the relaxation phase at the start of the simulation. The two main contributions come from the convective and radiative fluxes, with the kinetic energy flux being of much smaller amplitude, but systematically negative. Breaking down the different contributions to the convective flux, however (see middle panel), it is apparent that all contributions account for a significant portion of the total, with no one contribution standing out as dominant.
        
        We note that the energy conservation breaks down in the two narrow bands located at the top and bottom of the simulation -- grey bands in Fig.~\ref{fig:energy_flux_budget} --, where the particle-based internal energy equation given by Eq.~\ref{eq:SDE_energy} is not solved, but instead an internal energy profile is imposed (see Sect.~\ref{sec:SPH}).
        
\begin{figure}
            \centering
            \includegraphics[width=\linewidth]{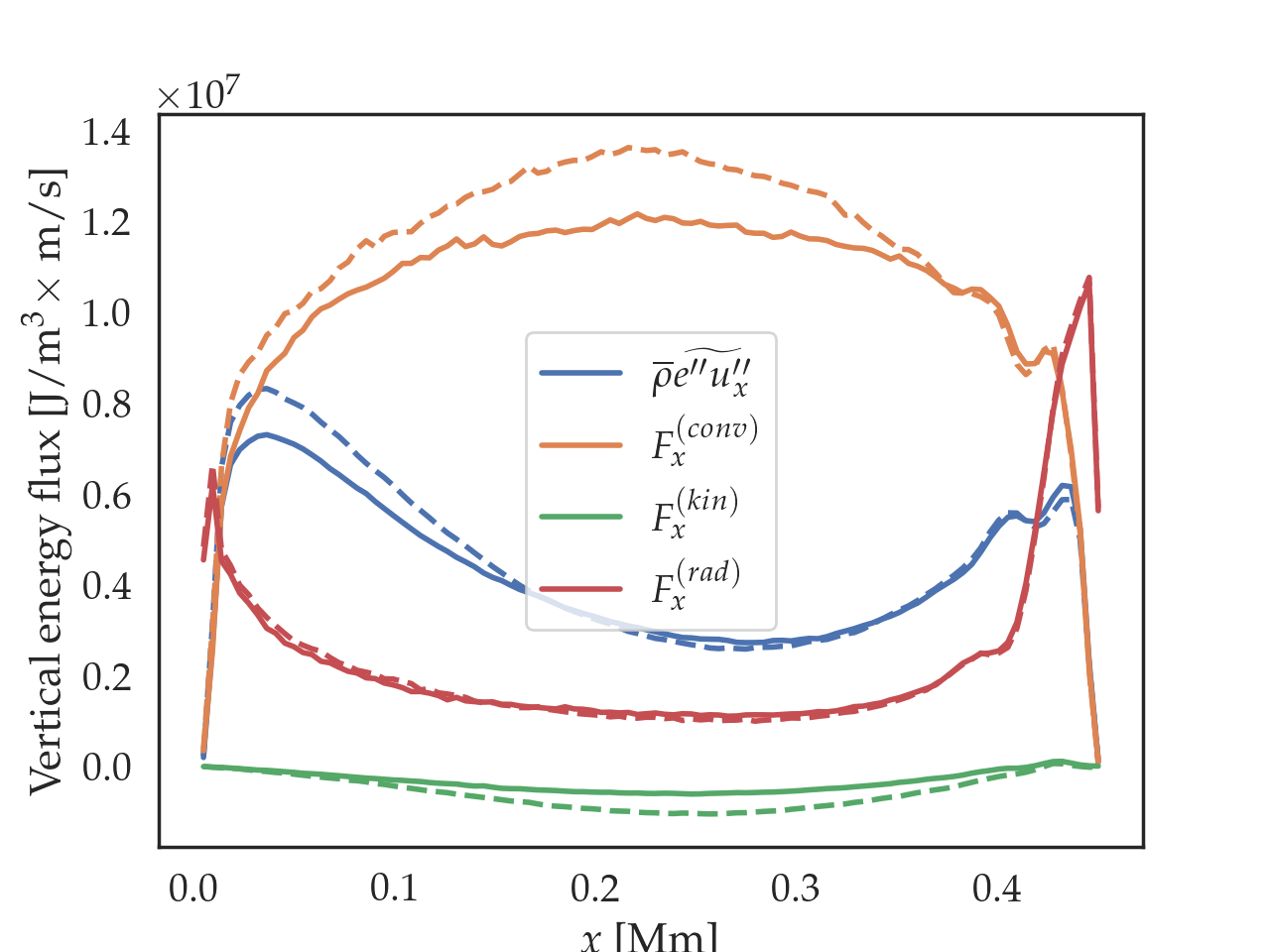}
            \caption{Comparison between the energy flux budgets of the two simulations described in the text, the only modified parameters being the constant $C_1$ in the Lagrangian stochastic model (solid lines: $C_1 = 3.5$; dashed lines: $C_1 = 2.0$). The different colours represent different terms in the energy flux budget, as indicated on the plot.}
            \label{fig:fluxes_vs_C1}
        \end{figure}

        The constant $C_1$ introduced in the Lagrangian stochastic model was chosen to be equal to the Kolmogorov constant $C_0$. This choice, however, is arbitrary. Since the constant $C_1$ controls the decay rate of internal energy fluctuations, changing its value can be expected to modify the second order moments of turbulent convection, and in particular the energy flux budget. In order to investigate this, we ran a simulation entirely identical to the one described in rest of this section, but adopting $C_1 = 2.0 \neq C_0$. The energy flux budgets are compared in Fig.~\ref{fig:fluxes_vs_C1}. Modifications to the energy flux budget remain small, with the quantity most impacted (in terms of relative values) being the turbulent kinetic energy flux.
        
    \subsection{Up- and down-flow asymmetry}
    
        Because of the presence of upflows and downdrafts, the vertical velocity and internal energy distributions are not expected to be Gaussian \citep[e.g.][]{Kupka2007}. To measure the departure from Gaussian distributions, it is customary to define the skewness $S_w$ (resp. $S_\theta$) and kurtosis $K_w$ (resp. $K_\theta$) of the vertical velocity (resp. internal energy) distribution as
        \begin{align*}
            & S_w = \left. \overline{u_x''^3} ~ \middle/ ~ \overline{u_x''^2}^{3/2} \right. ~, \qquad \left. K_w = \overline{u_x''^4} ~ \middle/ ~ \overline{u_x''^2}^2 \right. ~, \\
            & S_\theta = \left. \overline{e''^3} ~ \middle/ ~ \overline{e''^2}^{3/2} \right. ~, \qquad \left. K_\theta = \overline{e''^4} ~ \middle/ ~ \overline{e''^2}^2 \right. ~.
        \end{align*}
        where $u_x'' = u_x - \widetilde{u_x}$ and $e'' = e - \widetilde{e}$. The velocity skewness gives a measure of the asymmetry between the flows directed upwards and downwards (positive skewness means that upflows take up less space but are stronger, negative skewness means the opposite), while the kurtosis measures the rarity of intermittent, stronger-than-normal events (large kurtosis means that such events are frequent, small kurtosis means that they are rare, and a Gaussian distribution has a kurtosis of $3$). The internal energy skewness and kurtosis have a similar meaning for the asymmetry between hot and cold flows.
        
\begin{figure}
            \centering
            \begin{tabular}{c}
                \includegraphics[width=\linewidth]{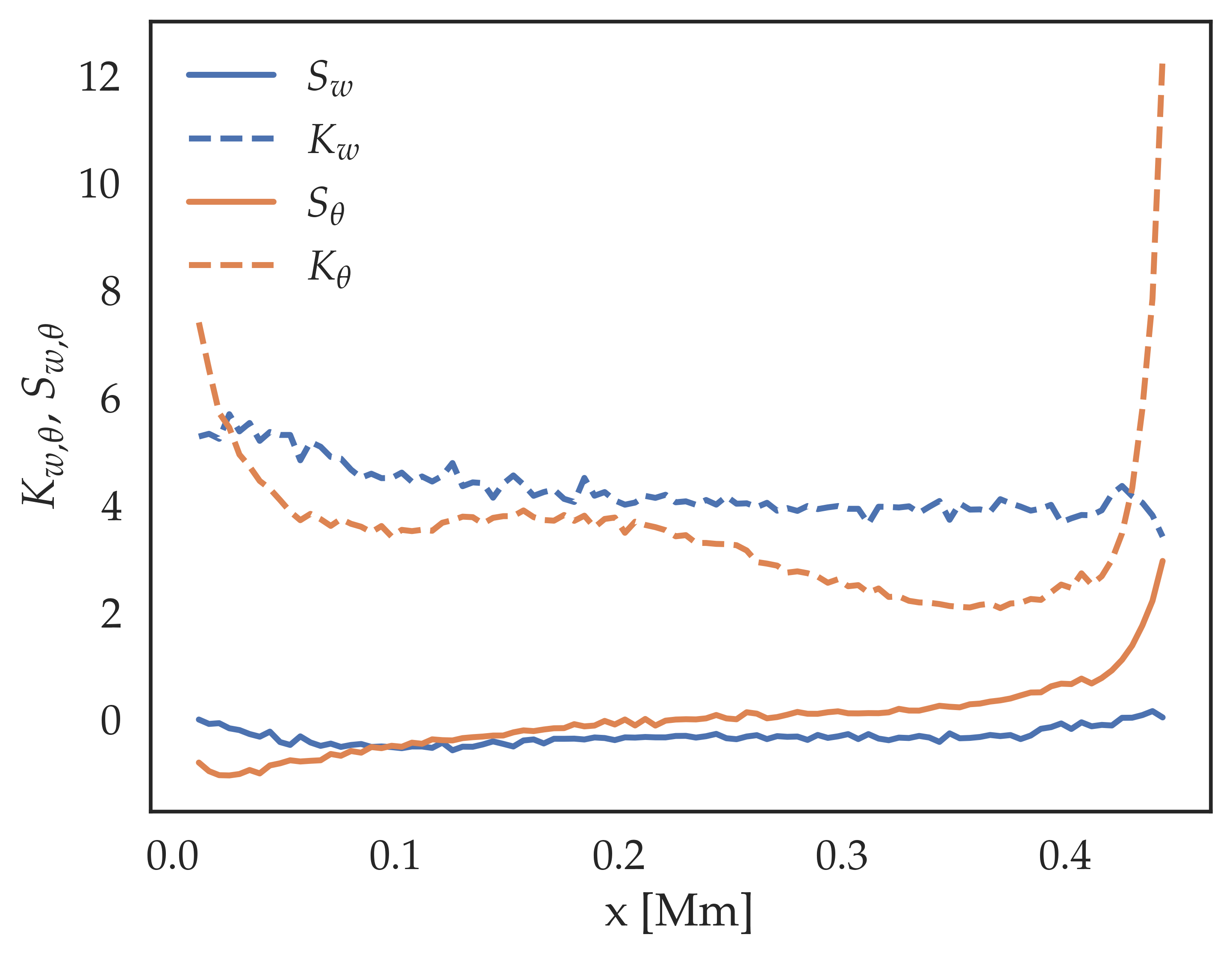} \\
                \includegraphics[width=\linewidth]{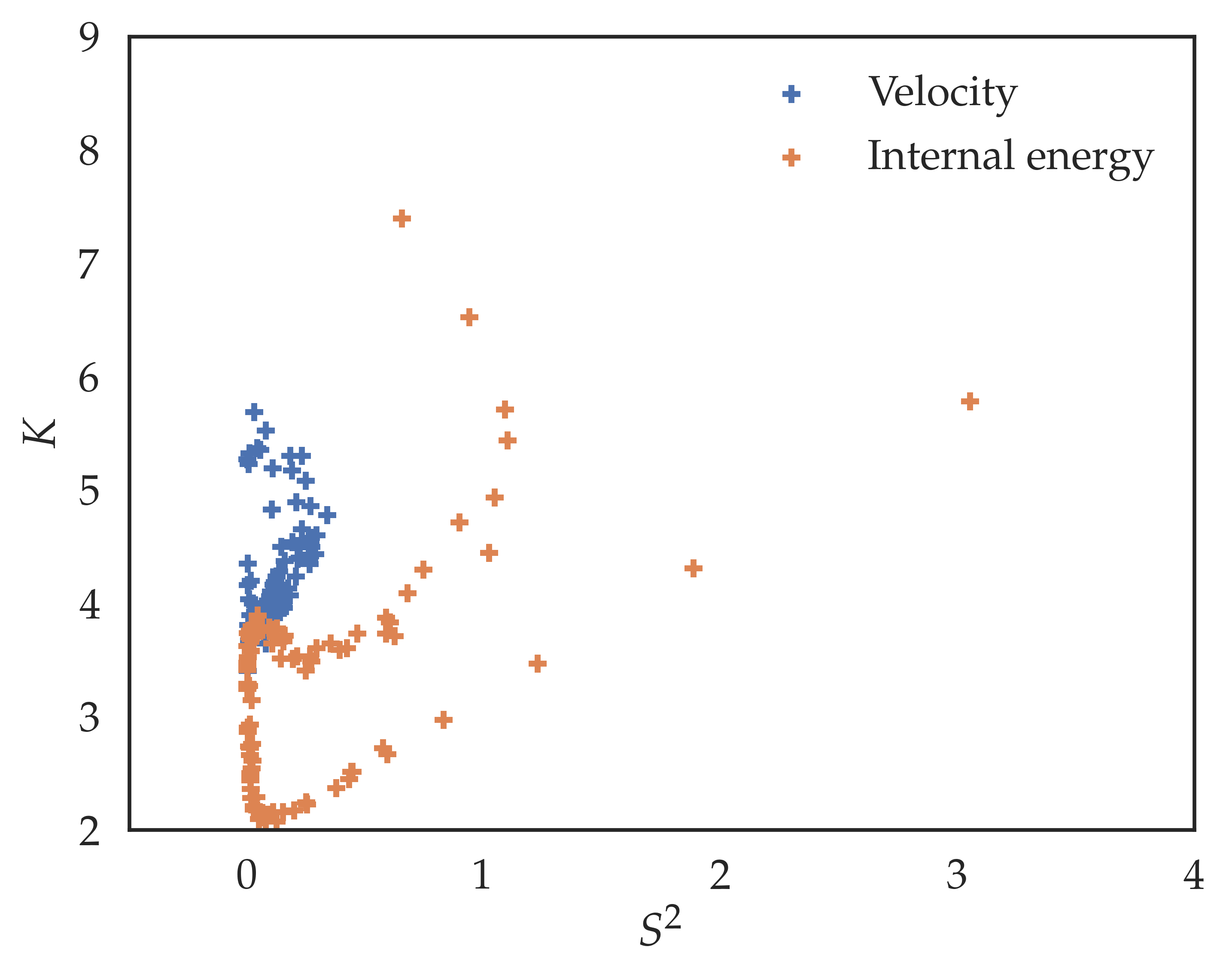}
            \end{tabular}
            \caption{\textbf{Top:} Skewness and kurtosis of the vertical velocity distribution and internal energy distribution, as a function of height. \textbf{Bottom:} Same, but in the $(K, S^2)$ plane. Each cross corresponds to one horizontal layer in the simulation.}
            \label{fig:bimodal_parameters}
        \end{figure}

        We sliced the domain into horizontal layers spanning $1 \%$ of the domain each, and for each layer we extracted the vertical velocity and internal energy of all notional particles located in that slice, thus reconstructing the velocity and energy distributions as a function of height. We show the skewness and kurtosis of each distribution, as a function of height, in Fig.~\ref{fig:bimodal_parameters}. As expected, the velocity distribution is skewed towards downwards velocities, pointing to the downdrafts taking less space than the upflows, but being more turbulent (as visible also from the top panel of Fig.~\ref{fig:snapshots_mean}). Furthermore, while the kurtosis of the energy distribution remains close to the Gaussian value of $3$ in the bulk of simulation, the velocity distribution is distincly platikurtic throughout -- meaning that its tails are less prominant than those of a normal distribution. These results -- and in particular the $(K, S^2)$ diagram shown in the bottom panel of Fig.~\ref{fig:bimodal_parameters} -- are in good agreement with results from Large-Eddy Simulations of solar convection, both 2D \citep{Kupka2007} and 3D \citep{Cai2018}.
        
\begin{figure}
            \centering
            \includegraphics[width=\linewidth]{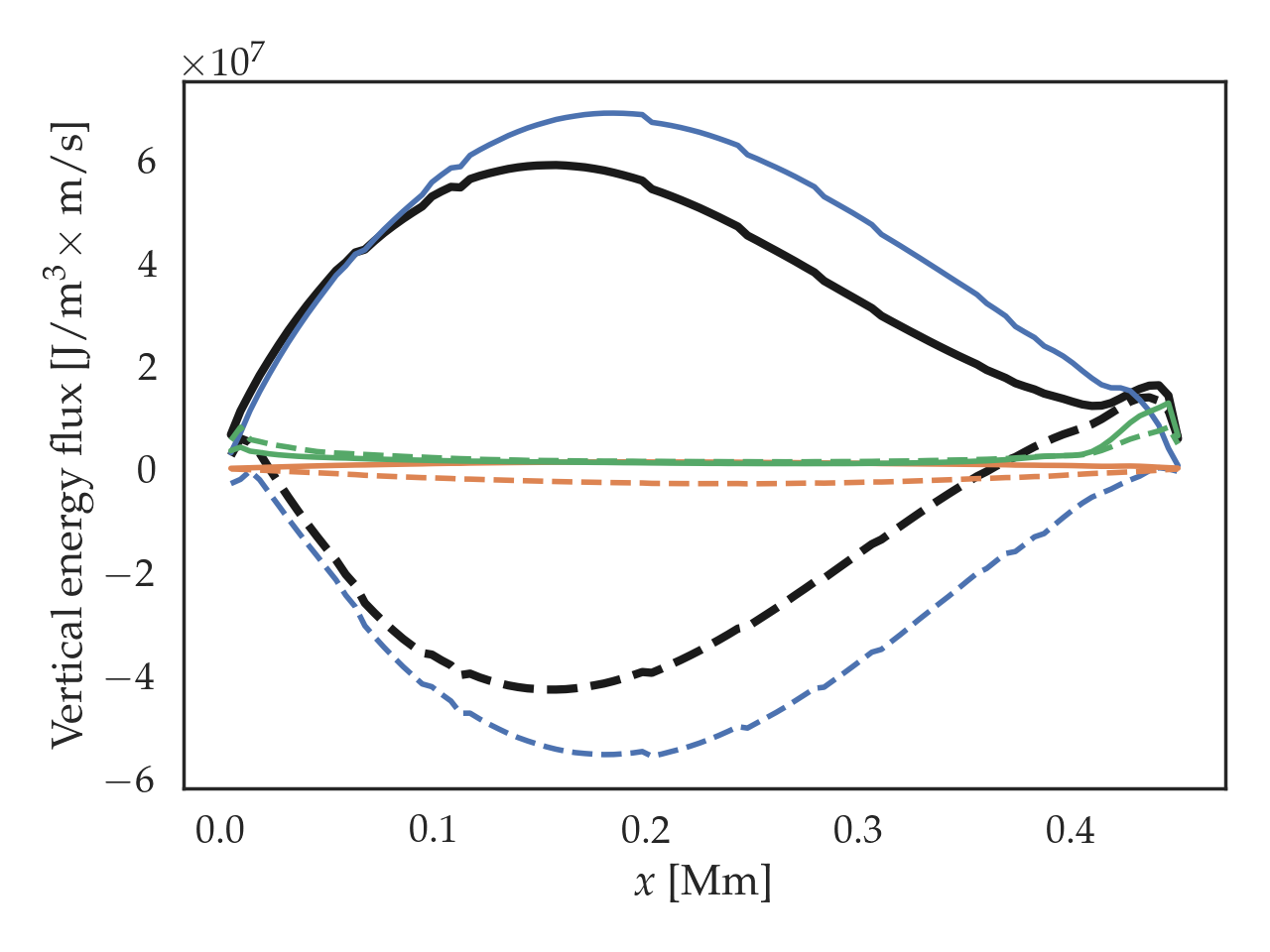}
            \caption{Breakdown of the energy flux budget into upflow (solid lines) and downflow (dashed lines) energy fluxes. Blue, green, orange and black lines respectively show the convective flux $F^\mathrm{(e)}$, the radiative flux $F^\mathrm{(rad)}$, the kinetic energy flux $F^\mathrm{(kin)}$ and the total energy flux $F^\mathrm{(tot)}$, as defined by Eq.~\ref{eq:budget_total}.}
            \label{fig:up_down_budget}
        \end{figure}

        The fact that the kernel estimation scheme provides time- and space-dependent ensemble averages also allows to compare vertical profiles of various mean quantities in up-flows and down-flows. We show in Fig.~\ref{fig:up_down_budget} how the energy flux budget presented in Sect.~\ref{sec:energy_flux_budget} breaks down into upwards and downwards energy flow. The most striking feature is the strong positive (resp. negative) convective flux in the upflows (resp. downflows), which cancel out when the total convective flux is computed (see Fig.~\ref{fig:energy_flux_budget}). It can also be seen that the downwards negative kinetic energy flux more than cancels out the upwards positive kinetic energy flux, thus yielding a negative total kinetic energy flux.
        
    \subsection{Turbulent diffusivity\label{sec:dispersion}}
    
\begin{figure}
            \centering
            \includegraphics[width=\linewidth]{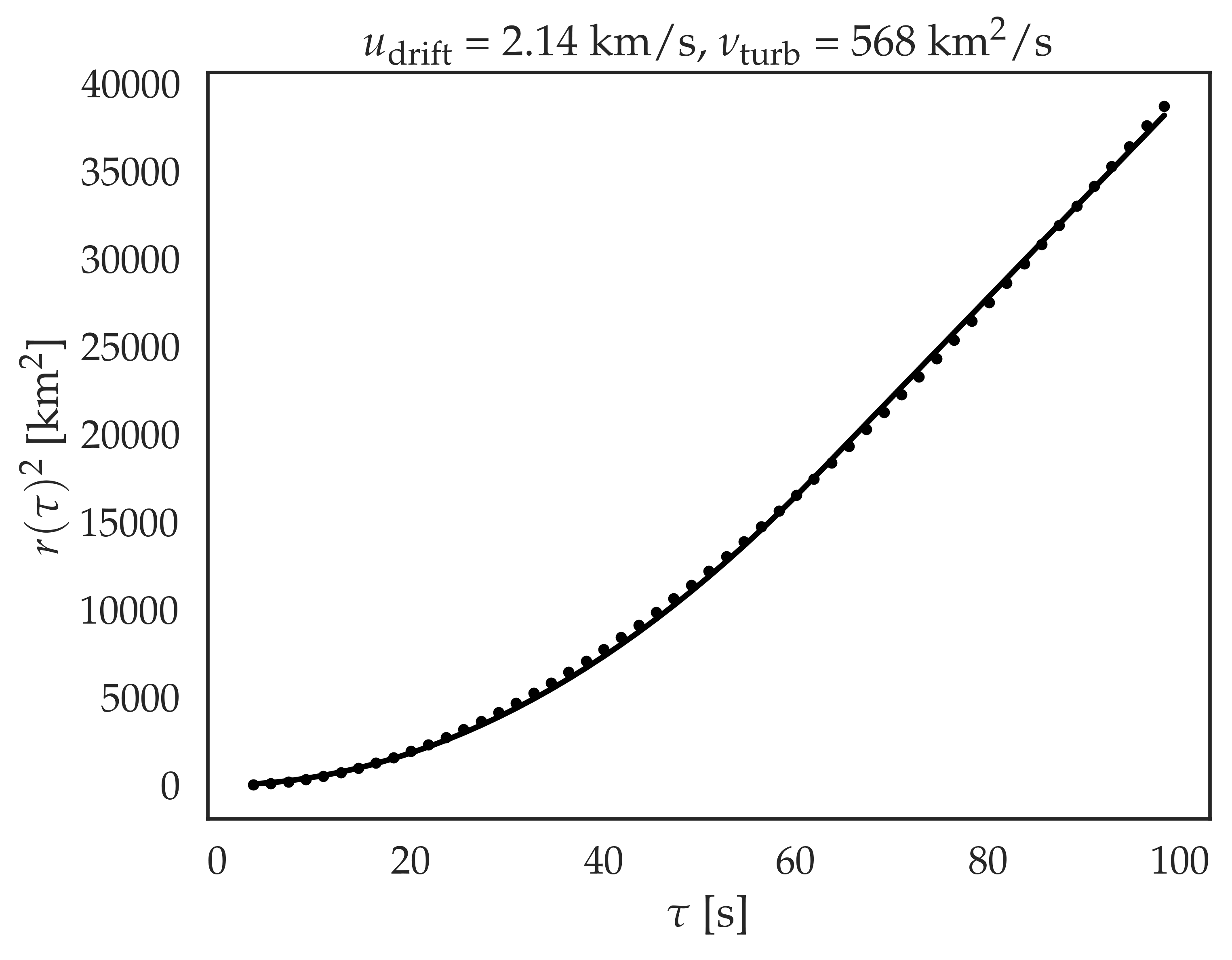}
            \caption{Squared particle dispersion distance $r^2$ as a function of time increment $\tau$, averaged over all particles trajectories. The black dots show the simulation data, and the solid black line shows a fit to the data, using the model described in the text. The RMS particle velocity $u_\mathrm{drift}$ and turbulent diffusion coefficient $\nu_\mathrm{turb}$ estimated from the fit are indicated on top.}
            \label{fig:travel_distance}
        \end{figure}

        The Lagrangian framework adopted in our simulations also allows us to measure particle dispersion, and therefore turbulent viscosity, much more easily and accurately than in Large-Eddy Simulations. Subdividing the simulation into time series of $100$ seconds each, we extracted the travel distance of each particle as a function of time increment
        \begin{equation}
            r(\tau)^2 \equiv \left( x^{\ast}\left( t_0 + \tau \right) - x^{\ast}\left( t_0 \right) \right)^2 + \left( y^{\ast}\left( t_0 + \tau \right) - y^{\ast}\left( t_0 \right) \right)^2 ~,
        \end{equation}
        where $t_0$ refers to the initial time of the time series. We cut each individual particle trajectory when the particle gets too close to one of the domain boundaries (to avoid the effect of boundary conditions on particle dispersion). We then averaged $r(\tau)$ over all particles, for every available value of the time increment $\tau$, regardless of initial particle position. The result is shown in Fig.~\ref{fig:travel_distance}, where one can clearly make out the two expected regimes \citep{Taylor1922}: for short travel times, particle dispersion is dominated by the initial velocity, which translates to an advective regime where $r(\tau) \propto \tau$, and the proportionality constant is the initial particle velocity $u_\mathrm{drift}$ (because we average over different particle trajectories, the value $u_\mathrm{drift}$ inferred from particle dispersion in the simulation must be taken to represent the RMS value of initial particle velocity). On the other hand, for long travel times, it is dominated by turbulence, which translates to a diffusive regime where $r^2(\tau) \propto \tau$, and the proportionality constant is the turbulent diffusivity $\nu_\mathrm{turb}$.
        
        We fit the following model to the data
        \begin{equation*}
            r(\tau)^2 =
            \begin{cases}
                u_\mathrm{drift}^2 \tau^2 & \mathrm{ if } ~~ \tau < \tau_0 ~, \\
                \nu_\mathrm{turb} \left(\tau - \dfrac{\tau_0}{2} \right) & \mathrm{ if } ~~ \tau > \tau_0 ~,
            \end{cases}
        \end{equation*}
        
        where $\tau_0 \equiv \nu_\mathrm{turb} / (2 u_\mathrm{drift}^2)$ marks the transition between the two regimes. We find a typical initial velocity of $u_\mathrm{drift} \sim 2$ km/s, and a turbulent diffusivity of $\nu_\mathrm{turb} \sim 500$ km$^2$/s. Those are in agreement with the typical values of the convective velocity and turbulent viscosity at the top of the solar convective zone \citep[e.g.][]{Rincon2025}. Alternatively, we can estimate the turbulent Reynolds number $\mathrm{Re}_\mathrm{turb}$, defined by
        \begin{equation*}
            \mathrm{Re}_\mathrm{turb} \equiv \dfrac{u_\mathrm{conv} L}{\nu_\mathrm{turb}} ~,
        \end{equation*}
        where $L$ is the typical length scale of the flow, which we take to be the vertical extent of the simulation ($L = 0.45$ Mm). With these values, we find $\mathrm{Re}_\mathrm{turb} \sim 2$. We also find that the transition between the advective and diffusive regimes occurs for $\tau_0 \sim 1$ minute, corresponding to the observed convective turnover time at the solar surface.

\begin{figure}
            \centering
            \includegraphics[width=\linewidth]{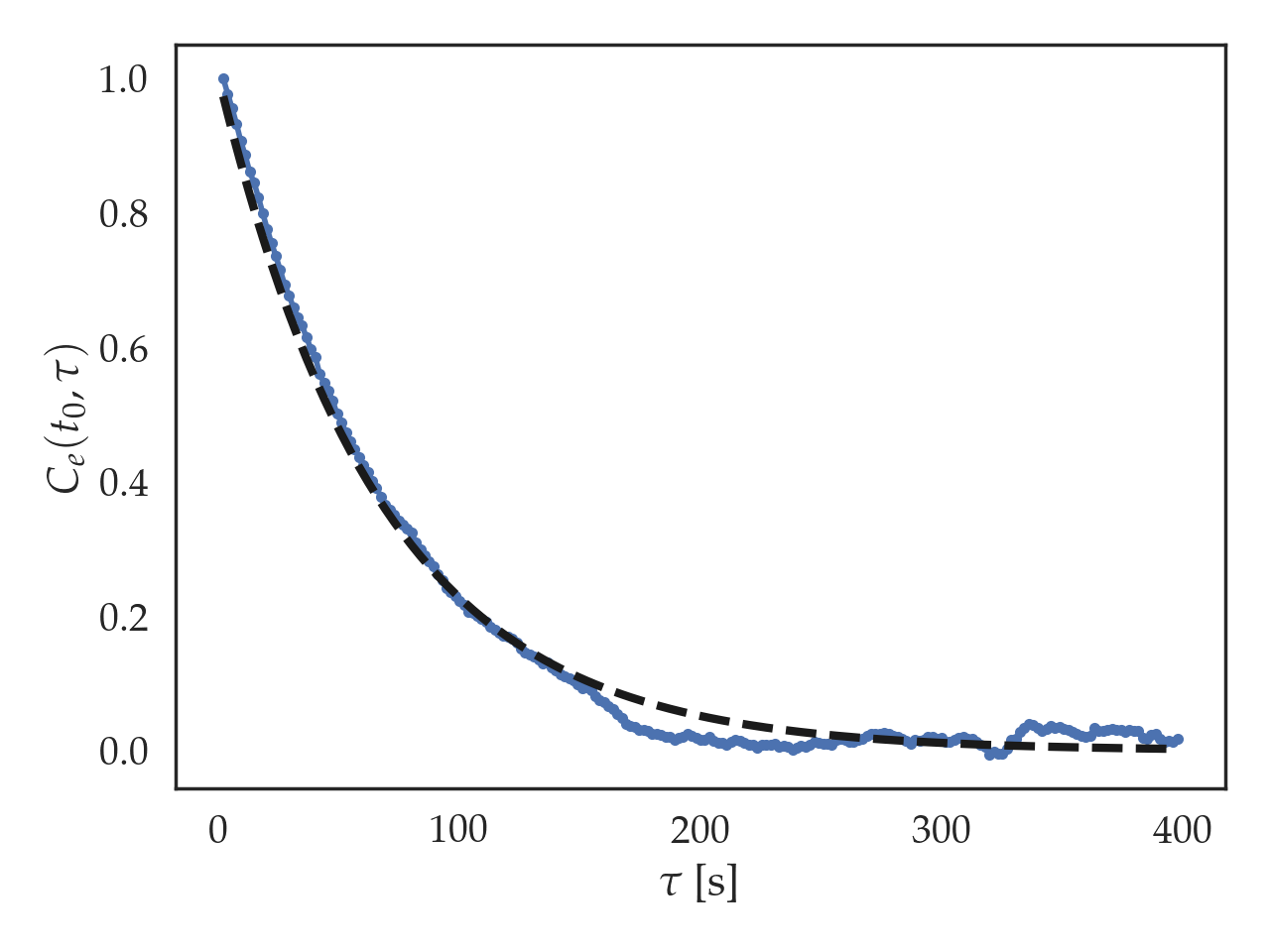}
            \caption{Autocorrelation function of particle energy fluctuations, as a function of time increment (blue), averaged over all particles trajectories. The dashed black line shows an exponential fit to the curve.}
            \label{fig:energy_autocorrelation}
        \end{figure}

        We can estimate the turbulent thermal diffusive time scale in a similar manner. Instead of extracting the travel distance of each particle, we extract the autocorrelation function of the internal energy fluctuations of each particle, as a function of time increment
        \begin{equation}
            C_e(t, \tau) \equiv \dfrac{\left\langle \left( e^\ast(t) - \widetilde{e}\left( \vphantom{p_p^d} \mathbf{x}^\ast(t)\right) \right) \left( e^\ast(t+\tau) - \widetilde{e}\left( \vphantom{p_p^d} \mathbf{x}^\ast(t+\tau) \right) \right) \right\rangle }{\left\langle \left( e^\ast(t) - \widetilde{e}(\mathbf{x}^\ast(t)) \right)^2 \right\rangle} ~.
        \end{equation}
        We then average it over all particles. Figure~\ref{fig:energy_autocorrelation} clearly shows that the autocorrelation function takes the expected form of a decaying exponential function. The decay time represents the integral Lagrangian time of the energy fluctuations (i.e. the typical time over which the thermal memory of the particle is lost). It represents the turbulent thermal diffusive time scale $\tau_\mathrm{th}$: the fit shown in Fig.~\ref{fig:energy_autocorrelation} provides $\tau_\mathrm{th} = 65 \pm 5 ~ s$, depending on which sub-timeseries of the simulation we considered. This is remarkably close to the momentum diffusivity time scale $\tau_0$ determined above, thus yielding a value $\mathrm{Pr} \sim 1$ for the turbulent Prandtl number. This is in accordance with what is expected in the regime of strong turbulence, in line with the assumptions made in the Reynolds analogy for instance.
        
\section{Conclusion\label{sec:conclusion}}

    In this paper, we developed a Lagrangian stochastic model for the one-point statistics of turbulent convection in stars. This model uses a particle representation of the flow, and is centered around evolution equations for the position, velocity, internal energy and turbulent frequency of each particle, as a function of time only, in the form of Eqs.~\ref{eq:SDE_position} to \ref{eq:SDE_frequency}. This set of particle-based stochastic equations is statistically equivalent to a Fokker-Planck equation for the Probability Density Function (PDF) of all relevant turbulent variables, which itself implies transport equations for the means and cross-variances of the velocity, internal energy, and turbulent frequency of the flow. We established this Lagrangian stochastic model in such a way that these transport equations are identical to the exact transport equations stemming from first principles, using only physically motivated assumptions to close the equations. The main advantage of this approach is that, thanks to the adopted Lagrangian point of view, it models the advective terms exactly, so that closure relations are not needed there.
    
    In a second part of the paper, we then developed a new numerical code to implement this Lagrangian stochastic model in a 2D setting. The code solves the set of stochastic differential equations (SDE) determined in the first part of the paper, for a large number of notional particles (several hundreds of thousands to several millions). These notional particles act as realisations of the same flow, and interact with each other through the mean flow fields, which are determined from the set of particle realisations themselves, through a filtering kernel estimation procedure, akin to Smoothed Particle Hydrodynamics. In classical hydrodynamic simulations, statistical averages of turbulent properties can only be studied through the usual proxies of horizontal or temporal averages. By contrast, this code models actual ensemble averages of turbulent flow properties, that is to say both means (density, velocity, internal energy, turbulent frequency) and second-order moments (turbulent kinetic energy, Reynolds stress tensor, internal energy variance, internal energy flux) as a function of time and all spatial coordinates. It is therefore much more suited to study how the statistical properties of stellar turbulent convection depends on time, and especially how they affect the propagation of waves. In addition, this code also includes an equation for the turbulent dissipation, which is modelled as a \textit{bona fide} turbulent variable, which contrasts with Eulerian grid-based hydrodynamic simulations where turbulent dissipation is dominated by artificial viscosity, and cannot be modelled physically.
    
    The example presented in the second part of this paper aims at showcasing the possibilities offered by this new code. In subsequent studies, we aim to use this code for several purposes. First, because statistical averages can be obtained without having to resort to proxies such as horizontal averages or time averages, this code is much more suited to study the interaction between waves and convection. It will be possible to extract oscillating modes directly from the simulation, and to study its imprint on the various second-order moments of turbulence. In particular, we will be able to extract directly the wave-induced fluctuation of turbulent pressure, convective flux, internal energy variance, etc. This will be of paramount importance to quantify surface effects (and particularly the modal part of surface effects) without having to rely on Mixing Length Theories.
    
    Second, as showcased in Sect.~\ref{sec:dispersion}, the purely Lagrangian nature of the code makes it perfect to study particle dispersion. In classical hydrodynamic simulations, Lagrangian trackers must be injected in order to measure turbulent diffusion, and the results usually suffer from the finite grid resolution. Here, Lagrangian trajectories of flow realisations are already followed by construction, and the determination of diffusion coefficients from these trajectories does not rely on any Eulerian grid. In particular, this code allows, as will be investigated in a future study, to assess the efficiency of turbulent transport through convective/radiative interfaces, and therefore of convective penetration in the stellar context.
    
    Finally, it will be possible to extend the code to 3D rather than 2D. While this is not conceptually complicated, it will be considerably more demanding in computational power and time. More specifically, while the appropriate number of notional particles needed to represent the flow in 2D is of the order of several hundreds of thousands, this number will need to be increased to tens or hundreds of millions in 3D.

\begin{acknowledgements}
The authors wish to thank the anonymous referee for useful comments that led to improvement of the manuscript. The authors acknowledge support from the "Action Thématique de Physique Stellaire" (ATPS) of CNRS/INSU PN Astro co-funded by CEA and CNES.
\end{acknowledgements}

\bibliographystyle{aa}
\bibliography{biblio}

\begin{thebibliography}{76}
\expandafter\ifx\csname natexlab\endcsname\relax\def\natexlab#1{#1}\fi

\bibitem[{{Ahlborn} {et~al.}(2022){Ahlborn}, {Kupka}, {Weiss}, \& {Flaskamp}}]{Ahlborn2022}
{Ahlborn}, F., {Kupka}, F., {Weiss}, A., \& {Flaskamp}, M. 2022, \aap, 667, A97

\bibitem[{{Almeida} \& {Navarro-Martinez}(2021)}]{Almeida2021}
{Almeida}, Y. \& {Navarro-Martinez}, S. 2021, Physics of Fluids, 33, 035155

\bibitem[{{Baglin} {et~al.}(2006{\natexlab{a}}){Baglin}, {Auvergne}, {Barge}, {Deleuil}, {Catala}, {Michel}, {Weiss}, \& {COROT Team}}]{Baglin2006b}
{Baglin}, A., {Auvergne}, M., {Barge}, P., {et~al.} 2006{\natexlab{a}}, in ESA Special Publication, Vol. 1306, The CoRoT Mission Pre-Launch Status - Stellar Seismology and Planet Finding, ed. M.~{Fridlund}, A.~{Baglin}, J.~{Lochard}, \& L.~{Conroy}, 33

\bibitem[{{Baglin} {et~al.}(2006{\natexlab{b}}){Baglin}, {Auvergne}, {Boisnard}, {Lam-Trong}, {Barge}, {Catala}, {Deleuil}, {Michel}, \& {Weiss}}]{Baglin2006a}
{Baglin}, A., {Auvergne}, M., {Boisnard}, L., {et~al.} 2006{\natexlab{b}}, in 36th COSPAR Scientific Assembly, Vol.~36, 3749

\bibitem[{{Bakosi} {et~al.}(2008){Bakosi}, {Franzese}, \& {Boybeyi}}]{Bakosi2008}
{Bakosi}, J., {Franzese}, P., \& {Boybeyi}, Z. 2008, Journal of Computational Physics, 227, 5896

\bibitem[{{Balmforth}(1992{\natexlab{a}})}]{Balmforth1992a}
{Balmforth}, N.~J. 1992{\natexlab{a}}, \mnras, 255, 603

\bibitem[{{Balmforth}(1992{\natexlab{b}})}]{Balmforth1992b}
{Balmforth}, N.~J. 1992{\natexlab{b}}, \mnras, 255, 632

\bibitem[{{Belkacem} {et~al.}(2012){Belkacem}, {Dupret}, {Baudin}, {Appourchaux}, {Marques}, \& {Samadi}}]{Belkacem2012}
{Belkacem}, K., {Dupret}, M.~A., {Baudin}, F., {et~al.} 2012, \aap, 540, L7

\bibitem[{{Belkacem} {et~al.}(2021){Belkacem}, {Kupka}, {Philidet}, \& {Samadi}}]{Belkacem2021}
{Belkacem}, K., {Kupka}, F., {Philidet}, J., \& {Samadi}, R. 2021, \aap, 646, L5

\bibitem[{{Belkacem} {et~al.}(2019){Belkacem}, {Kupka}, {Samadi}, \& {Grimm-Strele}}]{Belkacem2019}
{Belkacem}, K., {Kupka}, F., {Samadi}, R., \& {Grimm-Strele}, H. 2019, \aap, 625, A20

\bibitem[{{Belkacem} \& {Samadi}(2013)}]{Belkacem2013}
{Belkacem}, K. \& {Samadi}, R. 2013, in Lecture Notes in Physics, Berlin Springer Verlag, ed. M.~{Goupil}, K.~{Belkacem}, C.~{Neiner}, F.~{Ligni{\`e}res}, \& J.~J. {Green}, Vol. 865 (Springer), 179

\bibitem[{{B{\"o}hm-Vitense}(1958)}]{BohmVitense1958}
{B{\"o}hm-Vitense}, E. 1958, \zap, 46, 108

\bibitem[{{B{\"o}hm-Vitense}(1992)}]{BohmVitense1992}
{B{\"o}hm-Vitense}, E. 1992, {Introduction to stellar astrophysics. Volume 3. Stellar structure and evolution.}, Vol.~3 (Cambridge University Press)

\bibitem[{{Borucki} {et~al.}(2010){Borucki}, {Koch}, {Basri}, {Batalha}, {Brown}, {Caldwell}, {Caldwell}, {Christensen-Dalsgaard}, {Cochran}, {DeVore}, {Dunham}, {Dupree}, {Gautier}, {Geary}, {Gilliland}, {Gould}, {Howell}, {Jenkins}, {Kondo}, {Latham}, {Marcy}, {Meibom}, {Kjeldsen}, {Lissauer}, {Monet}, {Morrison}, {Sasselov}, {Tarter}, {Boss}, {Brownlee}, {Owen}, {Buzasi}, {Charbonneau}, {Doyle}, {Fortney}, {Ford}, {Holman}, {Seager}, {Steffen}, {Welsh}, {Rowe}, {Anderson}, {Buchhave}, {Ciardi}, {Walkowicz}, {Sherry}, {Horch}, {Isaacson}, {Everett}, {Fischer}, {Torres}, {Johnson}, {Endl}, {MacQueen}, {Bryson}, {Dotson}, {Haas}, {Kolodziejczak}, {Van Cleve}, {Chandrasekaran}, {Twicken}, {Quintana}, {Clarke}, {Allen}, {Li}, {Wu}, {Tenenbaum}, {Verner}, {Bruhweiler}, {Barnes}, \& {Prsa}}]{Borucki2010}
{Borucki}, W.~J., {Koch}, D., {Basri}, G., {et~al.} 2010, Science, 327, 977

\bibitem[{{Borue} \& {Orszag}(1995)}]{Borue1995}
{Borue}, V. \& {Orszag}, S.~A. 1995, \pre, 51, R856

\bibitem[{{Cai}(2018)}]{Cai2018}
{Cai}, T. 2018, \apj, 868, 12

\bibitem[{{Canuto}(1997)}]{Canuto1997}
{Canuto}, V.~M. 1997, \apj, 482, 827

\bibitem[{{Christensen-Dalsgaard} \& {Thompson}(1997)}]{JCD1997}
{Christensen-Dalsgaard}, J. \& {Thompson}, M.~J. 1997, \mnras, 284, 527

\bibitem[{{Das} \& {Durbin}(2005)}]{Das2005}
{Das}, S.~K. \& {Durbin}, P.~A. 2005, Physics of Fluids, 17, 025109

\bibitem[{{Delarue} \& {Pope}(1997)}]{Delarue1997}
{Delarue}, B.~J. \& {Pope}, S.~B. 1997, Physics of Fluids, 9, 2704

\bibitem[{{Delarue} \& {Pope}(1998)}]{Delarue1998}
{Delarue}, B.~J. \& {Pope}, S.~B. 1998, Physics of Fluids, 10, 487

\bibitem[{{Dopazo}(1975)}]{Dopazo1975}
{Dopazo}, C. 1975, Physics of Fluids, 18, 397

\bibitem[{{Dreeben} \& {Pope}(1998)}]{Dreeben1998}
{Dreeben}, T.~D. \& {Pope}, S.~B. 1998, Journal of Fluid Mechanics, 357, 141

\bibitem[{{Gough}(1977)}]{Gough1977a}
{Gough}, D.~O. 1977, \apj, 214, 196

\bibitem[{{Grigahc{\`e}ne} {et~al.}(2005){Grigahc{\`e}ne}, {Dupret}, {Gabriel}, {Garrido}, \& {Scuflaire}}]{Grigahcene2005}
{Grigahc{\`e}ne}, A., {Dupret}, M.~A., {Gabriel}, M., {Garrido}, R., \& {Scuflaire}, R. 2005, \aap, 434, 1055

\bibitem[{{Grigahc{\`e}ne} {et~al.}(2012){Grigahc{\`e}ne}, {Dupret}, {Sousa}, {Monteiro}, {Garrido}, {Scuflaire}, \& {Gabriel}}]{Grigahcene2012}
{Grigahc{\`e}ne}, A., {Dupret}, M.~A., {Sousa}, S.~G., {et~al.} 2012, \mnras, 422, L43

\bibitem[{Haworth(2010)}]{Haworth2010}
Haworth, D.~C. 2010, Progress in Energy and combustion Science, 36, 168

\bibitem[{{Haworth} \& {Pope}(1987)}]{Haworth1987}
{Haworth}, D.~C. \& {Pope}, S.~B. 1987, Physics of Fluids, 30, 1026

\bibitem[{Heinz(2003)}]{Heinz2003}
Heinz, S. 2003, Statistical mechanics of turbulent flows (Springer)

\bibitem[{{Houdek} \& {Dupret}(2015)}]{Houdek2015}
{Houdek}, G. \& {Dupret}, M.-A. 2015, Living Reviews in Solar Physics, 12, 8

\bibitem[{{Houdek} {et~al.}(2019){Houdek}, {Lund}, {Trampedach}, {Christensen-Dalsgaard}, {Handberg}, \& {Appourchaux}}]{Houdek2019}
{Houdek}, G., {Lund}, M.~N., {Trampedach}, R., {et~al.} 2019, \mnras, 487, 595

\bibitem[{{Houdek} {et~al.}(2017){Houdek}, {Trampedach}, {Aarslev}, \& {Christensen-Dalsgaard}}]{Houdek2017}
{Houdek}, G., {Trampedach}, R., {Aarslev}, M.~J., \& {Christensen-Dalsgaard}, J. 2017, \mnras, 464, L124

\bibitem[{{Janicka} {et~al.}(1979){Janicka}, {Kolbe}, \& {Kollmann}}]{Janicka1979}
{Janicka}, J., {Kolbe}, W., \& {Kollmann}, W. 1979, Journal of Non Equilibrium Thermodynamics, 4, 47

\bibitem[{{Kippenhahn} {et~al.}(1967){Kippenhahn}, {Weigert}, \& {Hofmeister}}]{Kippenhahn1967}
{Kippenhahn}, R., {Weigert}, A., \& {Hofmeister}, E. 1967, Methods in Computational Physics, 7, 129

\bibitem[{Korzilius {et~al.}(2016)Korzilius, Schilders, \& Anthonissen}]{Korzilius2016}
Korzilius, S., Schilders, W.~H., \& Anthonissen, M.~J. 2016, Journal of Applied Mathematics and Physics, 5, 168

\bibitem[{{Kupka} {et~al.}(2022){Kupka}, {Ahlborn}, \& {Weiss}}]{Kupka2022}
{Kupka}, F., {Ahlborn}, F., \& {Weiss}, A. 2022, \aap, 667, A96

\bibitem[{{Kupka} \& {Muthsam}(2017)}]{Kupka2017}
{Kupka}, F. \& {Muthsam}, H.~J. 2017, Living Reviews in Computational Astrophysics, 3, 1

\bibitem[{{Kupka} \& {Robinson}(2007)}]{Kupka2007}
{Kupka}, F. \& {Robinson}, F.~J. 2007, \mnras, 374, 305

\bibitem[{Launder \& Spalding(1974)}]{Launder1974a}
Launder, B. \& Spalding, D. 1974, Computer Methods in Applied Mechanics and Engineering

\bibitem[{{Launder} \& {Sharma}(1974)}]{Launder1974b}
{Launder}, B.~E. \& {Sharma}, B.~I. 1974, Letters Heat Mass Transfer, 1, 131

\bibitem[{{Lebreton} {et~al.}(2014{\natexlab{a}}){Lebreton}, {Goupil}, \& {Montalb{\'a}n}}]{Lebreton2014a}
{Lebreton}, Y., {Goupil}, M.~J., \& {Montalb{\'a}n}, J. 2014{\natexlab{a}}, in EAS Publications Series, Vol.~65, EAS Publications Series, ed. Y.~{Lebreton}, D.~{Valls-Gabaud}, \& C.~{Charbonnel}, 99--176

\bibitem[{{Lebreton} {et~al.}(2014{\natexlab{b}}){Lebreton}, {Goupil}, \& {Montalb{\'a}n}}]{Lebreton2014b}
{Lebreton}, Y., {Goupil}, M.~J., \& {Montalb{\'a}n}, J. 2014{\natexlab{b}}, in EAS Publications Series, Vol.~65, EAS Publications Series, ed. Y.~{Lebreton}, D.~{Valls-Gabaud}, \& C.~{Charbonnel}, 177--223

\bibitem[{{Liu} {et~al.}(2004){Liu}, {Liu}, \& {Li}}]{Liu2004}
{Liu}, G.~R., {Liu}, M.~B., \& {Li}, S. 2004, Computational Mechanics, 33, 491

\bibitem[{{Mosser} {et~al.}(2012){Mosser}, {Goupil}, {Belkacem}, {Marques}, {Beck}, {Bloemen}, {De Ridder}, {Barban}, {Deheuvels}, {Elsworth}, {Hekker}, {Kallinger}, {Ouazzani}, {Pinsonneault}, {Samadi}, {Stello}, {Garc{\'\i}a}, {Klaus}, {Li}, {Mathur}, \& {Morris}}]{Mosser2012a}
{Mosser}, B., {Goupil}, M.~J., {Belkacem}, K., {et~al.} 2012, \aap, 548, A10

\bibitem[{{Ot{\'\i} Floranes} {et~al.}(2005){Ot{\'\i} Floranes}, {Christensen-Dalsgaard}, \& {Thompson}}]{OtiFloranes2005}
{Ot{\'\i} Floranes}, H., {Christensen-Dalsgaard}, J., \& {Thompson}, M.~J. 2005, \mnras, 356, 671

\bibitem[{{Philidet} {et~al.}(2021){Philidet}, {Belkacem}, \& {Goupil}}]{Philidet2021}
{Philidet}, J., {Belkacem}, K., \& {Goupil}, M.~J. 2021, \aap, 656, A95

\bibitem[{{Philidet} {et~al.}(2022){Philidet}, {Belkacem}, \& {Goupil}}]{Philidet2022}
{Philidet}, J., {Belkacem}, K., \& {Goupil}, M.~J. 2022, \aap, 664, A164

\bibitem[{Platen(1995)}]{Platen1995}
Platen, E. 1995, Mathematics and computers in simulation, 38, 69

\bibitem[{Pope(1979)}]{Pope1979}
Pope, S. 1979, Combustion and Flame, 35, 41

\bibitem[{{Pope}(1981)}]{Pope1981}
{Pope}, S.~B. 1981, Physics of Fluids, 24, 588

\bibitem[{{Pope}(1991)}]{Pope1991}
{Pope}, S.~B. 1991, Physics of Fluids A, 3, 1947

\bibitem[{{Pope}(1994{\natexlab{a}})}]{Pope1994a}
{Pope}, S.~B. 1994{\natexlab{a}}, Annual Review of Fluid Mechanics, 26, 23

\bibitem[{{Pope}(1994{\natexlab{b}})}]{Pope1994b}
{Pope}, S.~B. 1994{\natexlab{b}}, Physics of Fluids, 6, 973

\bibitem[{{Pope}(2000)}]{Pope2000}
{Pope}, S.~B. 2000, {Turbulent Flows} (Springer)

\bibitem[{{Pope} \& {Chen}(1990)}]{Pope1990}
{Pope}, S.~B. \& {Chen}, Y.~L. 1990, Physics of Fluids A, 2, 1437

\bibitem[{{Rauer} {et~al.}(2025){Rauer}, {Aerts}, {Cabrera}, {Deleuil}, {Erikson}, {Gizon}, {Goupil}, {Heras}, {Walloschek}, {Lorenzo-Alvarez}, {Marliani}, {Martin-Garcia}, {Mas-Hesse}, {O'Rourke}, {Osborn}, {Pagano}, {Piotto}, {Pollacco}, {Ragazzoni}, {Ramsay}, {Udry}, {Appourchaux}, {Benz}, {Brandeker}, {G{\"u}del}, {Janot-Pacheco}, {Kabath}, {Kjeldsen}, {Min}, {Santos}, {Smith}, {Suarez}, {Werner}, {Aboudan}, {Abreu}, {Acu{\~n}a}, {Adams}, {Adibekyan}, {Affer}, {Agneray}, {Agnor}, {Aguirre B{\o}rsen-Koch}, {Ahmed}, {Aigrain}, {Al-Bahlawan}, {Alcacera Gil}, {Alei}, {Alencar}, {Alexander}, {Alfonso-Garz{\'o}n}, {Alibert}, {Allende Prieto}, {Almeida}, {Alonso Sobrino}, {Altavilla}, {Althaus}, {Alvarez Trujillo}, {Amarsi}, {Ammler-von Eiff}, {Am{\^o}res}, {Andrade}, {Antoniadis-Karnavas}, {Ant{\'o}nio}, {Aparicio del Moral}, {Appolloni}, {Arena}, {Armstrong}, {Aroca Aliaga}, {Asplund}, {Audenaert}, {Auricchio}, {Avelino}, {Baeke}, {Bailli{\'e}}, {Balado}, {Ballber Balaguer{\'o}}, {Balestra}, {Ball}, {Ballans}, {Ballot}, {Barban}, {Barbary}, {Barbieri}, {Barcel{\'o} Forteza}, {Barker}, {Barklem}, {Barnes}, {Barrado Navascues}, {Barragan}, {Baruteau}, {Basu}, {Baudin}, {Baumeister}, {Bayliss}, {Bazot}, {Beck}, {Belkacem}, {Bellinger}, {Benatti}, {Benomar}, {B{\'e}rard}, {Bergemann}, {Bergomi}, {Bernardo}, {Biazzo}, {Bignamini}, {Bigot}, {Billot}, {Binet}, {Biondi}, {Biondi}, {Birch}, {Bitsch}, {Bluhm Ceballos}, {B{\'o}di}, {Bogn{\'a}r}, {Boisse}, {Bolmont}, {Bonanno}, {Bonavita}, {Bonfanti}, {Bonfils}, {Bonito}, {Bonomo}, {B{\"o}rner}, {Boro Saikia}, {Borreguero Mart{\'\i}n}, {Borsa}, {Borsato}, {Bossini}, {Bouchy}, {Bou{\'e}}, {Boufleur}, {Boumier}, {Bourrier}, {Bowman}, {Bozzo}, {Bradley}, {Bray}, {Bressan}, {Breton}, {Brienza}, {Brito}, {Brogi}, {Brown}, {Brown}, {Brun}, {Bruno}, {Bruns}, {Buchhave}, {Bugnet}, {Buldgen}, {Burgess}, {Busatta}, {Busso}, {Buzasi}, {Caballero}, {Cabral}, {Cabrero Gomez}, {Calderone}, {Cameron}, {Cameron}, {Campante}, {Campos Gestal}, {Canto Martins}, {Cara}, {Carone}, {Carrasco}, {Casagrande}, {Casewell}, {Cassisi}, {Castellani}, {Castro}, {Catala}, {Catal{\'a}n Fern{\'a}ndez}, {Catelan}, {Cegla}, {Cerruti}, {Cessa}, {Chadid}, {Chaplin}, {Charpinet}, {Chiappini}, {Chiarucci}, {Chiavassa}, {Chinellato}, {Chirulli}, {Christensen-Dalsgaard}, {Church}, {Claret}, {Clarke}, {Claudi}, {Clermont}, {Coelho}, {Coelho}, {Cogato}, {Colom{\'e}}, {Condamin}, {Conde Garc{\'\i}a}, \& {Conseil}}]{Rauer2025}
{Rauer}, H., {Aerts}, C., {Cabrera}, J., {et~al.} 2025, Experimental Astronomy, 59, 26

\bibitem[{{Rincon} {et~al.}(2025){Rincon}, {Barr{\`e}re}, \& {Roudier}}]{Rincon2025}
{Rincon}, F., {Barr{\`e}re}, P., \& {Roudier}, T. 2025, \aap, 696, A143

\bibitem[{Rodean(1996)}]{Rodean1996}
Rodean, H.~C. 1996, Stochastic Lagrangian models of turbulent diffusion, Vol.~45 (Springer)

\bibitem[{Rodi(1980)}]{Rodi1980}
Rodi, W. 1980, State of the Arts Paper, IAHR

\bibitem[{{Rosenthal} {et~al.}(1999){Rosenthal}, {Christensen-Dalsgaard}, {Nordlund}, {Stein}, \& {Trampedach}}]{Rosenthal1999}
{Rosenthal}, C.~S., {Christensen-Dalsgaard}, J., {Nordlund}, {\r{A}}., {Stein}, R.~F., \& {Trampedach}, R. 1999, \aap, 351, 689

\bibitem[{{Roxburgh} \& {Vorontsov}(2003)}]{Roxburgh2003}
{Roxburgh}, I.~W. \& {Vorontsov}, S.~V. 2003, \aap, 411, 215

\bibitem[{{Samadi} {et~al.}(2015){Samadi}, {Belkacem}, \& {Sonoi}}]{Samadi2015}
{Samadi}, R., {Belkacem}, K., \& {Sonoi}, T. 2015, in EAS Publications Series, Vol. 73-74, EAS Publications Series, 111--191

\bibitem[{{Samadi} \& {Goupil}(2001)}]{Samadi2001a}
{Samadi}, R. \& {Goupil}, M.~J. 2001, \aap, 370, 136

\bibitem[{{Samadi} {et~al.}(2001){Samadi}, {Goupil}, \& {Lebreton}}]{Samadi2001b}
{Samadi}, R., {Goupil}, M.~J., \& {Lebreton}, Y. 2001, \aap, 370, 147

\bibitem[{{Schou} \& {Birch}(2020)}]{Schou2020}
{Schou}, J. \& {Birch}, A.~C. 2020, \aap, 638, A51

\bibitem[{{Sonoi} {et~al.}(2017){Sonoi}, {Belkacem}, {Dupret}, {Samadi}, {Ludwig}, {Caffau}, \& {Mosser}}]{Sonoi2017}
{Sonoi}, T., {Belkacem}, K., {Dupret}, M.~A., {et~al.} 2017, \aap, 600, A31

\bibitem[{Taylor(1922)}]{Taylor1922}
Taylor, G.~I. 1922, Proceedings of the london mathematical society, 2, 196

\bibitem[{{Unno}(1967)}]{Unno1967}
{Unno}, W. 1967, \pasj, 19, 140

\bibitem[{{Van Slooten} \& {Jayesh}(1998)}]{VanSlooten1998}
{Van Slooten}, P.~R. \& {Jayesh}, Pope, S.~B. 1998, Physics of Fluids, 10, 246

\bibitem[{{Welton}(1998)}]{Welton1998}
{Welton}, W.~C. 1998, Journal of Computational Physics, 139, 410

\bibitem[{{Welton} \& {Pope}(1997)}]{Welton1997}
{Welton}, W.~C. \& {Pope}, S.~B. 1997, Journal of Computational Physics, 134, 150

\bibitem[{{Xiong} {et~al.}(1997){Xiong}, {Cheng}, \& {Deng}}]{Xiong1997}
{Xiong}, D.~R., {Cheng}, Q.~L., \& {Deng}, L. 1997, \apjs, 108, 529

\bibitem[{{Xiong} {et~al.}(2000){Xiong}, {Cheng}, \& {Deng}}]{Xiong2000}
{Xiong}, D.~R., {Cheng}, Q.~L., \& {Deng}, L. 2000, \mnras, 319, 1079

\bibitem[{{Yeung} \& {Pope}(1989)}]{Yeung1989}
{Yeung}, P.~K. \& {Pope}, S.~B. 1989, Journal of Fluid Mechanics, 207, 531

\bibitem[{{Zahn}(1991)}]{Zahn1991}
{Zahn}, J.~P. 1991, \aap, 252, 179

\bibitem[{{Zhou} {et~al.}(2019){Zhou}, {Asplund}, \& {Collet}}]{Zhou2019}
{Zhou}, Y., {Asplund}, M., \& {Collet}, R. 2019, \apj, 880, 13

\end{thebibliography}

\begin{appendix}

\section{Derivation of the Lagrangian stochastic model \label{app:lagrangian_model}}

    We proceed in two steps: first, in App.~\ref{app:lagrangian_model_velocity_energy}, we assume a fixed turbulent frequency $\omega$ for all the fluid particles, and we derive the coefficients in the velocity and energy equations Eqs.~\ref{eq:SDE_velocity} and \ref{eq:SDE_energy}. Then, in App.~\ref{app:lagrangian_model_omega}, we introduce fluctuations of the turbulent frequency, derive the coefficients in its equation Eq.~\ref{eq:SDE_frequency}, and determine how the coefficients in the velocity and energy equations should be modified as a result.

    \subsection{Lagrangian model with fixed turbulent frequency \label{app:lagrangian_model_velocity_energy}}
        
        In order to determine the coefficients in Eqs.~\ref{eq:SDE_velocity} and \ref{eq:SDE_energy}, we first derive the equivalent Fokker-Planck equation (i.e. the transport equation for the joint velocity-energy PDF of the flow) in App.~\ref{app:equivalent_fokker_planck}. Then, in App.~\ref{app:equivalent_reynolds_stress}, we deduce the equivalent transport equations for all the means (mean density, velocity and energy) and all the second-order moments (Reynolds stress tensor, internal energy variance and internal energy flux) of the turbulent flow. Finally, in App.~\ref{app:comparison_mean_equations}, we compare them to the exact equations derived from first principles, which constrains the form of the coefficients that should be adopted in our Lagrangian stochastic model.

        \subsubsection{Equivalent Fokker-Planck equation \label{app:equivalent_fokker_planck}}
    
            Formally, the system of stochastic differential equations can be written
            \begin{equation}
                \d X_i = a_i(\mathbf{X},t) ~ \d t + b_{ij}(\mathbf{X},t) ~ \d W_j ~,
            \end{equation}
            where $\mathbf{X}$ is the multivariate stochastic process composed of the position $\mathbf{x}^\ast$ of the particle, its velocity $\mathbf{u}^\ast$, and its specific internal energy $e^\ast$. The equivalent Fokker-Planck equation is
            \begin{equation}
                \dfrac{\partial f_L^\ast}{\partial t} = -\dfrac{\partial A_i f_L^\ast}{\partial X_i} + \dfrac{1}{2} \dfrac{\partial^2 B_{ij} f_L^\ast}{\partial X_i \partial X_j} ~,
            \end{equation}
            where $A_i = a_i$, $B_{ij} = b_{ik} b_{kj}$. Using Eqs.~\ref{eq:SDE_position}, \ref{eq:SDE_velocity} and \ref{eq:SDE_energy}, and assuming that the drift coefficients in the latter two can be put in the general form of Eqs.~\ref{eq:drift_velocity} and \ref{eq:drift_energy}, the Lagrangian Fokker-Planck equation reads
            \begin{multline}
                \dfrac{\partial f_L^\ast}{\partial t} = -\dfrac{\partial}{\partial x_i^\ast} \left( u_i^\ast f_L^\ast \right) - \dfrac{\partial}{\partial u_i^\ast} \left( \left[ G_{0i} + G_{ij}\left( u_j^\ast - \widetilde{u_j} \right) + G_{ei} \left( e^\ast - \widetilde{e} \right) \right] f_L^\ast \right) \\
                - \dfrac{\partial}{\partial e^\ast} \left( \left[ K_0 + K_j \left( u_j^\ast - \widetilde{u_j} \right) + K_e \left( e^\ast - \widetilde{e} \right) \right] f_L^\ast \right) \\
                + \dfrac{1}{2} \dfrac{\partial^2}{\partial u_i^\ast \partial u_i^\ast} \left( b_u^2 f_L^\ast \right) + \dfrac{1}{2} \dfrac{\partial^2}{\partial e^\ast \partial e^\ast} \left( b_e^2 f_L^\ast \right) ~.
                \label{eq:lagrangian_fokker_planck}
            \end{multline}
            The function $f_L^\ast(\mathbf{x}^\ast, \mathbf{u}^\ast, e^\ast, t | \mathbf{x_0}, t_0)$ is defined as the PDF of the position, velocity and specific energy at time $t$ of the fluid particle which was located at $\mathbf{x_0}$ at the initial time $t_0$. Since the marginal PDF of the initial position of the particle is simply equal to the initial gas density $\rho(\mathbf{x_0}, t_0)$, the non-conditional Lagrangian PDF $f_L(\mathbf{x}^\ast, \mathbf{u}^\ast, e^\ast, t)$ is given by
            \begin{equation}
                f_L(\mathbf{x}^\ast, \mathbf{u}^\ast, e^\ast, t) = \dint \d^3\mathbf{x_0} ~ \rho(\mathbf{x_0}, t_0) f_L^\ast(\mathbf{x}^\ast, \mathbf{u}^\ast, e^\ast, t | \mathbf{x_0}, t_0) ~,
            \end{equation}
            which means that the Lagrangian Fokker-Planck equation on $f_L$ is obtained by multiplying Eq.~\ref{eq:lagrangian_fokker_planck} by $\rho(\mathbf{x_0}, t_0)$ and integrating over $\mathbf{x_0}$.

            It can be shown \citep[e.g.][]{Pope2000} that the Lagrangian PDF is related to the Eulerian PDF $f_E(\rho, \mathbf{u}, e, t ; (\mathbf{x}, t))$ through
            \begin{multline}
                \dint \d \rho' ~ \rho' f_E(\rho', \mathbf{u}, e ; (\mathbf{x}, t)) \\
                = \dint \d^3\mathbf{x_0} ~ \rho(\mathbf{x_0}, t_0) f_L^\ast(\mathbf{x}, \mathbf{u}, e, t | \mathbf{x_0}, t_0) ~.
            \end{multline}
            We note that the density variable is described by the Eulerian PDF but not by the Lagrangian PDF, where the information concerning the local density of the flow is contained in the dependence of $f_L^\ast$ on $\mathbf{x}^\ast$. The Fokker-Planck equation for the Eulerian PDF of the flow then reads
            \begin{multline}
                \dfrac{\partial}{\partial t} \left( \dint \d \rho' \rho' f_E \right) = -\dfrac{\partial}{\partial x_i} \left( \dint \d \rho' \rho' u_i f_E \right) \\
                - \dfrac{\partial}{\partial u_i} \left( \dint \d \rho' \rho' \left[ G_{0i} + G_{ij}\left( u_j^\ast - \widetilde{u_j} \right) + G_{ei} \left( e^\ast - \widetilde{e} \right) \right] f_E \right) \\
                - \dfrac{\partial}{\partial e} \left( \dint \d \rho' \rho' \left[ K_0 + K_j \left( u_j^\ast - \widetilde{u_j} \right) + K_e \left( e^\ast - \widetilde{e} \right) \right] f_E \right) \\
                + \dfrac{1}{2} \dfrac{\partial^2}{\partial u_i \partial u_i} \left( \dint \d \rho' \rho' b_u^2 f_E \right) + \dfrac{1}{2} \dfrac{\partial^2}{\partial e^2} \left( \dint \d\rho' ~ \rho' b_e^2 f_E \right) ~.
                \label{eq:eulerian_fokker_planck}
            \end{multline}
    
        \subsubsection{Equivalent mean equations \label{app:equivalent_reynolds_stress}}

            By definition, the mean density, mean density-weighted velocity, mean density-weighted specific internal energy, Reynolds-stress tensor, energy variance, and internal energy flux are given by
            \begin{align}
                & \rhom = \dint \d\rho \d^3\mathbf{u} \d e ~ \rho f_E ~, \\
                & \widetilde{\mathbf{u}} = \dfrac{1}{\rhom} \dint \d\rho \d^3\mathbf{u} \d e ~ \rho \mathbf{u} f_E ~, \\
                & R_{ij} = \dfrac{1}{\rhom} \dint \d\rho \d^3\mathbf{u} \d e ~ \rho \left( u_i - \widetilde{u_i} \right) \left( u_j - \widetilde{u_j} \right) f_E ~, \\
                & k_e = \dfrac{1}{\rhom} \dint \d\rho \d^3\mathbf{u} \d e ~ \rho \left( e - \widetilde{e} \right)^2 f_E ~, \\
                & \mathbf{\fe} = \dfrac{1}{\rhom} \dint \d\rho \d^3\mathbf{u} \d e ~ \rho \left( u_i - \widetilde{u_i} \right) \left( e - \widetilde{e} \right) f_E ~.
            \end{align}
            Transport equations for each of these are obtained by multiplying Eq.~\ref{eq:eulerian_fokker_planck} by the appropriate function of $\mathbf{u}$ and $e$, and integrating over all values thereof. We find
            \begin{equation}
                \dfrac{\partial \rhom}{\partial t} + \dfrac{\partial \rhom \um_i}{\partial x_i} = 0 ~,
                \label{eq:continuity}
            \end{equation}
            \begin{equation}
                \dfrac{\partial \rhom \um_i}{\partial t} + \dfrac{\partial \rhom \um_i \um_j}{\partial x_j} + \dfrac{\partial \rhom R_{ij}}{\partial x_j} = \rhom G_{0i} ~,
                \label{eq:mean_velocity}
            \end{equation}
            \begin{equation}
                \dfrac{\partial \rhom \enm}{\partial t} + \dfrac{\partial \rhom \enm \um_i}{\partial x_i} + \dfrac{\partial \rhom \fe}{\partial x_i} = \rhom K_0 ~,
                \label{eq:mean_energy}
            \end{equation}
            \begin{multline}
                \dfrac{\partial \rhom R_{ij}}{\partial t} + \dfrac{\partial \rhom \um_k R_{ij}}{\partial x_k} + \dfrac{\partial \rhom \widetilde{u_i''u_j''u_k''}}{\partial x_k} + \rhom R_{ik}\dfrac{\partial \um_j}{\partial x_k} + \rhom R_{jk}\dfrac{\partial \um_i}{\partial x_k} = \rhom G_{ik} R_{jk} \\
                + \rhom G_{jk} R_{ik} + \rhom G_{ei} \fe_j + \rhom G_{ej} \fe_i + \rhom b_u^2 \delta_{ij} ~,
                \label{eq:reynolds_stress}
            \end{multline}
            \begin{multline}
                \dfrac{\partial \rhom k_e}{\partial t} + \dfrac{\partial \rhom \um_j k_e}{\partial x_j} + \dfrac{\partial \rhom \widetilde{e''^2 u_j''}}{\partial x_j} + 2 \rhom \fe_j \dfrac{\partial \enm}{\partial x_j} \\
                = 2 \rhom K_e k_e + 2 \rhom K_j \fe_j + \rhom b_e^2 ~,
                \label{eq:energy_variance}
            \end{multline}
            and
            \begin{multline}
                \dfrac{\partial \rhom \fe_i}{\partial t} + \dfrac{\partial \rhom \um_j \fe_i}{\partial x_j} + \dfrac{\partial \rhom \widetilde{u_i''e''u_j''}}{\partial x_j} + \rhom R_{ij}\dfrac{\partial \enm}{\partial x_j} + \rhom \fe_j \dfrac{\partial \um_i}{\partial x_j} \\
                = \rhom G_{ij} \fe_j + \rhom G_{ei} k_e + \rhom K_j R_{ij} + \rhom K_e \fe_i ~,
                \label{eq:convective_flux}
            \end{multline}
            where $\mathbf{u}'' \equiv \mathbf{u} - \mathbf{\um}$ and $e'' \equiv e - \enm$.

            The diffusion coefficient in the velocity SDE Eq.~\ref{eq:SDE_velocity} is related to the Lagrangian structure function
            \begin{equation}
                D_L(\Delta t) \equiv \left\langle |\mathbf{u}^\ast(t+\Delta t) - \mathbf{u}^\ast(t)|^2 \right\rangle
                \label{eq:definition_structure_function}
            \end{equation}
            through $D_L(\Delta t) = b_u^2 \Delta t$. But under the Kolmogorov hypotheses, the structure function is predicted to be
            \begin{equation}
                D_L(\Delta t) = C_0 \epsilon \Delta t ~,
                \label{eq:kolmogorov_structure_function}
            \end{equation}
            where $C_0$ is the Kolmogorov constant, and $\epsilon \equiv \omega k$ is the turbulent dissipation rate ($k \equiv R_{ii} / 2$ is the turbulent kinetic energy). For the moment, $\omega$ is assumed constant (the effect of a fluctuating turbulent frequency is introduced in App.~\ref{app:lagrangian_model_omega}). The correct structure function is recovered by the Lagrangian stochastic model if the velocity diffusion coefficient is
            \begin{equation}
                b_u = \sqrt{C_0 \omega k} ~.
                \label{eqapp:bu_omegafixed}
            \end{equation}
            By analogy, we also adopt a similar expression for the energy diffusion coefficient
            \begin{equation}
                b_e = \sqrt{C_1 \omega k_e} ~,
                \label{eqapp:be_omegafixed}
            \end{equation}
            where $C_1$ is another dimensionless constant.
    
        \subsubsection{Constraints from first-principles mean equations \label{app:comparison_mean_equations}}
            
            The exact mean equations, derived from the equations of hydrodynamics, are
            \begin{equation}
                \dfrac{\partial \rhom}{\partial t} + \dfrac{\partial \rhom \um_i}{\partial x_i} = 0 ~,
                \label{eq:continuity_exact}
            \end{equation}
            \begin{equation}
                \dfrac{\partial \rhom \um_i}{\partial t} + \dfrac{\partial \rhom \um_i \um_j}{\partial x_j} + \dfrac{\partial \rhom R_{ij}}{\partial x_j} = -\dfrac{\partial \pgm}{\partial x_i} + g_i ~,
                \label{eq:mean_velocity_exact}
            \end{equation}
            \begin{multline}
                \dfrac{\partial \rhom \enm}{\partial t} + \dfrac{\partial \rhom \enm \um_i}{\partial x_i} + \dfrac{\partial \rhom \fe_i}{\partial x_i} \\
                = -\pgm \dfrac{\partial \um_i}{\partial x_i} - \pgm \dfrac{\partial \overline{u_i''}}{\partial x_i} - \overline{p' \dfrac{\partial u_i''}{\partial x_i}} + \rhom \kappa \dfrac{\partial^2 \enm}{\partial x_i \partial x_i} ~,
                \label{eq:mean_energy_exact}
            \end{multline}
            \begin{multline}
                \dfrac{\partial \rhom R_{ij}}{\partial t} + \dfrac{\partial \rhom \um_k R_{ij}}{\partial x_k} + \dfrac{\partial \rhom \widetilde{u_i''u_j''u_k''}}{\partial x_k} + \rhom R_{ik}\dfrac{\partial \um_j}{\partial x_k} + \rhom R_{jk}\dfrac{\partial \um_i}{\partial x_k} \\
                = -\left( \overline{u_i''} \dfrac{\partial \pgm}{\partial x_j} + \overline{u_j''} \dfrac{\partial \pgm}{\partial x_i} + \overline{u_j'' \dfrac{\partial p'}{\partial x_i}} + \overline{u_i'' \dfrac{\partial p'}{\partial x_j}} \right) \\
                + \overline{u_j''} \dfrac{\partial \overline{\sigma_{ik}}}{\partial x_k} + \overline{u_i''} \dfrac{\partial \overline{\sigma_{jk}}}{\partial x_k} + \overline{u_j'' \dfrac{\partial \sigma_{ik}'}{\partial x_k}} + \overline{u_i'' \dfrac{\partial \sigma_{jk}'}{\partial x_k}} ~,
                \label{eq:reynolds_stress_exact}
            \end{multline}
            \begin{multline}
                \dfrac{\partial \rhom k_e}{\partial t} + \dfrac{\partial \rhom \um_i k_e}{\partial x_i} + \dfrac{\partial \rhom \widetilde{e''^2 u_i''}}{\partial x_i} + 2 \rhom \fe_i \dfrac{\partial \enm}{\partial x_i} = -2 \overline{e''} \pgm \dfrac{\partial \um_i}{\partial x_i} \\
                - 2 \pgm \overline{e'' \dfrac{\partial u_i''}{\partial x_i}} - 2 \overline{p' e''} \dfrac{\partial \um_i}{\partial x_i} - 2 \overline{e'' p' \dfrac{\partial u_j''}{\partial x_j}} + 2 \overline{e''} \overline{\sigma_{ij}} \dfrac{\partial \um_i}{\partial x_j} \\
                + 2 \overline{e'' \sigma_{ij}'} \dfrac{\partial \um_i}{\partial x_j} + 2 \overline{\sigma_{ij}} \overline{e'' \dfrac{\partial u_i''}{\partial x_j}} + 2 \overline{e'' \sigma_{ij}' \dfrac{\partial u_i''}{\partial x_j}} ~,
                \label{eq:energy_variance_exact}
            \end{multline}
            and
            \begin{multline}
                \dfrac{\partial \rhom \fe_i}{\partial t} + \dfrac{\partial \rhom \um_j \fe_i}{\partial x_j} + \dfrac{\partial \rhom \widetilde{e'' u_i'' u_j''}}{\partial x_j} + \rhom R_{ij}\dfrac{\partial \enm}{\partial x_j} + \rhom \fe_j \dfrac{\partial \um_i}{\partial x_j} \\
                = - \overline{e''} \dfrac{\partial \pgm}{\partial x_i} - \overline{e'' \dfrac{\partial p'}{\partial x_i}} + \overline{e''} \dfrac{\overline{\sigma_{ij}}}{\partial x_j} + \overline{e'' \dfrac{\partial \sigma_{ij}'}{\partial x_j}} \\
                - \pgm \overline{u_i''} \dfrac{\partial \um_j}{\partial x_j} - \overline{p' u_i''} \dfrac{\partial \um_j}{\partial x_j} - \pgm \overline{u_i'' \dfrac{\partial u_j''}{\partial x_j}} - \overline{p' u_i'' \dfrac{\partial u_j''}{\partial x_j}} \\
                + \overline{u_i''} \overline{\sigma_{jk}} \dfrac{\partial \um_j}{\partial x_k} + \overline{\sigma_{jk}} \overline{u_i'' \dfrac{\partial u_j''}{\partial x_k}} + \overline{u_i'' \sigma_{jk}'} \dfrac{\partial \um_j}{\partial x_k} + \overline{u_i'' \sigma_{jk} \dfrac{\partial u_j''}{\partial x_k}} ~,
                \label{eq:convective_flux_exact}
            \end{multline}
            where $\kappa$ is the radiative diffusivity, and $\sigma_{ij}$ is the viscous tensor. While the mean radiative flux appears in Eq.~\ref{eq:mean_energy_exact}, we assumed its correlation with the velocity and energy fluctuations to be negligible, so that it does not appear in Eqs.~\ref{eq:energy_variance_exact} and \ref{eq:convective_flux_exact}.

            The continuity equation is exactly recovered by the Lagrangian stochastic model (i.e. Eqs.~\ref{eq:continuity} and \ref{eq:continuity_exact} are identical). This is because in a Lagrangian description, mass is naturally conserved, and there is no need for a dedicated equation to impose that constraint. By the same token, all the advection terms are recovered exactly (i.e. the left-hand sides of Eqs.~\ref{eq:mean_velocity}, \ref{eq:mean_energy}, \ref{eq:reynolds_stress}, \ref{eq:energy_variance} and \ref{eq:convective_flux} are identical to those of Eqs.~\ref{eq:mean_velocity_exact}, \ref{eq:mean_energy_exact}, \ref{eq:reynolds_stress_exact}, \ref{eq:energy_variance_exact} and \ref{eq:convective_flux_exact}).

            In the following, we make three important assumptions. First, we will assume that the turbulence is only weakly compressible, so that we may neglect the correlation of $\partial u_i'' / \partial x_i$ with other turbulent fluctuations like $p'$ or $e''$. Second, we will assume that the viscous tensor can be neglected everywhere except when correlated with the shear tensor. This defines the turbulent dissipation rate $\epsilon$, through
            \begin{equation}
                \overline{\sigma_{ik}' \dfrac{\partial u_j''}{\partial x_k}} = - \overline{u_j'' \dfrac{\partial \sigma_{ik}'}{\partial x_k}} = \dfrac{1}{3} \rhom \epsilon ~ \delta_{ij} ~.
            \end{equation}
            Finally, we assume that the fluctuations of density $\rho'$, gas pressure $p'$ and specific internal energy $e''$ are related to each other through a polytropic relation
            \begin{equation}
                \dfrac{p'}{\pgm} = n \dfrac{\rho'}{\rhom} = \dfrac{n}{n-1} \dfrac{\rho e''}{\rhom \enm} ~,
                \label{eq:polytropic_relation}
            \end{equation}
            where $n$ is the polytropic index. This polytropic relation directly entails
            \begin{align}
                & \overline{u_i''} = - \dfrac{\overline{\rho' u_i''}}{\rhom} = -\dfrac{1}{n-1} \dfrac{\fe_i}{\enm} ~, \\
                & \overline{e''} = - \dfrac{\overline{\rho' e''}}{\rhom} = -\dfrac{1}{n-1} \dfrac{k_e}{\enm} ~, \\
                & \overline{p' u_i''} = \dfrac{n}{n-1} \dfrac{\pgm}{\enm} \fe_i = \dfrac{n(\Gamma_1 - 1)}{n - 1} \rhom \fe_i ~, \\
                & \overline{p' e''} = \dfrac{n}{n-1} \dfrac{\pgm}{\enm} k_e = \dfrac{n(\Gamma_1 - 1)}{n - 1} \rhom k_e ~,
            \end{align}
            where we used the Reynolds average of the ideal gas law
            \begin{equation}
                \pgm = (\Gamma_1 - 1) \rhom \enm ~,
            \end{equation}
            and $\Gamma_1$ is the first adiabatic exponent.
            
            Under these assumptions, comparing the right-hand sides of Eqs.~\ref{eq:mean_velocity}, \ref{eq:mean_energy}, \ref{eq:reynolds_stress}, \ref{eq:energy_variance} and \ref{eq:convective_flux} with those of Eqs.~\ref{eq:mean_velocity_exact}, \ref{eq:mean_energy_exact}, \ref{eq:reynolds_stress_exact}, \ref{eq:energy_variance_exact} and \ref{eq:convective_flux_exact} provides the following constraints for the coefficients of the Lagrangian stochastic model
            \begin{equation}
                \rhom G_{0i} = -\dfrac{\partial \pgm}{\partial x_i} + \rhom g_i ~,
                \label{eq:constraint_mean_velocity}
            \end{equation}
            \begin{multline}
                \rhom K_0 = -\pgm \dfrac{\partial \um_i}{\partial x_i} + \dfrac{\Gamma_1 - 1}{n - 1} \left( \dfrac{\partial \rhom \fe_i}{\partial x_i} - \dfrac{\rhom \fe_i}{\pgm} \dfrac{\partial \pgm}{\partial x_i} \right) \\
                + \rhom \kappa \dfrac{\partial^2 \enm}{\partial x_i \partial x_i} + \rhom \omega k ~,
                \label{eq:constraint_mean_energy}
            \end{multline}
            \begin{multline}
                \rhom G_{ik} R_{jk} + \rhom G_{jk} R_{ik} + \rhom G_{ei} \fe_j + \rhom G_{ej} \fe_i + \rhom C_0 \omega k \delta_{ij} \\
                = \dfrac{1}{\enm (n - 1)} \left( \fe_i \dfrac{\partial \pgm}{\partial x_j} + \fe_j \dfrac{\partial \pgm}{\partial x_i} \right) \\
                - \overline{u_i'' \dfrac{\partial p'}{\partial x_j}} - \overline{u_j'' \dfrac{\partial p'}{\partial x_i}} - \dfrac{2}{3} \rhom \omega k \delta_{ij} ~,
                \label{eq:constraint_reynolds_stress}
            \end{multline}
            \begin{multline}
                2 \rhom K_e k_e + 2 \rhom K_i \fe_i + \rhom C_1 \omega k_e \\
                = - 2 (\Gamma_1 - 1) \rhom k_e \dfrac{\partial \um_i}{\partial x_i} - 2 \rhom \dfrac{2 \rhom \omega k}{\enm (n - 1)} k_e ~,
                \label{eq:constraint_energy_variance}
            \end{multline}
            and
            \begin{multline}
                \rhom G_{ij} \fe_j + \rhom G_{ei} k_e + \rhom K_j R_{ij} + \rhom K_e \fe_i = \dfrac{k_e}{\enm (n - 1)} \dfrac{\partial \pgm}{\partial x_i} \\
                - \overline{e'' \dfrac{\partial p'}{\partial x_i}} - (\Gamma_1 - 1) \rhom \fe_i \dfrac{\partial \um_j}{\partial x_j} - \dfrac{\rhom \omega k}{\enm (n - 1)} \fe_i ~.
                \label{eq:constraint_convective_flux}
            \end{multline}

        \subsubsection{Choice of velocity and energy coefficients \label{app:lagrangian_coefficients}}
            
            The coefficients $G_{0i}$ and $K_0$ are uniquely determined by Eqs.~\ref{eq:constraint_mean_velocity} and \ref{eq:constraint_mean_energy}. By contrast, there are multiple choices for $G_{ij}$, $G_{ei}$, $K_e$ and $K_j$ that satisfy Eqs.~\ref{eq:constraint_reynolds_stress}, \ref{eq:constraint_energy_variance} and \ref{eq:constraint_convective_flux}.
            
            First, let us treat Eqs.~\ref{eq:constraint_reynolds_stress}: one possible solution is to split it into the classical constraint \citep[e.g.][]{Pope1994b}
            \begin{multline}
                \rhom G_{ik} R_{jk} + \rhom G_{jk} R_{ik} + \rhom C_0 \omega k \delta_{ij} \\
                = - \overline{u_i'' \dfrac{\partial p'}{\partial x_j}} - \overline{u_j'' \dfrac{\partial p'}{\partial x_i}} - \dfrac{2}{3} \rhom \omega k \delta_{ij} ~,
                \label{eq:constraint_gij}
            \end{multline}
            and the following constraint for $G_{ei}$
            \begin{equation}
                \rhom G_{ei} \fe_j + \rhom G_{ej} \fe_i = \dfrac{1}{\enm (n - 1)} \left( \fe_i \dfrac{\partial \pgm}{\partial x_j} + \fe_j \dfrac{\partial \pgm}{\partial x_i} \right) ~.
                \label{eq:constraint_gei}
            \end{equation}
            Taking the trace of Eq.~\ref{eq:constraint_gij}, we get
            \begin{equation}
                2 \rhom G_{ij} R_{ij} + 3 \rhom C_0 \omega k = -2 \overline{u_i'' \dfrac{\partial p'}{\partial x_i}} - 2 \rhom \omega k ~.
            \end{equation}
            Under our previous assumptions, the velocity-pressure-gradient tensor can be neglected, because in weakly compressible turbulence, the divergence of the acoustic flux $\partial \overline{p' u_i''} / \partial x_i$ is a minor source of energy transport, and we already neglected $\overline{p' \partial u_i'' / \partial x_i}$. The simplest choice of $G_{ij}$ is to assume that it is isotropic, in which case the only possibility is
            \begin{equation}
                G_{ij} = -\left( \dfrac{1}{2} + \dfrac{3}{4} C_0 \right) \omega \delta_{ij} ~,
            \end{equation}
            which corresponds to the Simplified Langevin Model \citep[e.g.][]{Pope1994b}. It can be seen from Eq.~\ref{eq:constraint_gij} that this is equivalent to the Rotta model for the pressure-rate-of-strain tensor
            \begin{equation}
                \overline{p' \left( \dfrac{\partial u_i''}{\partial x_j} + \dfrac{\partial u_j''}{\partial x_i} \right)} = -C_R \left( R_{ij} - \dfrac{2}{3} k \delta_{ij} \right) ~,
            \end{equation}
            with $C_R = 1 + 3 C_0 / 2$. As for Eq.~\ref{eq:constraint_gei}, it straightforwardly gives
            \begin{equation}
                G_{ei} = \dfrac{1}{\rhom \enm (n-1)} \dfrac{\partial \pgm}{\partial x_i} = \dfrac{\Gamma_1 - 1}{n - 1} \dfrac{1}{\pgm} \dfrac{\partial \pgm}{\partial x_i} ~.
            \end{equation}

            Let us now turn to Eq.~\ref{eq:constraint_energy_variance}. All the terms are proportional to $k_e$, except for the second term on the left-hand side. A natural choice, then, is
            \begin{align}
                & K_i = 0 ~, \\
                & K_e = -\dfrac{1}{2} C_1 \omega - (\Gamma_1 - 1) \dfrac{\partial \um_i}{\partial x_i} - \dfrac{\omega k}{\enm (n-1)} ~.
            \end{align}

            Finally, using these expressions for $G_{ij}$, $G_{ei}$, $K_e$ and $K_i$, we see that Eq.~\ref{eq:constraint_convective_flux} reduces to the following closure relation for the energy-pressure-gradient tensor
            \begin{equation}
                \overline{e'' \dfrac{\partial p'}{\partial x_i}} = - C \rhom \omega \fe_i ~,
            \end{equation}
            with $C = 3 C_0 / 4 + (1-C_1)/2$.

    \subsection{Lagrangian model including turbulent frequency \label{app:lagrangian_model_omega}}

        We now lift the constraint that the value of the turbulent frequency $\omega$ be fixed, and we add a stochastic equation for $\omega^\ast$. We follow \citet{VanSlooten1998} and adopt the following stochastic differential equation
        \begin{equation}
            \d \omega^\ast = -\Omega \left( \omega^\ast - \om \right) \d t - \omega^\ast \om S_\omega \d t + \sqrt{2 \sigma^2 \omega^\ast \om \Omega} \d W ~,
        \end{equation}
        where the conditional mean turbulent frequency $\Omega$ and the source term $S_\omega$ are given by Eq.~\ref{eq:conditional_turbulent_frequency} and Eq.~\ref{eq:turbulent_frequency_source} respectively.

        The refined Kolmogorov hypotheses dictate \citep[e.g.][]{Pope1990} that the structure function $D_L(\Delta t)$ defined by Eq.~\ref{eq:definition_structure_function} be proportional to the instantaneous turbulent dissipation rate $\epsilon^\ast = k \omega^\ast$ instead of the mean dissipation rate. As a result, the diffusion coefficients $b_u$ and $b_e$ in Eq.~\ref{eq:SDE_velocity} and Eq.~\ref{eq:SDE_energy} should be given not by Eq.~\ref{eqapp:bu_omegafixed} and Eq.~\ref{eqapp:be_omegafixed}, but by
        \begin{align}
            & b_u^2 = C_0 \omega^\ast k ~, \\
            & b_e^2 = C_1 \omega^\ast k_e ~.
        \end{align}
        In order for the equivalent Fokker-Planck equation to remain identical, extra terms $A_i$ and $A_e$ should be added respectively to the coefficients $a_i$ and $a_e$ in Eqs.~\ref{eq:SDE_velocity} and \ref{eq:SDE_energy}. The corresponding extra terms in the Fokker-Planck Eq.~\ref{eq:lagrangian_fokker_planck} read
        \begin{multline}
            \Delta \mathrm{FP} = -\dfrac{\partial A_i f_L^\ast}{\partial u_i^\ast} + \dfrac{1}{2} C_0 \left( \omega^\ast - \om \right) k \dfrac{\partial^2 f_L^\ast}{\partial u_i^\ast \partial u_i^\ast} \\
            - \dfrac{\partial A_e f_L^\ast}{\partial e^\ast} + \dfrac{1}{2} C_1 \left( \omega^\ast - \om \right) k_e \dfrac{\partial^2 f_L^\ast}{\partial e^\ast \partial e^\ast} ~.
            \label{eq:delta_fokker_planck_general}
        \end{multline}
        Since we want the velocity-energy part of the Fokker-Planck equation to remain unaffected by the modification of $b_u$ and $b_e$, this must be zero. In order to determine the expressions of $A_i$ and $A_e$ leading to $\Delta \mathrm{FP} = 0$, we will make the assumption that the local, one-time, marginal $(e'', \mathbf{u''})$ PDF is joint-normal, with zero mean and a covariance matrix
        \begin{equation}
            \mathcal{C} =
            \begin{bmatrix}
                k_e & \fe_x & \fe_y & \fe_z \\
                \fe_x & R_{xx} & R_{xy} & R_{xz} \\
                \fe_y & R_{xy} & R_{yy} & R_{yz} \\
                \fe_z & R_{xz} & R_{yz} & R_{zz}
            \end{bmatrix}
            ~.
            \label{eq:covariance_matrix}
        \end{equation}
        Then we have
        \begin{equation}
            \dfrac{\partial f_L^\ast}{\partial X_i} = -\cmat_{ij} X_j f_L^\ast ~,
        \end{equation}
        where $\mathbf{X} \equiv (e'', \mathbf{u}'')$, and Eq.~\ref{eq:delta_fokker_planck_general} becomes
        \begin{multline}
            \Delta \mathrm{FP} = -\dfrac{\partial}{\partial u_i} \left( A_i f_L^\ast + \dfrac{1}{2} C_0 \left( \omega^\ast - \om \right) k \cmat_{ij} X_j f_L^\ast \right) \\
            - \dfrac{\partial}{\partial e} \left( A_e f_L^\ast + \dfrac{1}{2} C_1 \left( \omega^\ast - \om \right) k_e \cmat_{0j} X_j f_L^\ast \right) ~.
        \end{multline}
        The condition that this must be zero yields the expression of the extra terms in the velocity and energy SDEs
        \begin{align}
            & A_i = - \dfrac{1}{2} C_0 \left( \omega^\ast - \om \right) k \left( \cmat_{i0} \left( e^\ast - \enm \right) + \cmat_{ij} \left( u_j^\ast - \um_j \right) \right) ~, \\
            & A_e = - \dfrac{1}{2} C_1 \left( \omega^\ast - \om \right) k_e \left( \cmat_{00} \left( e^\ast - \enm \right) + \cmat_{0j} \left( u_j^\ast - \um_j \right) \right) ~.
        \end{align}
        These four extra terms are added respectively to the expressions of $G_{ie}$, $G_{ij}$, $K_e$ and $K_i$ given in App.~\ref{app:lagrangian_coefficients}.

\section{Corrective SPH scheme \label{app:corrected_SPH}}

    The corrective Smoothed Particle Method (CSPM) adopted in our code works in three steps. For a given location $\mathbf{x}_i$ within the domain, we first compute the following sums through the standard SPH scheme Eq.~\ref{eq:SPH_Reynolds_average}
    \begin{align}
        & W = \overline{1} ~, \qquad W_\alpha = \overline{1}^\alpha ~, \qquad W_{\alpha\beta} = \overline{1}^{\alpha\beta} ~, \\
        & X = \overline{x} ~, \qquad X_\alpha = \overline{x}^\alpha ~, \qquad X_{\alpha\beta} = \overline{x}^{\alpha\beta} ~, \\
        & Y = \overline{y} ~, \qquad Y_\alpha = \overline{y}^\alpha ~, \qquad Y_{\alpha\beta} = \overline{y}^{\alpha\beta} ~, \\
        & XX = \overline{x^2} ~, \qquad XX_\alpha = \overline{x^2}^\alpha ~, \qquad XX_{\alpha\beta} = \overline{x^2}^{\alpha\beta} ~, \\
        & XY = \overline{xy} ~, \qquad XY_\alpha = \overline{xy}^\alpha ~, \qquad XY_{\alpha\beta} = \overline{xy}^{\alpha\beta} ~, \\
        & YY = \overline{y^2} ~, \qquad YY_\alpha = \overline{y^2}^\alpha ~, \qquad YY_{\alpha\beta} = \overline{y^2}^{\alpha\beta} ~,
    \end{align}
    where
    \begin{align}
        & \overline{Q} \equiv \sum_{j=1}^N \mathcal{V}^{\ast~(j)} Q^{\ast~(j)} K\left(\mathbf{x}_j - \mathbf{x}_i\right) ~, \label{eq:app_mean_Q} \\
        & \overline{Q}^\alpha \equiv \sum_{j=1}^N \mathcal{V}^{\ast~(j)} Q^{\ast~(j)} \left. \dfrac{\partial K}{\partial x_\alpha} \right|_{\mathbf{x}_j - \mathbf{x}_i} ~, \label{eq:app_grad_Q} \\
        & \overline{Q}^{\alpha\beta} \equiv \sum_{j=1}^N \mathcal{V}^{\ast~(j)} Q^{\ast~(j)} \left. \dfrac{\partial^2 K}{\partial x_\alpha \partial x_\beta} \right|_{\mathbf{x}_j - \mathbf{x}_i} ~. \label{eq:app_hess_Q}
    \end{align}
    They only need to be computed once per time step. In parallel, we compute the mean of the quantity that we want to average, as well as its gradient and second order derivatives, through the standard SPH scheme Eqs.~\ref{eq:app_mean_Q}, \ref{eq:app_grad_Q} and \ref{eq:app_hess_Q}. They need to be computed separately for each mean flow quantity.

    This first step requires the kernel estimation of means that are not density-weighted, so that we need to estimate the lumped volume $\vpart$ occupied by each notional particle. To do this, we do a first estimation of the mean density $\rhom^{(k)}$ at each particle location, and we assign them the volume $\vpart = \Delta m / \rhom^{(k)}$.

    In a second step, we correct the mean and gradient through
    \begin{equation}
        \begin{bmatrix}
            \overline{Q}_\mathrm{CSPM} \\
            \overline{Q}^x_\mathrm{CSPM} \\
            \overline{Q}^y_\mathrm{CSPM}
        \end{bmatrix}
        = M_\mathrm{grad}^{-1}
        \begin{bmatrix}
            \overline{Q} \\
            \overline{Q}^x \\
            \overline{Q}^y
        \end{bmatrix}~,
    \end{equation}
    where
    \begin{equation}
        M_\mathrm{grad} =
        \begin{bmatrix}
            W & X - x_i W & Y - y_i W \\
            W_x & X_x - x_i W_x & Y_x - y_i W_x \\
            W_y & X_y - x_i W_y & Y_y - y_i W_y
        \end{bmatrix} ~.
    \end{equation}

    In a third step, we correct the hessian matrix of $\overline{Q}$ through
    \begin{equation}
        \begin{bmatrix}
            \overline{Q}^{xx}_\mathrm{CSPM} \\
            \overline{Q}^{xy}_\mathrm{CSPM} \\
            \overline{Q}^{yy}_\mathrm{CSPM}
        \end{bmatrix}
        = M_\mathrm{hess}^{-1}
        \begin{bmatrix}
            \overline{Q}^{xx} - \overline{Q}_\mathrm{CSPM} W_{xx} - \overline{Q}^x_\mathrm{CSPM} \d X_{xx} - \overline{Q}^y_\mathrm{CSPM} \d Y_{xx} \\
            \overline{Q}^{xy} - \overline{Q}_\mathrm{CSPM} W_{xy} - \overline{Q}^x_\mathrm{CSPM} \d X_{xy} - \overline{Q}^y_\mathrm{CSPM} \d Y_{xy} \\
            \overline{Q}^{yy} - \overline{Q}_\mathrm{CSPM} W_{yy} - \overline{Q}^x_\mathrm{CSPM} \d X_{yy} - \overline{Q}^y_\mathrm{CSPM} \d Y_{yy}
        \end{bmatrix}~,
    \end{equation}
    where
    \begin{multline}
        M_\mathrm{hess} = \dfrac{1}{2}
        \begin{bmatrix}
            \d XX_{xx} & \d XY_{xx} & \d YY_{xx} \\
            \d XX_{xy} & \d XY_{xy} & \d YY_{xy} \\
            \d XX_{yy} & \d XY_{yy} & \d YY_{yy}
        \end{bmatrix}
        \\
        - \dfrac{1}{2}
        \begin{bmatrix}
            W_{xx} & \d X_{xx} & \d Y_{xx} \\
            W_{xy} & \d X_{xy} & \d Y_{xy} \\
            W_{yy} & \d X_{yy} & \d Y_{yy}
        \end{bmatrix}
        M_\mathrm{grad}^{-1}
        \begin{bmatrix}
            \d XX & \d XY & \d YY \\
            \d XX_x & \d XY_x & \d YY_x \\
            \d XX_y & \d XY_y & \d YY_y
        \end{bmatrix} ~,
    \end{multline}
    and we have defined
    \begin{align}
        & \d X_\alpha = X_\alpha - x_i W_\alpha ~, \\
        & \d Y_\alpha = Y_\alpha - y_i W_\alpha ~, \\
        & \d XX_\alpha = XX_\alpha - 2 x_i X_\alpha + x_i^2 W_\alpha ~, \\
        & \d XY_\alpha = XY_\alpha - x_i Y_\alpha - y_i X_\alpha + x_i y_i W_\alpha ~, \\
        & \d YY_\alpha = YY_\alpha - 2 y_i Y_\alpha + y_i^2 W_\alpha ~.
    \end{align}

\end{appendix}

\end{document}